%% file: main.tex
\documentclass[10pt,journal,compsoc]{IEEEtran}
\ifCLASSOPTIONcompsoc
  \usepackage[nocompress]{cite}
\else
  \usepackage{cite}
\fi
\ifCLASSINFOpdf
   \usepackage[pdftex]{graphicx}
   \graphicspath{{./img/}}
\else
   \usepackage[dvips]{graphicx}
   \graphicspath{{../img/}}
\fi
\usepackage{caption}
\usepackage{subcaption}

\usepackage{amsmath,amssymb,amsfonts, mathrsfs}
\usepackage[ruled,vlined]{algorithm2e}
\usepackage{framed}

\usepackage{url}
\usepackage{hyperref}

\newboolean{showcomments}

\usepackage[table]{xcolor}
\usepackage{booktabs}
\usepackage{xspace}

\newcommand{\secref}[1]{Section~\ref{#1}\xspace}
\newcommand{\figref}[1]{Fig.~\ref{#1}\xspace}
\newcommand{\algoref}[1]{Algorithm ~\ref{#1}\xspace}
\newcommand{\eqqref}[1]{Equation ~(\ref{#1})\xspace}
\newcommand{\toolName}{DeLag\xspace}
\newcommand{\tabref}[1]{Table~\ref{#1}\xspace}

\setboolean{showcomments}{true}

\ifthenelse{\boolean{showcomments}}
  {\newcommand{\nb}[2]{
    \fbox{\bfseries\sffamily\scriptsize#1}
    {\sf\small$\blacktriangleright$\textit{#2}$\blacktriangleleft$}
   }
  }
  {\newcommand{\nb}[2]{}
 }

\newcommand{\ie}{\textit{i.e.,}\xspace}
\newcommand{\eg}{\textit{e.g.,}\xspace}
\newcommand{\etc}{\textit{etc.}\xspace}
\newcommand{\etal}{\textit{et al.}\xspace}

\begin{document}
%
\title{\toolName: Using Multi-Objective Optimization to Enhance the Detection of Latency Degradation Patterns in Service-based Systems
\thanks{This is a pre-copy-editing, author-produced version of an article accepted for publication in IEEE Transactions on Software Engineering (TSE).
}
\thanks{Final authenticated version available at \href{https://doi.org/10.1109/TSE.2023.3266041}{DOI:10.1109/TSE.2023.3266041}.}
\thanks{Accepted: 6 April 2023.}
}
%
%
%
%

\author{Luca~Traini
        and~Vittorio~Cortellessa
\IEEEcompsocitemizethanks{\IEEEcompsocthanksitem L. Traini and V. Cortellessa are with the Department of Information Engineering, Computer Science and Mathematics, University of L’Aquila, L’Aquila,
Italy.\protect\\
E-mail: luca.traini@univaq.it, vittorio.cortellessa@univaq.it}
}

\IEEEtitleabstractindextext{%
\begin{abstract}
Performance debugging in production is a fundamental activity in modern service-based systems.
The diagnosis of performance issues is often time-consuming,
since it requires thorough inspection of large volumes of traces and performance indices.   
In this paper we present \toolName, a novel automated search-based approach for diagnosing performance issues in service-based systems.
\toolName identifies subsets of requests that show, in the combination of their Remote Procedure Call execution times,
symptoms of potentially relevant performance issues.
We call such symptoms \emph{Latency Degradation Patterns}.
\toolName simultaneously searches for multiple \emph{latency degradation patterns} while optimizing precision, recall and latency dissimilarity.
Experimentation on 700 datasets of requests generated from two microservice-based systems shows that our approach provides better and more stable effectiveness than three state-of-the-art approaches and general purpose machine learning clustering algorithms.
\toolName is more effective than all baseline techniques in at least one case study (with $p\leq 0.05$ and non-negligible effect size).
Moreover, \toolName outperforms in terms of efficiency the second and the third most effective baseline techniques on the largest datasets used in our evaluation (up to 22\%).
\end{abstract}

}

\maketitle

\IEEEdisplaynontitleabstractindextext

%
\IEEEpeerreviewmaketitle

\input{introduction}
\input{latency_degradation_patterns}
\input{usage_scenario}
\input{model}
\input{approach}
\input{evaluation}
\input{threats}

\subsection{Data availability}
To aid reproducibility we provide the data and scripts needed to replicate our findings \cite{replication}.
\input{discussion}
\input{related_work}
\input{conclusion}


%


\ifCLASSOPTIONcompsoc
  \section*{Acknowledgments}
\else
  \section*{Acknowledgment}
\fi

This work is supported by ``Territori Aperti'' (a project funded by Fondo Territori Lavoro e Conoscenza CGIL, CSIL and UIL), and by European Union - NextGenerationEU - National Recovery and Resilience Plan (Piano Nazionale di Ripresa e Resilienza, PNRR) - Project: ``SoBigData.it - Strengthening the Italian RI for Social Mining and Big Data Analytics'' - Prot. IR0000013 - Avviso n. 3264 del 28/12/2021. 

\ifCLASSOPTIONcaptionsoff
  \newpage
\fi



%
\bibliographystyle{IEEEtran}
\bibliography{references}

%

\begin{IEEEbiography}
[{\includegraphics[width=1in,height=1.25in,clip,keepaspectratio]{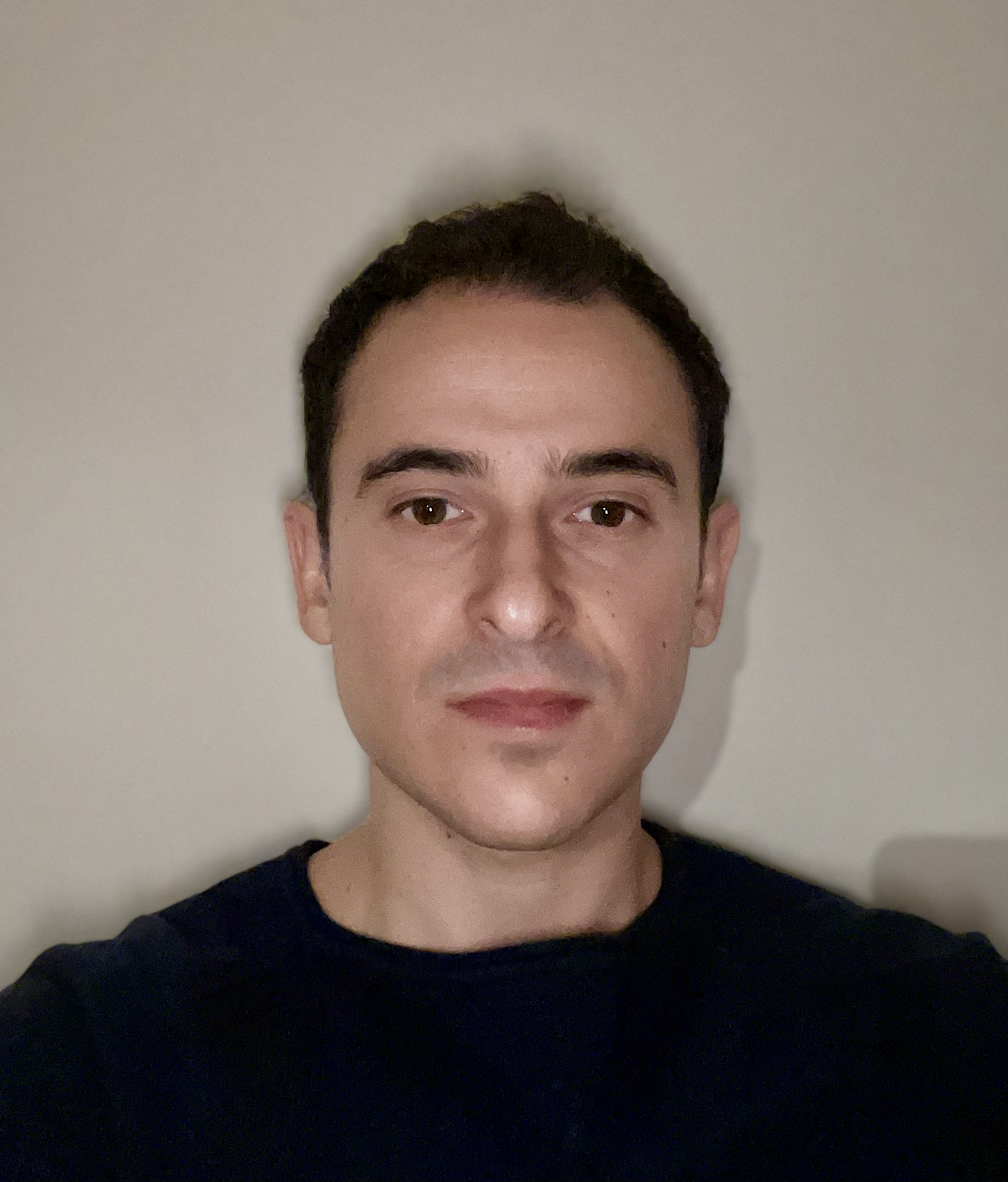}}]
{Luca Traini}
is a postdoctoral researcher at the Department of Computer Science and Engineering, and Mathematics of University of L’Aquila, and member of the SPENCER (Software PErformaNCe EngineeRing) laboratory. His research interests centre around software performance engineering, encompassing both human and technical aspects, with the goal of improving techniques and methodologies for software performance assurance. His current research is focused on performance assurance processes, performance testing and debugging, and the interplay between software maintenance and performance.\\
More information at \url{https://lucatraini.me}.

\end{IEEEbiography}

\begin{IEEEbiography}
[{\includegraphics[width=1in,height=1.25in,clip,keepaspectratio]{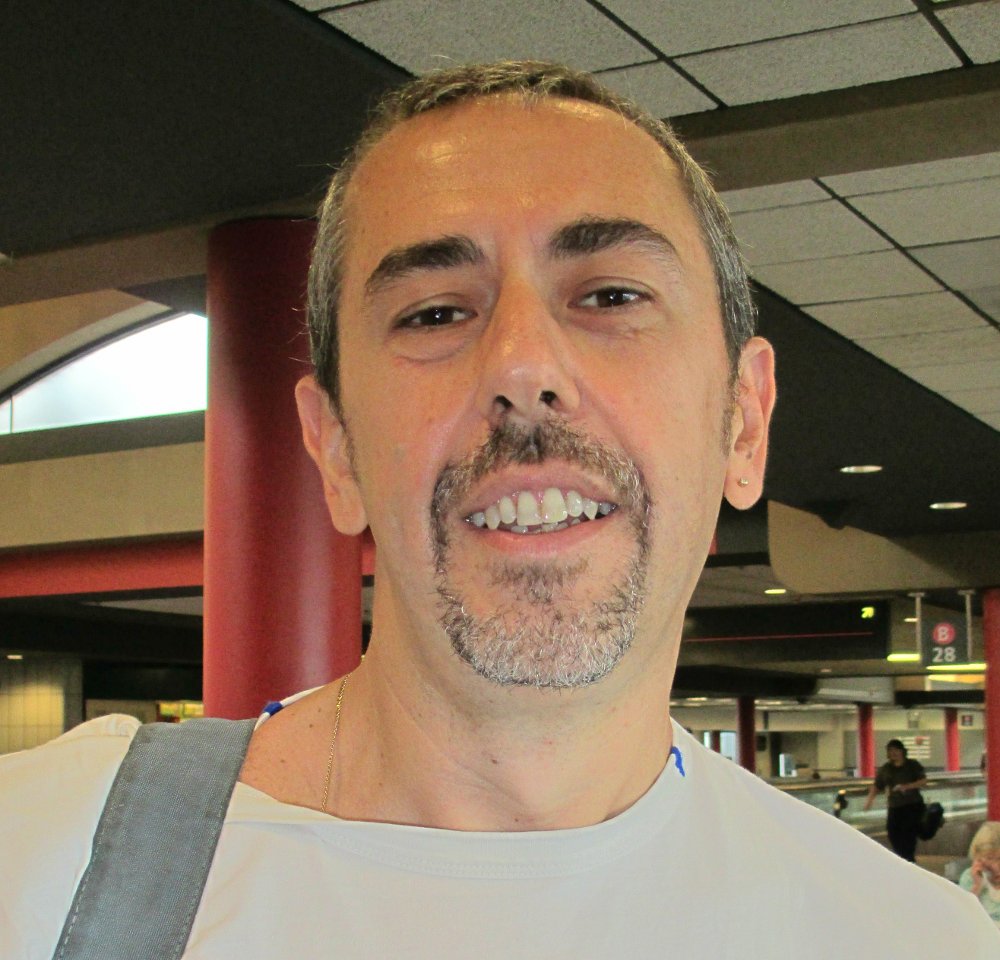}}]
{Vittorio Cortellessa}
Vittorio Cortellessa is a professor at the Department of Computer Science and Engineering, and Mathematics of University of L’Aquila, and Chair of the SPENCER (Software PErformaNCe EngineeRing) laboratory. He has received his Ph.D. degree in Computer Science at University of Roma Tor Vergata. He was postdoc at European Space Agency, and Research Assistant Professor at West Virginia University.
His main research interests are in the areas of Software Performance, Software Reliability, and Model-Driven Engineering.\\
More information at \url{http://people.disim.univaq.it/cortelle}.
\end{IEEEbiography}





\end{document}

%% file: introduction.tex
\IEEEraisesectionheading{\section{Introduction}\label{sec:introduction}}
\IEEEPARstart{M}{odern} high-tech companies deliver new software in production every day \cite{Feitelson2013} and perceive this capability as a key competitive advantage.
In order to support this fast-paced release cycle, IT organizations often employ several independent teams that are responsible ``from development to deploy'' \cite{OHanlon2006} of loosely coupled independently deployable services.
Unfortunately, frequent software releases often hamper the ability to deliver high quality software \cite{Rubin2016}. 
For example, widely used performance assurance techniques, like load testing \cite{Jiang2015}, are often too time-consuming for these contexts~\cite{traini2022}.
Also, given the complexity of these systems and their workloads \cite{Ardelean2018}, it is often unfeasible to proactively detect performance issues in a testing environment \cite{Veeraraghavan2016}.
For these reasons, today, the diagnosis of performance issues in production is a fundamental capability for maintaining high-quality service-based systems.

Service owners are usually responsible and accountable for meeting Service Level Objectives (SLOs) on Key Performance Indicators (KPIs).
Software engineers and performance analysts continuously monitor KPIs and execution traces on the run-time system to identify symptoms of potentially relevant performance issues that lead to SLOs violations.
The truly identification of such symptoms is often critical:
a request may involve several Remote Procedure Calls (RPC) and 
the number of performance traces and performance metrics to analyze can be huge.
According to a recent study on microservice-based systems \cite{Zhou2018}, software engineers spend days or even weeks to debug a software issue, 
and initial understanding, scoping and localization are among the most time-consuming phases during debugging.
Although several techniques have been introduced to provide automation in diagnosing performance issues in service-based systems \cite{Cohen2004,Malik2010,Kim2013,Nair2015,Krushevskaja2013,Cortellessa2020,Bansal2019}, the reduction of the manual effort and the time needed is still critical.

Techniques for automating performance issue diagnosis
rely on pattern mining to spot patterns in trace attributes (\eg request size, response size, RPCs execution times) correlated to latency degradation of requests.
The benefit provided by these techniques is threefold: 1) they provide evidences based on data on the existence of relevant performance issues, 2) they reduce the amount of traces to inspect and, 3) they provide useful information to effectively localize and debug performance issues.

Prior work on automated anomaly detection relies on association rule mining \cite{Brauckhoff2012, Han2012}.
These techniques extract patterns in traces that are correlated to anomalous system executions by using well known association rule mining algorithms \cite{Agrawal1993, Agrawal1994}.
Unfortunately, these techniques do not work with continuous data, which are highly relevant for the detection of performance issues (\eg execution time).
%
More recent techniques \cite{Krushevskaja2013, Cortellessa2020, Bansal2019} enable the detection of patterns correlated with latency degradation also on continuous trace attributes, for example, by using \emph{F1-score} optimization.
\emph{F1-score}-based techniques \cite{Krushevskaja2013, Cortellessa2020} search for the optimal partition of the latency range considered as degraded, while searching for each sub-interval the pattern that maximizes F1-score.
These techniques work fairly good as far as performance issues lead to clearly different latency behaviors,
but when they cause (partially or entirely) overlapping latency distributions, effectiveness may potentially decrease \cite{Krushevskaja2013}.
Other techniques suitable for detecting patterns in continuous trace attributes rely on machine learning techniques such as tree-augmented bayesian networks \cite{Cohen2004} or random forests \cite{Bansal2019}, as well as on clustering approaches \cite{Duan2009,Farshchi2015}.

Some of the previous techniques work with any types of trace attributes (either continuous or categorical) \cite{Krushevskaja2013, Bansal2019}, while others target only specific types of attributes such as categorical ones \cite{Yairi2001, Brauckhoff2012} or execution time \cite{Cortellessa2020}.
In this work, we explicitly target execution time of Remote Procedure Calls (RPC), given its relevance for the diagnosis of performance issues in service-based systems.
In particular, we focus on \emph{Latency Degradation Patterns} (LDPs), a concept we introduced in our recent work~\cite{Cortellessa2020}. In~\cite{Cortellessa2020} we defined LDPs as RPCs execution time behaviors that are likely to be related to relevant performance issues, and we proposed a \emph{F1-score}-based technique which aims at automatically identifying them.

In this paper we present \toolName: a novel approach that simultaneously searches for multiple LDPs.
\toolName searches for a whole set of patterns by maximizing precision and recall, and by simultaneously minimizing latency dissimilarity.
The optimal \emph{pattern set} is then selected from non-dominated Pareto-optimal solutions by using a decision making heuristic.

We evaluated \toolName on 700 datasets involving different combinations of LDPs from two service-based systems and compared \toolName to three state-of-the-art techniques for pattern detection in execution traces \cite{Krushevskaja2013, Cortellessa2020, Bansal2019} and two general-purpose clustering algorithms.
Datasets are generated by performing load testing sessions while injecting different performance issues on systems (hence different LDPs), for an overall time of $\sim$15.5 days\footnote{Note that 15.5 days only include the time needed to generate datasets, while they do not include the time spent to run DeLag and baseline techniques.}.
We found that \toolName provides better and more stable effectiveness than other approaches.
\toolName outperforms in terms of effectiveness all the baseline techniques in at least one case study (with $p\leq 0.05$ and non-negligible effect size),
and the effectiveness provided by \toolName is more stable than those provided by other techniques (the interquartile range for F1-scores of \toolName is smaller than those of other techniques).
Additionally, we found that \toolName effectiveness is not affected by similarity of latency distributions related to different patterns (contrariwise to F1-score-based techniques),
and it is also not affected by RPC execution time variation not correlated with latency degradation.
Moreover, we found that \toolName is more efficient on the largest datasets used in our evaluation than the second most effective technique (by 15\% in the first case study and by 22\% in the second case study) and the third most effective technique (by 15\% and 17\% respectively).\\

\emph{Novelty with respect to our previous work.} This work cannot be considered an extension of our previous work~\cite{Cortellessa2020}, though they both target the same problem.
Indeed, when compared to our prior technique, \toolName introduces a novel modeling approach, a new workflow, and a different set of search algorithms. In the following, we report in detail the main differences between \toolName and the technique introduced in our previous work~\cite{Cortellessa2020} (which we name here as CoTr):
\begin{itemize}
	\item CoTr searches for the optimal partition of the latency range while seeking the optimal pattern for each sub-interval, instead \toolName simultaneously searches multiple patterns for the whole latency range;
	\item \toolName models the problem of detecting LDPs using a novel multi-objective problem formulation, while CoTr reuses the single-objective formulation proposed by Krushevskaja and Sandler~\cite{Krushevskaja2013};
	\item CoTr solves the optimization problem using a single-objective genetic algorithm combined with dynamic programming, while \toolName uses a multi-objective genetic algorithm combined with a decision-making heuristic.
\end{itemize}

Besides these differences, in this paper we present an evaluation that is only barely comparable to the one presented in~\cite{Cortellessa2020}. Indeed, in the latter, CoTr is evaluated on 60 datasets generated from one system, while, in the former, \toolName is evaluated on 700 datasets generated from two systems including (to the best of our knowledge) the largest and most complex existing open-source microservice system. In addition, the datasets used for the evaluation are extremely different both in terms of involved workloads and sizes. The datasets used in our previous work involve only $\sim$1k requests, and they are generated from load tests that involve just one user. In this paper, instead, we use datasets with sizes ranging from $\sim$2.5k requests to $\sim$90.5k requests, which are generated from non-trivial workloads that involve 20 parallel users in an accelerated scenario.
Finally, in this paper we compare \toolName against the techniques considered in our previous work\footnote{We excluded here Mean shift algorithm from our evaluation because it has shown extremely poor effectiveness for LDPs detection~\cite{Cortellessa2020}. It is worth to notice that, although we exclude Mean Shift as baseline, we still use it as part of \toolName and CoTr to discretize RPC execution times (see \secref{sec:ssc} for details).} including CoTr, plus a recent state-of-the-art technique proposed by Bansal~\etal~\cite{Bansal2019}. 
The comparison with CoTr showed that \toolName is more effective in detecting LDPs, by providing better F1-scores with statistical significance ($p<0.05$) in both case study systems (with large and negligible effect sizes, respectively). Also, it is more resilient to overlapping LDPs than our previous approach, \ie closeness of latency distribution related to different patterns does not affect the effectiveness of \toolName. Finally, \toolName proved to be more efficient than CoTr on the largest datasets used in our evaluation by 15\% and 17\%.
\\

%

\emph{Paper structure.} The rest of the article is structured as follows. \secref{sec:background} describes the concept of Latency Degradation Pattern, and how LDP detection can be used in practice to automate performance issues diagnosis.
\secref{sec:model} describes how the problem of detecting LDPs is modeled as multi-objective optimization problem.
\secref{sec:approach} outlines the workflow used by \toolName to detect LDPs.
In \secref{sec:evaluation} we present our research questions along with experimental design, results and threats to validity.
\secref{sec:discussion} discusses some implications of our findings.  
\secref{sec:related} presents related work, and \secref{sec:conclusion} concludes this paper.


%% file: latency_degradation_patterns.tex
\section{Background}\label{sec:background}
In this Section, we first introduce the concept of Latency Degradation Patterns, then we give an intuition on how LDP detection can be integrated in a software development process to enable the automated diagnosis of performance issues.

\subsection{Latency Degradation Patterns}\label{sec:LDP}
Services are often subject to SLO on latency depending on the type of request. For example, requests that load the homepage of a website may have a specific SLO, while those loading the login page may have another.
Usually, a SLO on latency defines a range of acceptable values for a specific type of request, \ie $L\leq L_{SLO}$.
In this paper, we name the range of latency values that do not meet SLO expectations as the \emph{targeted latency range}, \ie $L>L_{SLO}$.

A request to a service-based system often involves several RPCs.
Each request is associated to a set of execution trace attributes (\ie RPC execution time).
In this paper, we denote a request $r$ as an ordered sequence of trace attributes $r=(e_0, e_1, ..., e_m, L)$,
where $e_j$ represents the execution time of a specific RPC $j$ triggered by the request.

\emph{Latency Degradation Patterns} (LDPs) \cite{Cortellessa2020}  are patterns in RPCs execution times correlated with SLO violation.
They can be represented as conjunctions of predicates over RPCs execution time.
Conjunctions of predicates are used, instead of single predicates, because several software issues in service-based systems lie in the interaction of multiple RPCs \cite{Zhou2018} rather than being rooted in the internal implementation of individual RPCs. Moreover, a single predicate alone is often not sufficient to capture the patterns of SLO violations~\cite{Cohen2004}.

An informal example of LDP could be:
\begin{framed}
The homepage latency exceeds $L_{SLO}$ when $Auth$ execution time is greater than 30 milliseconds and $getProfile$ execution time is between 20 and 50 milliseconds.
\end{framed}

More formally, a pattern $P$ is denoted as a set of predicates  $\{p_0, p_1, ..., p_k\}$ with $k\geq 0$.
A request $r$ satisfies ($\vartriangleleft$) a pattern $P$
if every predicate $p\in P$ is satisfied by the request $r$:
\begin{equation*}
	\begin{aligned}
 		& r\vartriangleleft P \iff \forall p\in P, & r\vartriangleleft p
 	\end{aligned}
\end{equation*}
Each predicate targets a specific RPC $j$ and is denoted as a triple $p=\langle j, e_{min}, e_{max} \rangle$,
where $[e_{min}, e_{max})$ represents a range of values on the RPC execution time.

We say that a request $r=(..., e_j, ...)$ satisfies $p$, denoted as $r\vartriangleleft  p$ , if:
\begin{equation*}
 e_{min}\leq e_j< e_{max}
 \end{equation*}
Previous approaches use F1-score \cite{Krushevskaja2013 ,Cortellessa2020} to measure the degree of correlation between patterns and latency degradation.
The idea is to partition the set $R$ of requests under analysis in two subsets $R_{POS}$ and $R_{NEG}$,
 namely the set of requests not meeting SLO (or positives) and the set of requests meeting SLO (or negatives)
 \begin{equation}
	\begin{aligned}
		& R_{pos}=\{r\in R \mid L > L_{SLO} \}\\
		& R_{neg}=\{r\in R \mid L \leq L_{SLO} \}\\
	\end{aligned}
\end{equation}
and to compute F1-score for a pattern $P$ accordingly:
\begin{equation}\label{eq:fscore}
	F1\text{-}score=2\cdot\frac{precision\cdot recall}{precision+recall}
\end{equation}
where precision and recall of the pattern are defined as follows:
\begin{equation}\label{eq:prec}
	\begin{aligned}
		precision= &\frac{\mid tp \mid}{\mid tp \mid + \mid fp \mid}&\\\\
	\end{aligned}
\end{equation}
\begin{equation}\label{eq:rec}
	\begin{aligned}
		recall= & \frac{\mid tp \mid}{\mid R_{pos} \mid }&		
	\end{aligned}
\end{equation}
 and true positives $tp$ and false positives $fp$ are defined as:
 \begin{equation}\label{eq:patternsets_tp_fp}
	\begin{aligned}
		& tp=\{r\in R_{pos} \mid r\vartriangleleft P \} \\
		& fp=\{r\in R_{neg} \mid r\vartriangleleft P \}\\
	\end{aligned}
\end{equation}

If a pattern shows high recall then it frequently appears in requests with latency falling in the \emph{targeted latency range}.
But, it does not provide any guarantees on its infrequency in requests meeting SLO.
On the other hand, a high value of precision indicates that most of the requests satisfied by the pattern do not meet SLO expectations, 
but the number of the involved requests may be negligible and not worth to investigate.
F1-score, which is the harmonic mean of precision and recall, provides a unique measure to evaluate the quality of a pattern while keeping into consideration both these aspects.

%% file: usage_scenario.tex
\subsection{LDP Detection in Practice}\label{sec:scenario}

\toolName is designed to be agnostic to the type of service-based system and its implementation details, the only assumption is the presence of a distributed tracing infrastructure \cite{Sambasivan2016} (\eg Zipkin\footnote{Zipkin. \url{https://zipkin.io}} , Jaeger\footnote{Jaeger. \url{https://www.jaegertracing.io}}, Dapper~\cite{Sigelman2010}, \etc).
Distributed tracing infrastructure captures the workflow of causally-related events (\ie work done to process a request) within and among the components of a service-based system \cite{Sambasivan2016}. For example, a distributed \emph{trace} can include the RPCs invocation flow of a request along with the related performance metrics (\eg RPC execution time).
Nowadays, distributed tracing is widely used by modern high-tech companies \cite{mace2017survey}, and many popular microservice frameworks provide out-of-the box support for it (\eg Spring Cloud \footnote{Spring Cloud.  \url{https://spring.io/projects/spring-cloud}}, Istio \footnote{Istio. \url{https://istio.io}}). Therefore, we consider the adoption of distributed tracing solution as a reasonable assumption for a modern service-based system.

\toolName is thought to be periodically executed (\eg every day) to automate the detection of symptoms of relevant performance issues in RPCs execution time.
Following the example of DeCaf \cite{Bansal2019}, patterns derived by \toolName can be stored into a database along with their F1-scores (see \eqqref{eq:fscore}) to build historical knowledge and to automatically diagnose different categories of potential performance issues:
\begin{enumerate}
\item New: A new LDP is identified that has never appeared in the past.
\item Regressed: F1-score of the LDP is substantially increased compared to the past.
\item Known: F1-score of the LDP is similar to the recent one.
\item Improved: F1-score of the LDP is substantially decreased compared to the past.
\item Resolved: LDPs that were previously detected do not appear anymore.
\end{enumerate}

Note that the description of each one of these categories is intentionally broad.
Their concrete definition highly depends on the context (\eg system characteristics and service owners needs),
hence they are reported only to provide intuition on how \toolName can be used in practice.

%% file: model.tex
\section{Multi-objective optimization model}\label{sec:model}
\begin{figure}
\centering
\begin{subfigure}[h]{1\linewidth}
  	\includegraphics[width=\linewidth]{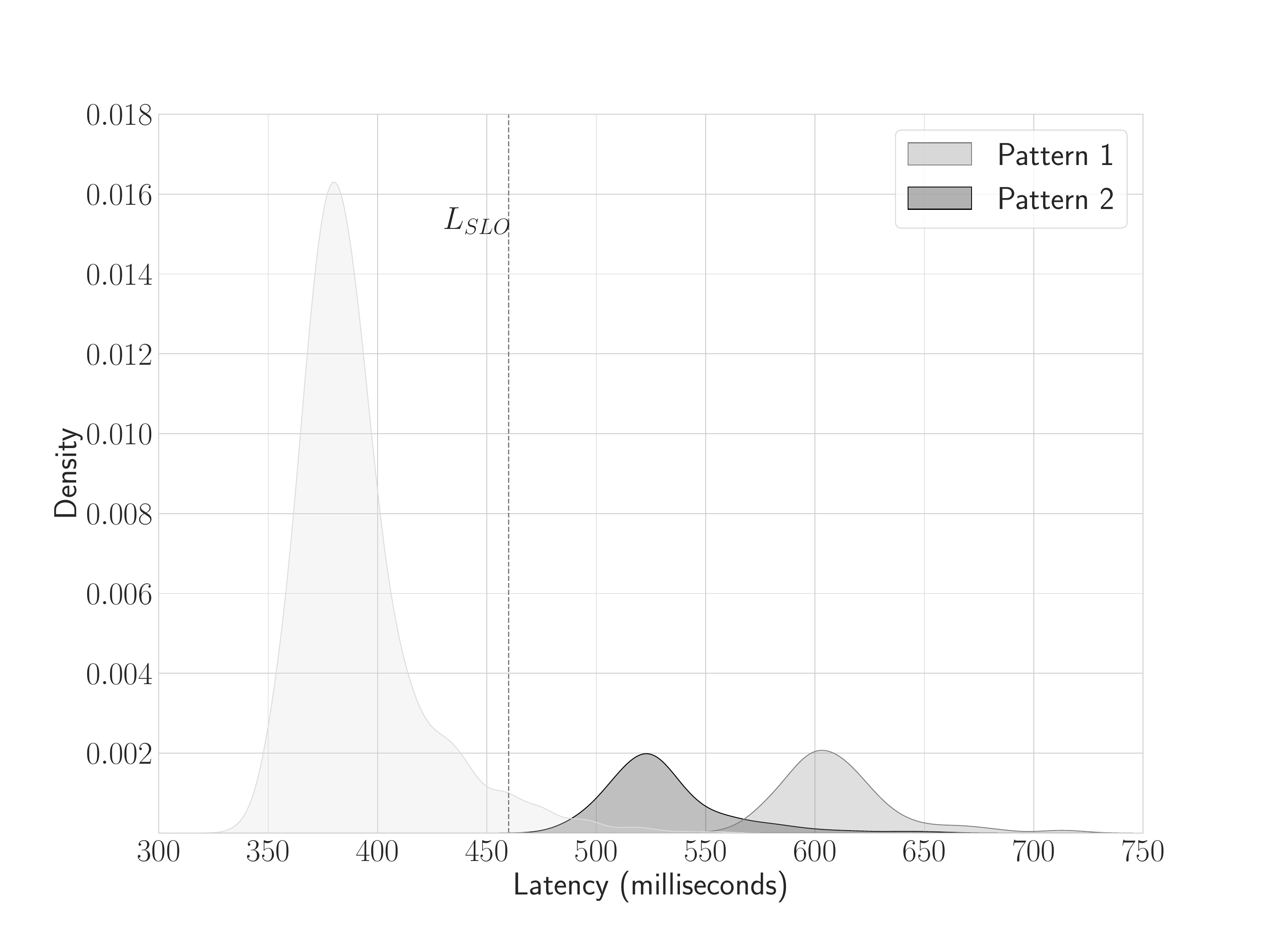}
  	\caption{Clearly separated LDPs}
  	\label{fig:separated_patterns}
\end{subfigure}

\begin{subfigure}[h]{1\linewidth}
  	\includegraphics[width=\linewidth]{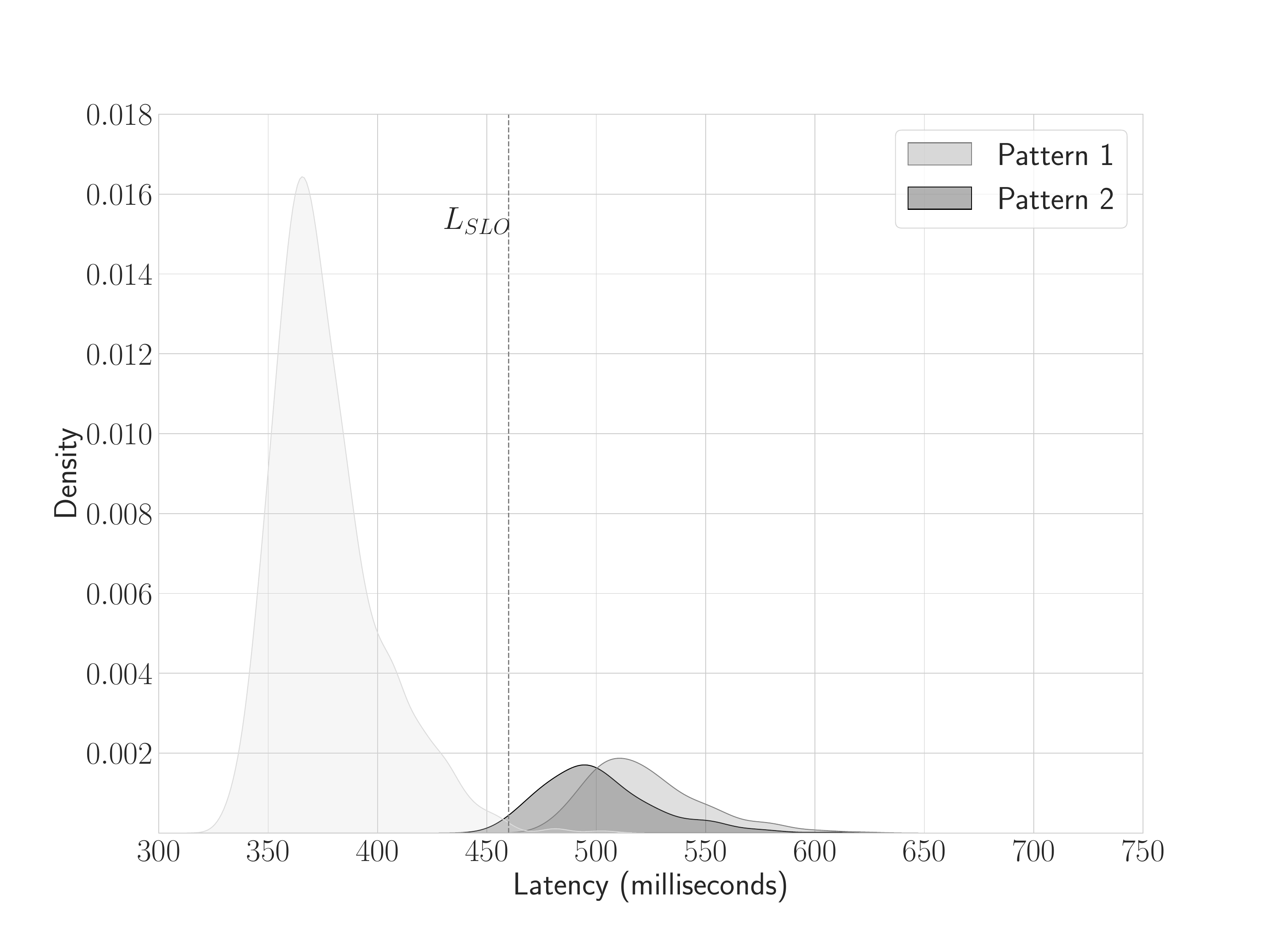}
  	\caption{Overlapping LDPs}
  	\label{fig:overlapping_patterns}
\end{subfigure}
\caption{Two different scenarios of request latency distribution with two LDPs}
\end{figure}

Existing approaches based on F1-Score optimization \cite{Cortellessa2020, Krushevskaja2013} search for optimal partitions of the \emph{targeted latency range} while considering a single optimal pattern for each sub-interval.
The sum of F1-scores is maximized in order to get the optimal set of patterns.
Although this technique works properly in situations where different performance issues lead to clearly distinguishable latency behaviors,
its  effectiveness decreases when latency degradations introduced by different issues are similar.
For example, \figref{fig:separated_patterns}
shows a scenario with two performance issues leading to two clearly separated latency distributions.
This is the ideal scenario for F1-score-based approaches, since the targeted latency range can be divided in a way that clearly separates the LDPs (\eg pattern $2$ for the $(460ms, 550ms)$ sub-interval and pattern $1$ for $(550ms, 750ms)$). 
However, there may be cases where it is difficult to partition the targeted latency range so that patterns are clearly separated (\eg \figref{fig:overlapping_patterns}).
This limitation was also highlighted in the work of Krushevskaja and Sandler \cite{Krushevskaja2013} by showing that, the more latency distributions (related to different patterns) are close one to another, the more the effectiveness of the approach decreases.

Our approach overcomes the latter problem by simultaneously searching multiple LDPs for the entire \emph{targeted latency range}.

In this section, we describe how we model the problem of detecting LDPs as a multi-objective optimization problem. First, we define the search space of our optimization problem. Then, we describe our optimization objectives.

\subsection{Search Space}\label{sec:ss}
\toolName simultaneously searches multiple patterns for the entire \emph{targeted latency range},
therefore each possible set of patterns $S=\{ P_1, P_2, ..., P_n\}$ is considered as a \emph{solution}.\\
As described in \secref{sec:LDP}, a pattern is a set of predicates  $P=\{ p_1, p_2, ..., p_m\}$ and each predicate is a triple $p=\langle j, e_{min}, e_{max}\rangle$ where $[e_{min}, e_{max})$ defines the execution time range for the RPC $j$.
RPC execution time is a continuous value, thus $e_{min}$ and $e_{max}$ can assume a wide range of possible values.
In order to exclude, from our search space, solutions with near-similar predicates as well as unrelevant predicates (\ie related to rare execution time behaviors), we identify (through clustering method), for each RPC $j$, a set of eligible values $E_j$.
Therefore, each predicate $p=\langle j, e_{min}, e_{max}\rangle$ in the solution space must be such that $e_{min}\in E_j$ and $e_{max}\in E_j$.
$E_j$, for a given RPC $j$, is defined by selecting values in the RPC execution time range that separate dense regions of the execution time distribution.
For example, a plausible set for the execution time of the RPC $Auth$, showed in \figref{fig:rpc_dist}, could be $E_{Auth}=\{ 25, 175, 250, 350\}$.\\
The key intuition of this search space reduction is that it allows to consider only patterns related to relevant RPC execution time behaviors,
while excluding from the search space those patterns related to rare transient execution time behaviors,
as well as patterns that are similar in terms of RPC execution time behavior.

\subsection{Optimization Objectives}\label{sec:obj}
\toolName optimizes \emph{pattern sets} by simultaneously maximizing \emph{precision} and \emph{recall}, and by minimizing \emph{latency dissimilarity}.

In \secref{sec:LDP}, we defined precision and recall to measure the quality of a single pattern, and
in the following we adapt these measures to a whole \emph{pattern set}.\\
We say that a request $r$ satisfies a \emph{pattern set} $S$ if at least one pattern $P$ in $S$ is satisfied by $r$:
\begin{equation*}
	\begin{aligned}
 		& r\vartriangleleft S \iff \exists P\in S, & r\vartriangleleft P
 	\end{aligned}
\end{equation*}
It is worth noting that a request $r$ can satisfy multiple patterns in the set $S$.
Nevertheless, we minimize the number of requests satisfied by multiple patterns by minimizing \emph{latency dissimilarity}, as it will be detailed later in this section.

True positives and false positives for the pattern set $S$ can be defined as follows:
 \begin{equation}
	\begin{aligned}
		& tp=\{r\in R_{pos} \mid r\vartriangleleft S \} \\
		& fp=\{r\in R_{neg} \mid r\vartriangleleft S \}\\
	\end{aligned}
\end{equation}

\textbf{Precision} measures the proportion of requests satisfied by $S$ having latency above $L_{SLO}$ (\ie $R_{pos}$),
 as described in \eqqref{eq:prec}.

\textbf{Recall}, instead, measures the proportion of requests that do not meet SLO (\ie $R_{pos}$) and satisfy $S$,
as described in \eqqref{eq:rec}.

\begin{figure}
\centering
\begin{subfigure}[h]{1\linewidth}
  	\includegraphics[width=\linewidth]{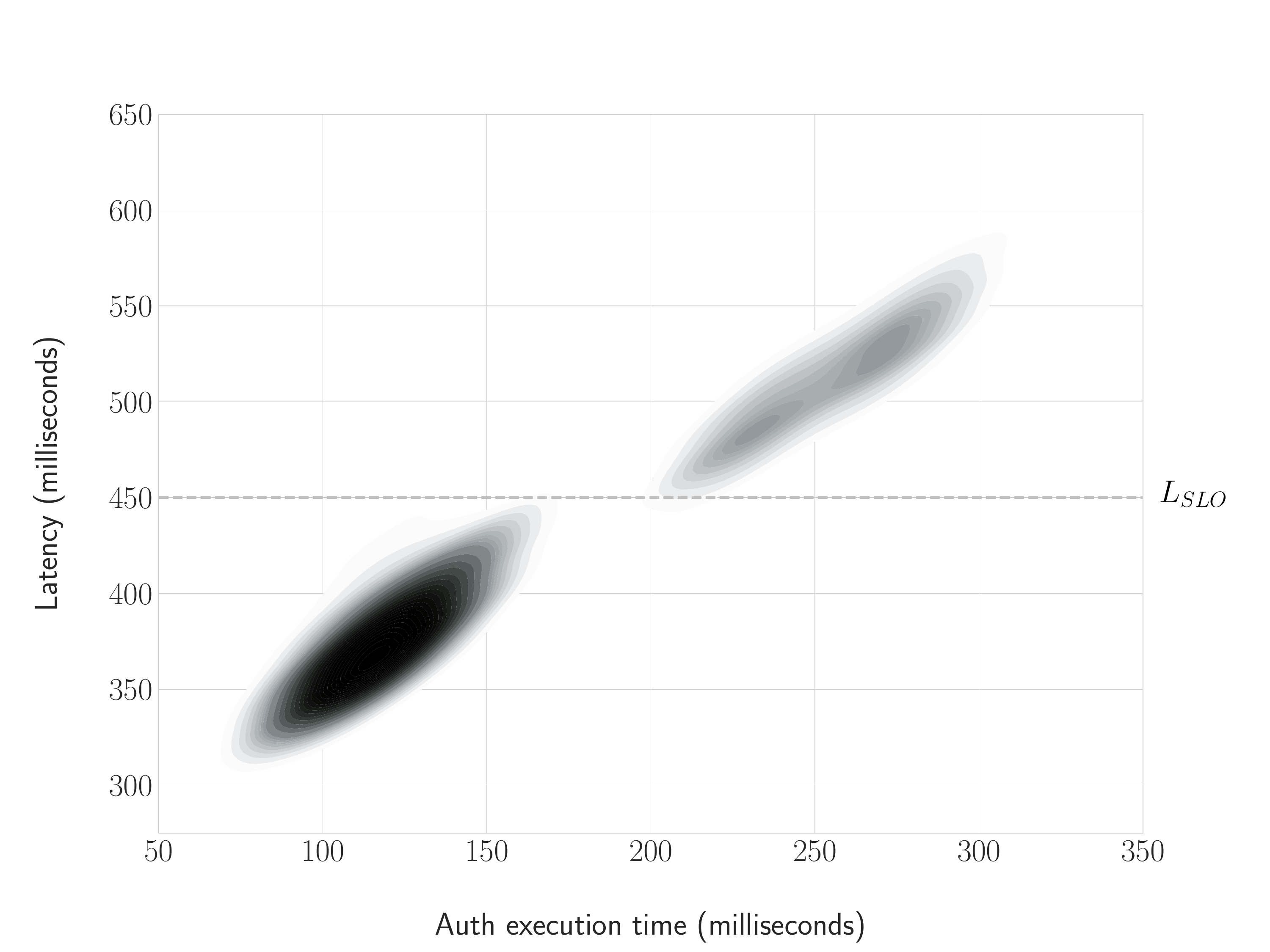}
  	\caption{Bivariate distribution of request latency and Auth execution time.
  	Darker parts of plot denote higher density.}
  	\label{fig:rpc_vs_lat}
\end{subfigure}
\hfill
\begin{subfigure}[h]{1\linewidth}
  	\includegraphics[width=\linewidth]{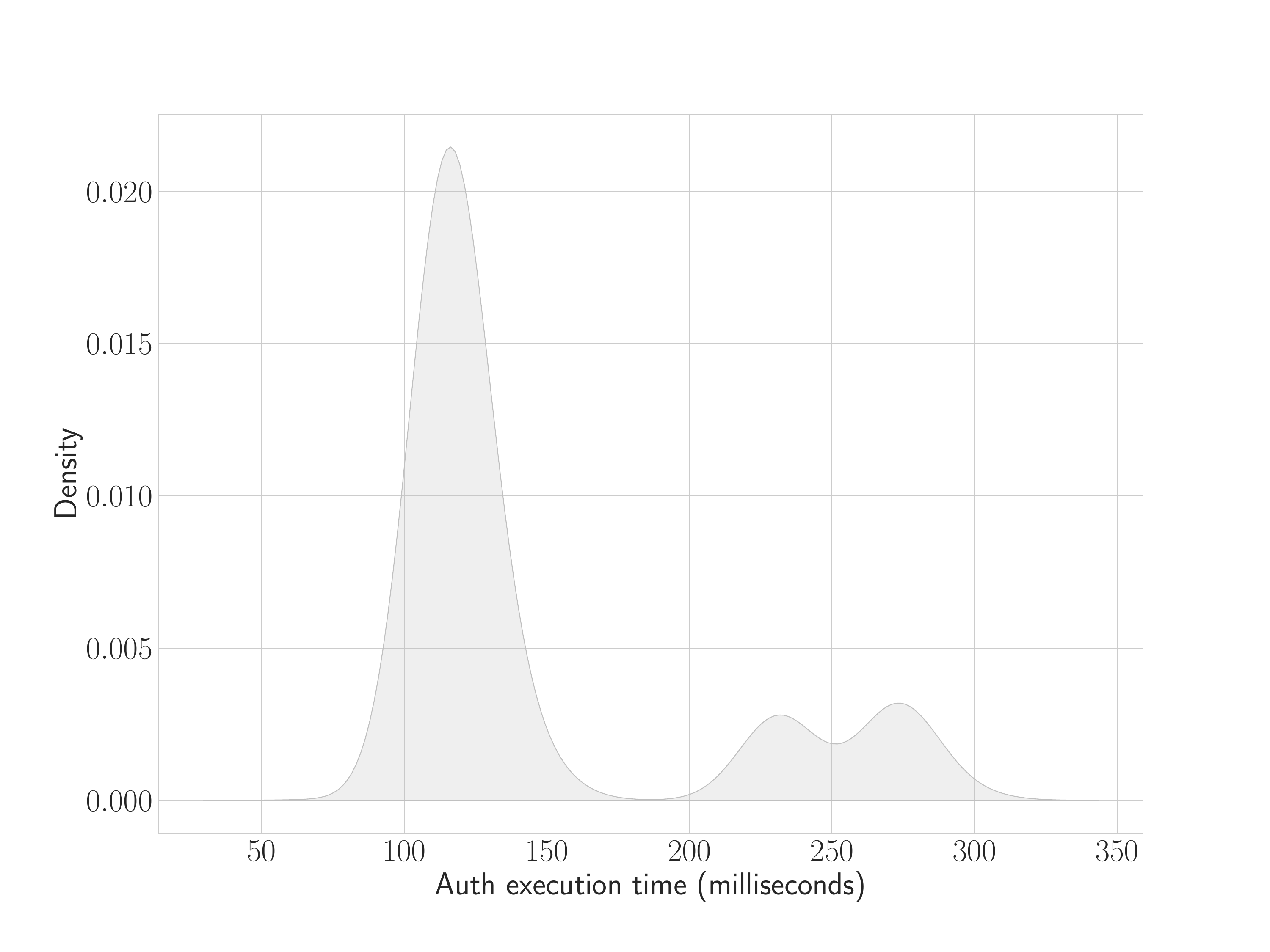}
  	\caption{Auth execution time distribution}
  	\label{fig:rpc_dist}
\end{subfigure}
\caption{Example of request latency distribution and execution time distribution of an invoked RPC (Auth).}
  	\end{figure}

However, only maximizing precision and recall may not be enough.
In the following we describe an exemplificative scenario,
where requests affected by two distict performance issues can be satisfied by a single pattern while reaching both the maximum precision and recall.
\figref{fig:rpc_vs_lat} shows the bivariate distribution of the latency for loading the homepage of a website
and the execution time of an invoked RPC, namely $Auth$.
Requests not meeting $SLO$ expectation are above the $L_{SLO}$ horizontal line.
The figure shows that every request with $Auth$ execution time above 175ms does not meet SLO expectations,
therefore a solution $S_1$ containing a single pattern $P=\{\langle Auth, 175 , 350\rangle  \}$ will have both the highest possible precision and the highest possible recall.
However, a closer look at the $Auth$ execution time distribution (see \figref{fig:rpc_dist}) shows that $Auth$ anomalous executions (\ie the ones with execution time $>175ms$) manifest two distinct behaviors, one defined by the $(175ms , 250ms)$ execution time range and the other one by $(250ms, 350ms)$.
These behaviors are also reflected in the request latency distribution (see \figref{fig:rpc_vs_lat}) and may be potential symptoms of two distinct performance issues.
Moreover, if $Auth$ invokes others RPCs, these issues may be even rooted in different RPCs.
A better solution would be $S_2=\{P_1, P_2\}$, where $P_1=\{ \langle Auth, 175 , 250\rangle  \}$ and $P_2=\{\langle Auth, 250 , 350\rangle \}$.
While keeping the same precision and recall,
$S_2$ provides a more informative view on the nature of the latency degradation.
Indeed, $P_1$ and $P_2$ identify two clusters of requests with different performance behaviors.\\
In order to avoid shallow solutions like $S_1$, we penalize \emph{pattern sets} where latencies of requests within each cluster are diverse, by minimizing \emph{latency dissimilarity}.

\textbf{Latency dissimilarity} is the sum of the average squared distance of latencies from the mean value, within each cluster of requests.
Each pattern $P\in S$ identifies a set of requests $C_P=\{r\in R \mid r\vartriangleleft P \}$.  
Latency dissimilarity, for a given pattern set $S$, can be computed as follows:
\begin{equation}\label{eq:latencysimilarity}
	\sum_{P\in S} \sum_{\substack{r\in C_P}} (L_r - \mu_{P})^2
\end{equation}
where $L_r$ is the latency for the request $r$ and $\mu_P$ is the average latency for requests satisfied by $P$:

\begin{equation}\label{eq:avglatency}
	\mu_P = \frac{\sum_{r\in C_P} L_r}{|C_P|}
\end{equation}
Furthermore, the minimization of latency dissimilarity reduces the chance that the same request satisfies multiple patterns in $S$. Indeed, if the same request satisfies multiple patterns   then latency dissimilarity tends to increase as the same request $r$ will contribute multiple times to the summation in \eqqref{eq:latencysimilarity}. 

Our optimization model involves three orthogonal objectives, \ie maximizing \emph{precision} and \emph{recall} while minimizing \emph{latency dissimilarity}.
We use Pareto optimality to plot the set of non-dominating solutions.
We decided to keep precision and recall as two separate objectives in our optimization problem rather than aggregating them into an individual objective (\ie F1-score). Indeed, this kind of aggregations may often result in a bias during search, where the maximization of one objective may be achieved at the expense of the potential maximization of another \cite{Harman2012}.
Nonetheless, solutions achieving optimal F1-scores are still included in the Pareto-optimal front that maximizes precision and recall.

%% file: approach.tex
\section{The \toolName Approach}\label{sec:approach}
\begin{figure}
  	\includegraphics[width=\linewidth]{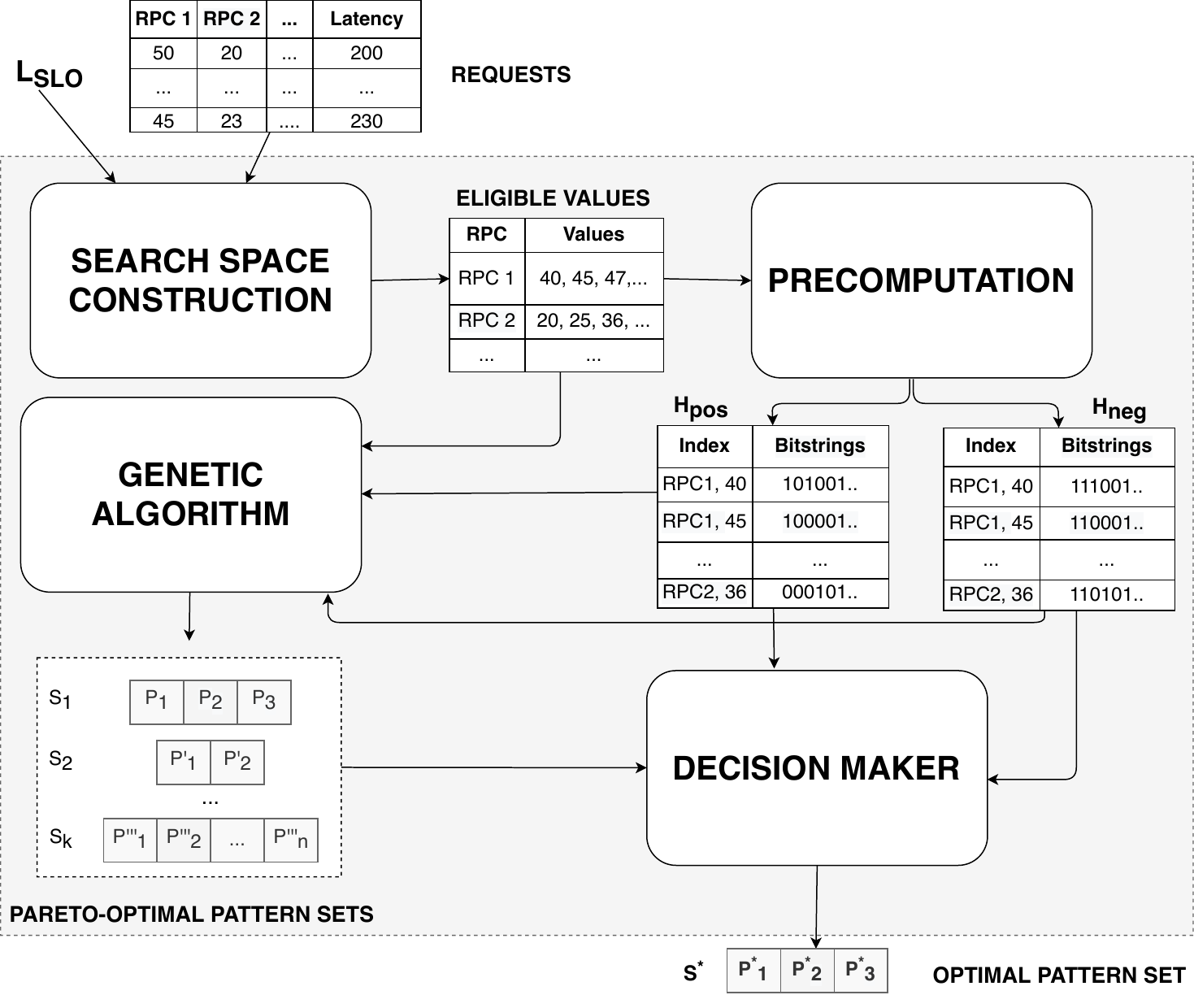}
  	\caption{\toolName workflow}
  	\label{fig:workflow}
\end{figure}
\toolName workflow is depicted in \figref{fig:workflow}.
Firstly, \toolName starts by constructing the search space of the optimization problem (\emph{Search Space Construction}).
Secondly, it precomputes results of inequality checks and stores them in lookup tables that will be then used to avoid repeated computation in fitness evaluation (\emph{Precomputation}).
Thirdly, it generates a set of non-dominated Pareto-optimal \emph{patterns sets} through a multi-objective evolutionary algorithms (\emph{Genetic Algorithm}).
Finally, it employs a heuristic to select a single \emph{pattern set} from the set of Pareto-optimal solutions as the final solution (\emph{Decision Maker}).

In the following we describe details of each workflow component.
\secref{sec:ssc} describes \emph{Search Space Construction}.
\secref{sec:ea} describes the main components of the \emph{Genetic Algorithm},
while \secref{sec:precomp} describes how \emph{Precomputation} improves the efficiency of the evolutionary process.
Finally, \secref{sec:decmak} outlines the \emph{Decision Maker}.

\subsection{Search Space Construction}\label{sec:ssc}
The key step for shaping the search space of our problem
involves the identification of highly dense regions of the RPC execution time.
Basically, a set $E_j$ of eligible values must be identified for each RPC $j$.
As in our previous work \cite{Cortellessa2020}, \toolName employs a Mean shift algorithm \cite{Comaniciu2002} to automatically identify high density intervals of the RPC execution time range.
Mean shift is a feature-space analysis technique for locating maxima of a density function \cite{Cheng1995}, and its application domains include cluster analysis in computer vision and image processing \cite{Comaniciu2002}.
We use the implementation provided by \texttt{scikit-learn}~\cite{Pedregosa2011}.
Contrariwise to other clustering algorithms, Mean shift does not require to specify the number of clusters beforehand, instead it requires only one parameter called bandwidth, which is used to control the level of (over/under)fitting of the algorithm. Intuitively, a small bandwidth tends to produce smaller clusters, conversely larger values of bandwidth induce larger clusters. We do not manually set bandwidth in our approach, instead we automatically determine it using the \texttt{estimate\_bandwidth} function of \texttt{scikit-learn}.
For each RPC $j$, Mean shift algorithm clusters requests according to their corresponding execution time.
We then infer split points $E_j$ according to the highly dense identified regions.
We discard clusters with size less than $|R_{pos}|\cdot 0.05$ to exclude execution time values rarely occurring in requests (further discussion on this point can be found in \secref{sec:threats}).

The Mean shift algorithm can be computationally expensive, given its quadratic time complexity $\mathcal{O}(n^2)$.
In order to mitigate this aspect, \toolName parallelizes eligible values selection.
In particular, since the selection of eligible values for each RPC is independent one by another,
\toolName selects $E_j$ for each RPCs $j$ in parallel to speed-up the process.

\subsection{Genetic Algorithm}\label{sec:ea}
The usage of a brute-force algorithm to solve the optimization problem defined in Section 3 would be computationally impractical, as it would require to compute precision, recall and dissimilarity for any feasible \emph{pattern set} $S$.
To the best of our knowledge, currently, there are no exact algorithms that can efficiently solve this optimization problem.
For this reason, \toolName relies on a genetic algorithm, \ie a biologically-inspired meta-heuristic that belongs to the larger class of evolutionary algorithms.
In particular, \toolName uses NSGA-II \cite{Deb2002} to build (successively-improved) Pareto-optimal solutions, while seeking new non-dominating \emph{pattern sets}.
We rely on NSGA-II because it has been shown to be effective on a wide variety of multi-objective search problems both within and outside the software engineering domain \cite{Tawosi2021,Mao2016,Ni2021,Harman2012,Deng2022,Hesami2019,Lee2021}. Using different search algorithms might potentially improve \toolName effectiveness, however, given the extensiveness of our experimental evaluation, we deem this investigation out of the scope of this work, and we leave it for future studies.

\algoref{alg:ea} presents the genetic algorithm used by \toolName to solve the optimization problem defined in \secref{sec:model}. The algorithm first generates a random initial population $P$.
Then, it performs $g_{max}$ generations while keeping track of the best individuals ever lived in the evolutionary process, namely Pareto front $\mathcal{PF}$. 
At each generation, a new population $Q$ is generated by performing crossover, mutation or reproduction on randomly selected individuals.
The population for the subsequent generation is then obtained by using the NSGA-II selection operator \cite{Deb2002} on the original population $P$ joined with the newly generated population $Q$.
At the end of the search, the algorithm returns the set of generated solutions found to be non-dominating $\mathcal{PF}$.

The \toolName genetic algorithm is implemented on top of the \emph{DEAP} evolutionary computation framework \cite{Fortin2012}.
In the following we describe the "key ingredients" of our genetic algorithm: representation, crossover, mutation, selection and fitness.
\input{algorithms/evolutionary_algorithm}

\subsubsection{Representation}
\begin{figure}
	\center
  	\includegraphics[width=0.8\linewidth]{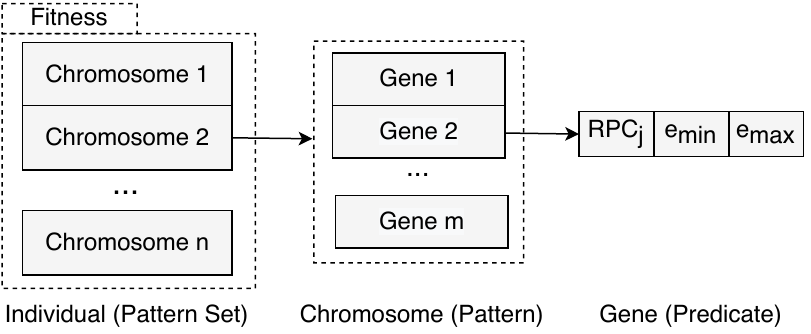}
  	\caption{Genetic Representation}
  	\label{fig:representation}
\end{figure}
The genetic algorithm simultaneously searches multiple LDPs, thus each individual corresponds to a whole \emph{pattern set}.
The representation of an individual is illustrated in \figref{fig:representation}.
Our approach generates a set of of these individuals, which corresponds to a population of individuals in the evolutionary algorithm.
Each individual consists of several chromosomes (patterns $\{ P_1, P_2, ..., P_n\}$), and each chromosome
contains multiple genes (predicates $\{ p_1, p_2, ..., p_m\}$), which
consist of random combinations of triples $\langle j, e_{min}, e_{max}\rangle$ where $e_{min},e_{max}\in E_j$ and $j$ denotes the RPC $j$.

\subsubsection{Crossover}
\begin{figure}
	\center
  	\includegraphics[width=0.8\linewidth]{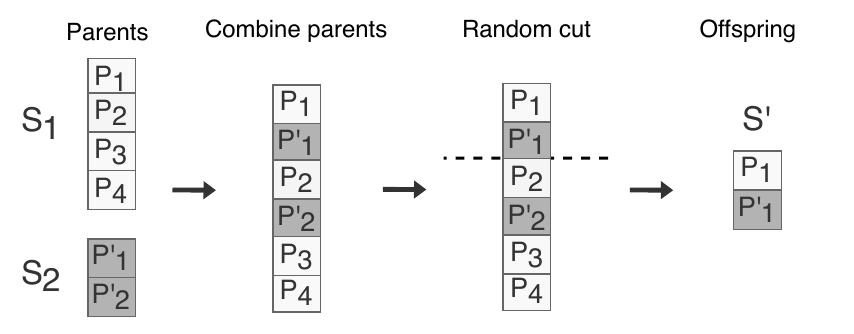}
  	\caption{Example of crossover operation}
  	\label{fig:crossover}
\end{figure}
The crossover operator is performed with probability $p$ at each new generation.
The operator randomly selects two individuals $S_1$ and $S_2$ that will be used to generate a new offspring individual $S^\prime$.
First, $S1$ and $S2$ are combined in an alternating fashion.
Then, a cut point for the combined individual is randomly chosen.
Finally, chromosomes at the left part of the cut point are chosen to form the new offspring individual $S^\prime$.
An example of a crossover operation is showed in \figref{fig:crossover}.
The time complexity of this operation is linear with respect to sizes of individuals  $\mathcal{O}(|S_1| + |S_2|)$.

\subsubsection{Mutation}
\input{algorithms/mutation}
\algoref{alg:mut} illustrates the mutation operator, which is performed with probability $q$ at each new generation on a randomly selected individual $S$.
Mutation is performed at two levels: individual and chromosome.

First, mutation is applied at individual level by performing one among three possible types of mutation with equal probability: \emph{add}, \emph{remove} or \emph{split}.
The \emph{add mutation} randomly adds a newly generated chromosome.
The \emph{remove mutation} removes a randomly chosen chromosome from the individual.
The \emph{split mutation} splits a randomly selected chromosome $P$ within $S$ in two novel chromosomes $P_1$ and $P_2$.
The latter operator first randomly selects a RPC $j$ and a threshold $t\in E_j$.
Then, $P_1$ and $P_2$ are created by partitioning requests that satisfy the randomly chosen $P$ in two parts (those having $e_j<t$ and those having $e_j\geq t$).
Finally, $P$ is replaced by $P_1$ and $P_2$ in $S$.
The detailed steps performed by the \emph{split mutation} operator are outlined in \algoref{alg:split}.
\input{algorithms/split_pattern}

Then, chromosome level mutation is performed on each chromosome, in turn, with probability $q$.
Similarly to individual level mutation, chromosome mutation applies one between two possible types of mutation with equal probability: \emph{add} or \emph{remove}.
The \emph{add mutation} randomly adds a newly generated gene,
while the \emph{remove mutation} removes a randomly chosen gene from the chromosome.

The time complexity of the mutation operator is linear with respect of the size of the individual $\mathcal{O}(|S|$).

\subsubsection{Selection}
At the end of each generation, the newly generated population and the population from the previous generation are joined together. The joint population is then sorted based on a well-defined elitism method, and the first $k$ individuals are kept for the subsequent generation, where $k$ denotes the initial population size.
The selection is performed using the widely used elitism method defined in NSGA-II \cite{Deb2002}, which sorts individuals based on their \emph{rank} and \emph{crowding-distance}. 
The \emph{rank} measures the level of non-dominance of an individual within the population.
The entire population is partitioned into fronts, where each front defines the level of non-domination of individuals, and therefore their rank.
For example, the first front represents the set of non-dominated individuals within the whole population (rank=1), the second front represents the set of non-dominated individuals while excluding the first front (rank=2), the third front represents the set of non-dominated individuals while excluding the first and second front (rank=3), and so on so forth.
On the other hand, \emph{crowding-distance} measures the degree of similarity of a particular individual with respect the whole population. The similarity is estimated through the density of individuals surrounding a particular individual based on their fitness values.
A high crowding-distance indicates that the individual provides ``peculiar'' fitness values. Conversely, a low crowding-distance indicates that there many ``similar'' individuals within the population. \\
Formally, the NSGA-II elitism method defines a comparison operator $\prec_c$ defined as follows. Given two pattern sets $S_1$ and $S_2$, we say $S_1 \prec_c S_2$ if and only if:
\begin{equation}\label{eq:nsgaii}
	S_1^{rank}<S_2^{rank} \vee (S_1^{rank}=S_2^{rank} \wedge	 S_1^{dist}>S_2^{dist})
\end{equation}
This selection policy favors individuals with smaller non-domination rank and, when the rank is equal, it favors the one with greater crowding distance (\ie less dense region) to increase diversity.

\begin{subsubsection}{Fitness}
The individual fitness value is recorded as a triple: precision, recall and latency dissimilarity.
In order to avoid overfitting, solutions containing patterns with recall $\leq 0.05$ are penalized by assigning them the worst fitness, \ie $\langle 0, 0, +\infty \rangle$ (further discussion on this point can be found in \secref{sec:threats}).

Fitness evaluation can be time consuming, as it requires to compute false positives, true positives and latency dissimilarity, for each \emph{pattern set} of the population. In particular, the time complexity for an individual fitness evaluation is $\mathcal{O}(n\cdot k)$, where $n$ denotes the total number of requests and $k$ the number of predicates involved in the individual (\ie\emph{pattern set}).
In order to speed-up this computationally expensive operation, \toolName
supports parallel fitness evaluation by assigning individuals to
multiple fitness evaluators, which may run on distributed machines (a single multicore machine was used in our experimental evaluation,
when comparing \toolName with other techniques).
In addition, \toolName precomputes results of inequality checks and stores them in a lookup table to avoid redundant computation during the evolutionary process (see \secref{sec:precomp} for details).
 \end{subsubsection}\\
 
In our experimental evaluation the initial population was created by generating 30 random individuals, where each individual is composed by a single chromosome, and each chromosome by a single gene. Genes are generated in three sequential steps, where each step randomly selects respectively the RPC $j$, $e_{min}$ and $e_{max}$.
Specifically, $j$ is chosen from the whole set of RPCs under analysis, $e_{min}$ is chosen from the set of eligible values $E_j$ (while excluding the maximum value in the set), and $e_{max}$ is randomly chosen from $E_j$ while excluding all values less than $e_{min}$.
 Crossover and mutation probability were fixed to 0.6 and 0.4, respectively, and the evolutionary process terminates after 300 generations (no early stopping criteria are employed).

\subsection{Precomputation}\label{sec:precomp}
The identification of true positives and false positives for a pattern set $S$ (see \eqqref{eq:patternsets_tp_fp}) is the key operation for computing precision and recall.
This operation requires to verify, for each request $r\in R$, whether $r\vartriangleleft S$.
This verification involves several \emph{inequality checks},
which are likely to be repeated several times during the evolution process.
\toolName reduces the fitness evaluation effort
by precomputing inequality check results, thus avoiding redundant computation.
Inequality checks are denoted as pairs $\langle j, t\rangle $, where $j$ is a RPC and $t\in E_j$ is an eligible value, hence an execution time threshold.
Inequality check results are denoted as ordered sequences of booleans $B_{\langle j, t\rangle }=\langle b_0^{\langle j, t\rangle }, b_1^{\langle j, t\rangle }, ..., b_n^{\langle j, t\rangle }\rangle $,
where $b_i^{\langle j, t\rangle }$ refers to the check result for the request $r_i\in R$.
A check result $b_i^{\langle j, t\rangle }$ for a given inequality check $\langle j, t\rangle $ and a request $r_i= (..., e_j, ...)$ is defined as:
\begin{equation*}
	b_i^{\langle j, t\rangle }=\begin{cases}
			True & \text{if $ e_j \geq t $}\\
    		False & \text{otherwise}
  		\end{cases}
\end{equation*}

\begin{figure}
  	\includegraphics[width=\linewidth]{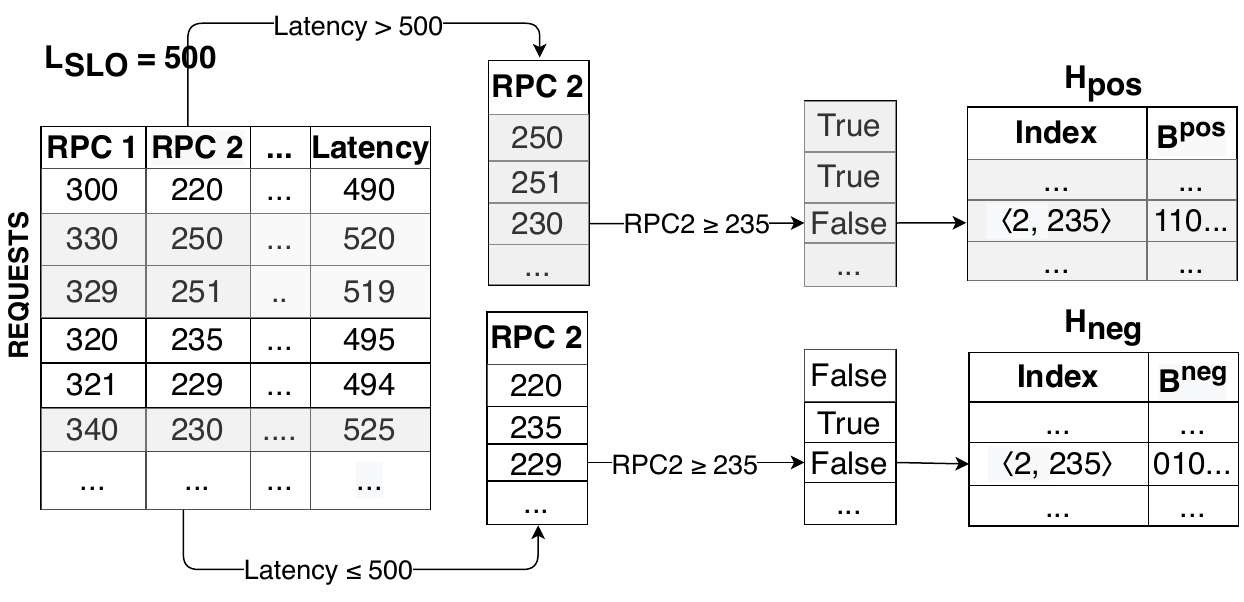}
  	\caption{Precomputation for $\langle j, t\rangle $ inequality check $\langle 2, 235\rangle $.}
  	\label{fig:precomputation}
\end{figure}

Two boolean sequences (namely $B^{pos}_{\langle j, t\rangle }$ and $B^{neg}_{\langle j, t\rangle }$) are precomputed for every pair $\langle j, t\rangle$,
which represent inequality check results, respectively, for positive and negative requests (\ie $R_{pos}$ and $R_{neg}$).
Boolean sequences are encoded as bitstrings and stored in two lookup tables, one for positives requests $H_{pos}$ and another one for negative requests $H_{neg}$, as shown in \figref{fig:workflow}.
The length of each bitstring in $H_{pos}$ (resp. $H_{neg}$) is equal to the number of requests in $R_{pos}$ (resp. $R_{neg}$) .
A bitstring can be retrieved by lookup tables using the corresponding index, \ie $\langle j, t\rangle $.
\figref{fig:precomputation} shows the process used to create $H_{pos}$ and $H_{neg}$.
The time complexity of this process is $\mathcal{O}(n\cdot m\cdot t)$, where $n$ denotes the total number of requests, $m$ the number of RPCs, and $t$ the number of eligible values.
Such process can be computationally intensive, but it induces an efficiency gain across the entire evolutionary process. In particular, it avoids a non-trivial amount of redundant inequality checks, which are, instead, precomputed and stored through efficient data structures, \ie bitstrings and lookup tables.

These data structures enable fast identification of true positives and false positives across multiple requests through bitwise operations.
A bitwise operation works on one or more bit strings at the level of their individual bits.
We use three common bitwise operators: \emph{and} ($\wedge$), \emph{or} ($\vee$) and \emph{not} ($\neg$).
A predicate $p=\langle j, e_{min} , e_{max} \rangle$, with $e_{min},e_{max}\in E_j$ (see \secref{sec:ssc}),
can be efficiently evaluated on positive requests as well as on negative requests by performing the following steps.
First, boolean sequences associated to inequality checks $\langle j, e_{min} \rangle$ and $\langle j, e_{max} \rangle$ are retrieved by lookup tables $H_{pos}$ and $H_{neg}$.
We denote them as $B_{\langle j, e_{min} \rangle}^{pos}$ and $B_{\langle j, e_{min} \rangle}^{neg}$, 
and $B_{\langle j, e_{max}\rangle}^{pos}$ and $B_{\langle j, e_{max}\rangle }^{neg}$.
Then, positive and negative requests that satisfy predicate $p$ are derived through bitwise operations:
\begin{equation*}
	B^{pos}_{p} = B_{\langle j, e_{min} \rangle}^{pos} \wedge \neg B_{\langle j, e_{max}\rangle }^{pos}
\end{equation*}
\begin{equation*}
	B^{neg}_{p} = B_{\langle j, e_{min} \rangle}^{neg} \wedge \neg B_{\langle j, e_{max}\rangle }^{neg}
\end{equation*}
Hence $b_i^p\in B^{pos}_{p}$ (resp. $b_i^p\in B^{neg}_{p}$) is equal to $True$ if the request $r_i\in R_{pos}$ (resp. $r_i\in R_{neg}$) satisfies the predicate $p$, \ie $r_i\vartriangleleft p$, and is equal to $False$ otherwise.\\
The same approach is also applied to check whether a request satisfies a pattern or not, $r_i\vartriangleleft P$:
\begin{equation*}
	\begin{aligned}
		& B^{pos}_{P}=\bigwedge\limits_{p\in P} B_p^{pos} \\
		& B^{neg}_{P}=\bigwedge\limits_{p\in P}B_p^{neg} \\
	\end{aligned}
\end{equation*}
And to verify whether a request satisfies the whole \emph{pattern set}, $r_i\vartriangleleft S$:
\begin{equation*}
	\begin{aligned}
		& B^{pos}_{S}=\bigvee \limits_{P\in S} B_P^{pos} \\
		& B^{neg}_{S}=\bigvee\limits_{P\in S}B_P^{neg} \\
	\end{aligned}
\end{equation*}

Number of true positives and false positives for a given pattern set $S$ are obtained by counting $True$ booleans (\ie number of 1 in the bit string) in both $B^{pos}_{S}$ and $B^{neg}_{S}$:
\begin{equation*}
	\begin{aligned}
		& |tp|= |\{b\in B^{pos}_{S} \mid b=True\}| \\
		& |fp|= |\{b\in B^{neg}_{S} \mid b=True\}| \\
	\end{aligned}
\end{equation*}
Finally, precision and recall can be derived through a simple numerical computation (see Equations (\ref{eq:prec}) and (\ref{eq:rec})).

\subsection{Decision Maker}\label{sec:decmak}
\input{algorithms/decision_maker}
The manual identification of a relevant single \emph{pattern set} from the Pareto-optimal set $\mathcal{PF}$ can be complex.
\toolName employs a heuristic to avoid this manual process, thus enabling full automation.
The decision making heuristic uses the generalized form of the F1-score formula:
\begin{equation}\label{eq:fbetascore}
	F_\beta\text{-}score=(1+\beta^2)\cdot\frac{precision\cdot recall}{(\beta^2\cdot precision)+recall}
\end{equation}
The F$_\beta$-score is a generalization of the F1-score that adds a configuration parameter called beta.
The F1-score uses a beta value of 1.0, which gives the same weight to both precision and recall.
A beta value less than 1 gives more weight to precision and less to recall, whereas a larger beta value gives less weight to precision and more weight to recall.
Maximizing precision implies the minimization of false positives, whereas maximizing recall implies the minimization of false negative.

We argue that minimization of false positives is likely to be more relevant than minimization of false negatives in the context of LDPs detection.
Indeed, patterns with non-negligible amounts of false positives are likely to be less meaningful (whatever the amount of false negatives is),
since they are not peculiar to requests not meeting SLO.
Conversely, patterns with non-negligible amount of false negatives can still be relevant if the number of false positives is low,
because they are peculiar to a portion of requests in $R_{pos}$,
therefore they are likely to be potential symptoms of performance issues. 
For these reasons, the decision-making heuristic sacrifices \emph{recall} in favor of \emph{precision} if this implies a gain in terms of \emph{latency dissimilarity}.
\algoref{alg:decmak} outlines the decision making heuristic.
The heuristic selects \emph{pattern sets} with maximum $F_{\beta\text{-}score}$ while $\beta$ ranges from 0.1 (precision is weighted 10 times as much as recall) to 1 (equally weighted).
Then, it chooses among the selected solutions the one with minimum \emph{latency dissimilarity}.
In other words, the heuristic considers multiple F$_\beta$-score formulations simultaneously, and it chooses the one that enables optimal \emph{latency dissimilarity}.
This aspect makes our decision-making heuristic much more flexible when compared to traditional F1-score optimization.

In order to speed up the process, the heuristic leverages lookup tables $H_{pos}$ and $H_{neg}$ generated in the precomputation phase. 

%% file: algorithms/evolutionary_algorithm.tex
\begin{algorithm}
\caption{Genetic Algorithm}
\small
\SetKwComment{Comment}{{$\triangleright$}\ }{}
\SetCommentSty{algcommentfont}
\label{alg:ea}
 \KwData{max generation $g_{max}$, crossover probability $p$, mutation probability $q$
}
 \KwResult{Pareto front $\mathcal{PF}$ }
 $\mathcal{PF} \leftarrow \emptyset$ \; 
 initialise population $\mathcal{P}$ \;
 evaluate fitness of $\mathcal{P}$ and update $\mathcal{PF}$ \;
 \For {$i$ in range(0, $g_{max}$)}{
  $Q \leftarrow \emptyset$ \;
  \For {$j$ in range(0, $|\mathcal{P}|$)}{
  $r \sim U(0,1) $ \;
  \uIf(\Comment*[f]{apply crossover}){$r < p$}{
   randomly select $S_1$ and $S_2$ from $\mathcal{P}$ \;
   $S^{\prime} \leftarrow crossover(S_1, S_2)$ \;
   }
   \uElseIf(\Comment*[f]{apply mutation}){$r < p+q$}{
   randomly select $S$ from $\mathcal{P}$ \;
   $S^{\prime} \leftarrow mutation(S, q)$ \;
  }\Else(\Comment*[f]{apply reproduction}){
   randomly select $S$ from $P$ \;
   $S^{\prime} \leftarrow S$ \;
   }
   $Q \leftarrow Q \cup \{S^{\prime}\}$\;
  }
  evaluate fitness of $Q$ and update $\mathcal{PF}$ \;
  
  $\mathcal{P}^\prime \leftarrow sorted(Q\cup \mathcal{P}, \prec_c)$  \Comment*[r]{see \eqqref{eq:nsgaii} for \small{$\prec_c$}}
  $\mathcal{P} \leftarrow  \mathcal{P}^\prime[0 : |\mathcal{P}|]$ \Comment*[r]{new population} 
  }
 return $\mathcal{PF}$ \; 
 
\end{algorithm}

%% file: algorithms/mutation.tex
\begin{algorithm}
\caption{Mutation}
\label{alg:mut}
\SetKwComment{Comment}{{$\triangleright$}\ }{}
\SetCommentSty{algcommentfont}
\small
 \KwData{individual $S$, mutation probability $q$
}
 \KwResult{ mutant $S^{\prime}$ }
 $r \sim U(0,3)$ \;
 \uIf(\Comment*[f]{apply remove mutation (individual)}){$r<1$}{
 	randomly select one pattern $P$ from $S$\;
 	$S^\prime \leftarrow S \setminus \{P\} $ \;
 }\uElseIf(\Comment*[f]{apply add mutation (individual)}){$r<2$}{
 	generate a new random pattern $P$ \;
 	$S^\prime \leftarrow S \cup \{P\} $ \;
 }\Else(\Comment*[f]{apply split mutation (individual)}){
  randomly select a pattern $P$ from $S$ \;
  $P_1, P_2 \leftarrow splitPattern(P)$ \;
  $S^\prime \leftarrow S \setminus \{P\}$ \;
  $S^\prime \leftarrow S^\prime \cup \{P_1, P_2\}$ \;
 }
 \For {each pattern $P$ in  $S^\prime$}{
 	$r \sim U(0,1)$ \;
 	\uIf{$r<q$}{
 		\uIf(\Comment*[f]{apply remove mutation (chromosome)}){$r< q/2$}{
 		 	randomly select one predicate $p$ from $P$\;
 		 	$P \leftarrow P \setminus \{p\} $ \;
 		 }
 		\Else(\Comment*[f]{apply add mutation (chromosome)}){
 			generate a new random predicate $p$ \;
 			$P \leftarrow P \cup \{p\} $ \;
 		}
 	}
 	
}
 return $S^{\prime}$ \; 

\end{algorithm}

%% file: algorithms/split_pattern.tex
\begin{algorithm}
\caption{splitPattern}
\label{alg:split}
\small
\SetKwComment{Comment}{{$\triangleright$}\ }{}
\SetCommentSty{algcommentfont}
 \KwData{pattern $P$}
 \KwResult{pattern $P_1$, pattern $P_2$}
 randomly select a RPC $j$ \;
 select a predicate $p=\langle k, e_{min}, e_{max}\rangle$ with $k=j$ \;
 \uIf{$p$ exists 
 }{
 	$e^\prime_{min} \leftarrow  e_{min}$ \;
 	$e^\prime_{max} \leftarrow  e_{max}$ \;
    $P^\prime \leftarrow P \setminus \{ p \}$ \;
 }\Else{
 	$e^\prime_{min} \leftarrow  min(E_j)$ \;
 	$e^\prime_{max} \leftarrow  max(E_j)$ \;
 	$P^\prime \leftarrow P$ \;
 }
 randomly select $t\in E_j$ s.t. $e^\prime_{min}<t<e^\prime_{max}$ \;
  $p_1 \leftarrow \langle j, e^\prime_{min}, t\rangle$ \;
  $p_2 \leftarrow \langle j, t, e^\prime_{max}\rangle$ \;
  $P_1 \leftarrow P \cup \{ p_1 \}$ \;
  $P_2 \leftarrow P \cup \{ p_2 \}$ \;

 return $P_1$, $P_2$ \; 
 \end{algorithm}

%% file: algorithms/decision_maker.tex
\begin{algorithm}
\caption{Decision maker}
\label{alg:decmak}
\small
 \KwData{Pareto front $\mathcal{PF}$}
 \KwResult{ Pattern set $S^*$ }
 $\beta \leftarrow 0.1$ \; 
 \While{$\beta \leq 1$}{
 	$\mathcal{S} \leftarrow \emptyset$ \;
 	select $S\in \mathcal{PF}$ with maximum $F_\beta$\text{-}score  \;
 	add $S$ to $\mathcal{S}$ \;
 	$\beta \leftarrow \beta + 0.01$ \;
 	
 }
 select $S^*\in\mathcal{S}$ with minimum \emph{latency dissimilarity}\;
 return $S^*$ \; 
\end{algorithm}

%% file: evaluation.tex
\section{Evaluation}\label{sec:evaluation}

We evaluated \toolName by performing a set of experiments aimed at answering the following research questions.
\begin{enumerate}

\item[RQ$_1$] \emph{Can \toolName \textbf{effectively} detect LDPs?}
In this RQ, the LDPs detected by \toolName are compared to the ones detected by prior work and general purpose clustering algorithms.
The effectiveness of the methods are compared on precision ($Q_{prec}$), recall ($Q_{rec}$) and F1-score ($Q_{F1}$), as they will be defined in \secref{sec:methodology}. In addition, we further investigate the effectiveness of \toolName with a complementary assessment on continuously-varying load mixtures.

\item[RQ$_2$] \emph{How do \textbf{overlapping patterns} affect \toolName effectiveness?}
F1-score-based techniques are less effective when distinct patterns lead to partially (or entirely) overlapping latency distributions \cite{Krushevskaja2013}.
With this RQ, we want to check whether \toolName overcomes this limitation. 
We evaluate how proximity of latency distributions (related to distinct LDPs) affects our effectiveness.
The approach is experimented on a variety of scenarios while controlling the proximity of latency distributions related to different patterns.

\item[RQ$_3$] \emph{How \textbf{non-critical RPC} execution time variation affects \toolName effectiveness?}
Not all execution time variations of RPCs produce effect on latency (\eg RPCs outside the critical path).
With this RQ, we want to determine whether deviations of execution times on non-critical RPCs decrease the effectiveness of \toolName.
The approach is experimented on a variety of scenarios while controlling the magnitude of increment of execution times on non-critical RPCs.

\item[RQ$_4$]  \emph{What is the \toolName \textbf{efficiency}?}
Modern service-based systems collect thousands of traces per day (or even more), therefore efficiency is a major concern for \toolName.
In this RQ, we evaluate the efficiency of \toolName on datasets by different sizes.

\item[RQ$_5$] \emph{How can \toolName support \textbf{performance analysis}?} With this RQ, we want to investigate the capabilities of \toolName in supporting performance analysis. To do so, we qualitatively analyze the results of \toolName , and we compare them against those provided by two machine learning based approaches for performance issue diagnosis.

\end{enumerate}

In order to assess the generality of \toolName, we carried out our experiments on two case studies based on open-source microservice-based systems.
The first one relies on \emph{E-Shopper}, an e-commerce web application, while the second one on \emph{Train Ticket}, a web-based booking system.
Both case studies are introduced in \secref{sec:case_studies}.
In \secref{sec:baselines} we describe the techniques used as baselines.
The methodology used for the evaluation is described in \ref{sec:methodology},
followed by descriptions of experiments carried out to address each research question, respectively, in Sections \ref{sec:rq1}, \ref{sec:rq2}, \ref{sec:rq3}, and \ref{sec:rq4}.

\subsection{Case studies}\label{sec:case_studies}
\subsubsection{E-Shopper}
The first case study is based on E-Shopper\footnote{\url{https://github.com/SEALABQualityGroup/E-Shopper}},
a small-size microservice-based system, already used in the evaluation of our previous works \cite{Cortellessa2020, Arcelli2019}.
E-Shopper is a typical e-commerce application, which is developed as a suite of small services, each running in its own Docker\footnote{\label{fn:docker}\url{https://www.docker.com}} container.
It is designed using microservice design principles.
Microservices are developed in Java and interactions among them are based on RESTful APIs.
The application produces execution traces that are reported and collected by Zipkin, \ie a popular distributed tracing system \cite{mace2017survey},
and stored in Elasticsearch\footnote{\label{fn:elasticsearch}\url{https://www.elastic.co/elasticsearch}}.
We focus our experimentation on requests loading the homepage,
which involve a number of 25 invocations of 7 unique RPCs spread across 5 microservices. 

\subsubsection{Train Ticket}
The second case study is based on Train Ticket\footnote{\url{https://github.com/FudanSELab/train-ticket}}, which is the largest and most complex open source microservice-based system (within our knowledge at the time of writing).
The system was already used in previous software engineering studies \cite{Zhou2018,Zhou2019} .
Train Ticket provides typical train ticket booking functionalities such as ticket enquiry, reservation, payment, change, and user notification.
It is designed using microservice design principles and covers different interaction modes such as synchronous and asynchronous invocations, and message queues.
The application produces execution traces that are reported and collected by Jaeger,
and stored in Elasticsearch.
The system contains 41 microservices related to business logic (uses four programming languages Java, Python, Node.js, and Go) with each service running in its own Docker container.
Our experimentation focuses on requests devoted to the ticket searching process,
which involve a number of 48 invocations of 14 unique RPCs spread across 9 microservices.

\subsubsection{Experimental setup}
As performance measurements are prone to confounding factors~\cite{Curtsinger2013, Maricq2018, Mytkowicz2009, Traini2023, Traini2021b}, we run our experiments on a dedicated environment while controlling as many sources of variability as possible.
The generation of datasets and the experimentation of techniques are performed on a bare-metal machine~\cite{Arif2018} running Linux Ubuntu 18.04.2 LTS on a dual Intel Xeon CPU E5-2650 v3 at 2.30GHz, with a total of 40 cores and 80GB of RAM.
All non-mandatory background processes except ssh are disabled, and we ensured that no other users interacted with the dedicated machine during our experiments.
Furthermore, in each load testing session, we discard performance data collected during the first 30 seconds to avoid performance fluctuations due to system warmup.

When subject to continuously varying workloads (see Section 5.3 for details), E-Shopper and Train Ticket reported a mean CPU utilization of 10\% (min: 9\%, max 24\%), and 38\% (min: 26\%, max: 52\%), respectively. We profiled CPU utilization using \texttt{psutil}\footnote{We profile CPU utilization using \cite{psutil} with a sampling rate of 30 seconds}~\cite{psutil}, a tool that has been widely used in prior software engineering research \cite{Ding2020, Chen2022, Chen2017}.

\subsection{Baselines techniques}\label{sec:baselines}
We compare \toolName against both domain-specific state-of-the-art approaches and general-purpose clustering algorithms.
The latters were considered because of their straightforward application to the subject problem, \ie the identification of clusters of requests that shows similar behavior in terms of RPCs execution time.
We considered two widely popular clustering algorithms, \ie \emph{K-Means}\cite{macqueen1967} and \emph{Hierarchical clustering}\cite{Rokach2005}, and for both of them we used the implementation provided by \texttt{scikit-learn}~\cite{Pedregosa2011}.

As domain-specific baselines we considered both F1-score-based methods (\ie our previous work \cite{Cortellessa2020} and the one proposed by Krushevskaja and Sandler \cite{Krushevskaja2013}), and a recently proposed random-forest-based approach \cite{Bansal2019}.
Unfortunately, only our previous work provides publicly available source code,
therefore we have re-implemented both the approach of Krushevskaja and Sandler \cite{Krushevskaja2013} and the one of Bansal~\etal~\cite{Bansal2019},
which are now publicly available in the replication package \cite{replication}.\\
In the following we provide descriptions of each baseline technique:

\textbf{K-Means}. The K-Means algorithm \cite{macqueen1967} clusters data by trying to separate samples in $k$ groups, while minimizing a criterion known as within-cluster sum-of-squares.
K-Means requires the number of clusters to be specified.
In our experimentation, we execute the algorithm with $k$ ranging from 2 to 10 and we pick the best solution among them, \ie the one that achieves the higher effectiveness ($Q_{F1}$).

\textbf{Hierachical clustering (HC)}.
Hierarchical clustering \cite{Rokach2005} is a general family of clustering algorithms that build nested clusters by progressively merging or splitting them.
We use an implementation based on a bottom-up approach: 
each observation starts in its own cluster, and clusters are successively merged together.
Also hierarchical clustering requires the number of clusters to be specified, therefore we employed the same approach used for the K-Means algorithm.

\textbf{Krushevskaja and Sandler (KrSa)}.
This approach\cite{Krushevskaja2013} models the problem of detecting patterns as a binary optimization problem and uses a branch-and-bound algorithm combined with a dynamic programming algorithm to maximize the sum of the F1-scores achieved by the patterns.
The approach requires the encoding of trace attributes to binary features and the selection of a set of split points to divide the targeted latency range.
Similarly to what has been done here in the search space construction phase (\secref{sec:ssc}),
our re-implementation of this approach uses Mean shift algorithm \cite{Comaniciu2002} for both encoding and split points selection.
In order to avoid overfitting, we force the Mean Shift algorithm to discard clusters of requests with size less or equal to $|R_{POS}|\cdot 0.05$.
We used the implementation of this approach already used in our previous work \cite{Cortellessa2020}.

\textbf{Cortellessa and Traini (CoTr)}.
Similarly to KrSa, our previous algorithm \cite{Cortellessa2020} searches for the optimal set of patterns that maximize the sum of the F1-scores.
The main difference relies on the technique used to search the optimal pattern for each sub-interval, which is based on a genetic algorithm.
More details on the approach can be found in \cite{Cortellessa2020}.
We used the same experimental setup used for KrSa for both encoding step and split point selection, 
while the size of population of the genetic algorithm is set to 30.
The genetic algorithm performs 300 iterations with
mutation and crossover probability set to 0.4 and 0.6, respectively.
In the experiments, we used the original implementation of the approach \cite{Cortellessa2020}.

\textbf{DeCaf}.
DeCaf\cite{Bansal2019} trains a random forest model \cite{Breiman2001} and then infers predicates correlated with anomalous behavior.
We used the implementation of the random forest algorithm based on classification trees provided by \texttt{scikit-learn}~\cite{Pedregosa2011}.
Predicates extraction and deduplication were re-implemented in Python.
In order to avoid overfitting, the minimum number of training data in a leaf is set to $|R_{POS}|\cdot 0.05$.
The number of trees and features sampling ratio are set to 50 and 0.6, respectively, as in the original paper \cite{Bansal2019}.
The output of the algorithm is a ranking of predicates based on their correlation scores.
In our evaluation we considered the top 10 scored predicates.

\subsection{Methodology}\label{sec:methodology}
We evaluate the effectiveness of \toolName on a variety of scenarios from two case study systems. We first generate different datasets representing different scenarios that involve distinct combinations of LDPs for both case studies. Then, we run \toolName on the generated datasets to evaluate its effectiveness.
In a nutshell, each dataset used in our evaluation was generated as follows.
Firstly, we altered the source code of the system to introduce delays on certain RPCs with certain probabilities, thus reproducing performance issues that lead to LDPs.
Then, we performed load testing on the altered system to simulate user traffic, thus producing a dataset of requests.

Delays simulate performance issues that repeatedly occur on the system and cause latency degradation for relevant portions of requests, thus producing LDPs.
We call these simulated performance issues Artificial Delay Combinations (ADCs).
ADCs are designed to simulate both performance issues that are rooted in the internal implementation of individual RPCs (\ie a single RPC is involved), and performance issues that arises from the interaction of multiple services \cite{Zhou2018} (\ie multiple RPCs are involved).
Specifically, each ADC involves from a minimum of one RPC to a maximum of three RPCs, and it is defined by a set of pairs $\langle j, d\rangle$,
where $d$ denotes the delay in milliseconds that is introduced in RPC  $j$.

\begin{figure}
	\center
  	\includegraphics[width=\linewidth]{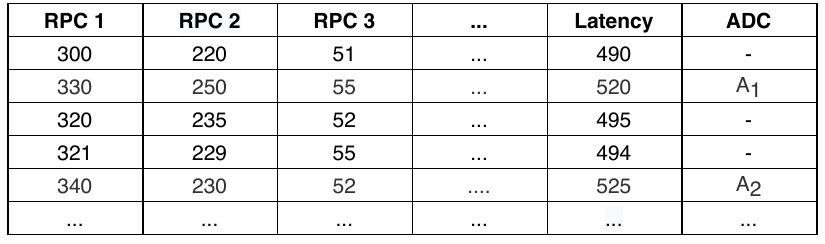}
  	\caption{Example of dataset}
  	\label{fig:dataset}
\end{figure}

We evaluated \toolName using a variety of scenarios from both case studies,
where each scenario involves two randomly generated ADCs (hence two LDPs).
We developed a process that enables us to automatically alter the source code of the system by injecting delays according to the generated ADCs.
Each request to the altered system is randomly assigned to one of the two ADCs with probability 0.1 (hence, with 0.8 probability is not subject to any delay) and delays are automatically introduced to the corresponding RPCs according to the ADC.
For each scenario, we perform a load testing session involving 20 synthetic users simulated by Locust\footnote{\url{https://locust.io/}},
where each user makes a request to the system and randomly waits 1 to 3 seconds for the next request\footnote{The random waiting time follows a uniform distribution. Note that a user waiting time of 1-3 seconds corresponds to an accelerated scenario when compared to traditional web benchmarks (\eg SPEC Web).}.
At the end of each session, the operational data collected by distributed tracing tools (\ie Jaeger and Zipkin) are processed and transformed in tabular format, thus producing a dataset.
\figref{fig:dataset} shows an example of dataset,
where each row represents a request and each column contains the cumulative execution time of a RPC.
It is cumulative because it represents the whole contribution, in terms of execution time, of the RPC to the whole request.
In other words, if a request involves multiple invocations of the same RPC, then the cell contains the sum of all invocation execution times. 
The $Latency$ column contains the overall request latency, while the $ADC$ column reports whether the request is assigned to an ADC ($A_1$ or $A_2$) or not (-).
Each dataset contains approximately 10\% of requests assigned to the first ADC $A_1$ (we denote this subset of requests as $R_{A_1}$),
10\% assigned to $A_2$ (resp. $R_{A_2}$) and 80\% of requests showing the non-altered RPCs execution times behavior.
For each scenario, $L_{SLO}$ is defined as the smallest latency for requests assigned to one of the two ADCs. 

We measure the \emph{effectiveness} of \toolName on a given scenario (\ie dataset) by using three quality measures: precision, recall and F1-score.
\toolName outputs a \emph{pattern set} $S^*$=$\{ P_1, ..., P_n \}$ which identifies a set of clusters of requests $\mathcal{C}^*$=$\{ C_{P_1}, ..., C_{P_n} \}$,  where each cluster $C_P\in\mathcal{C}^*$ identifies the set of requests satisfied by $P$, \ie $C_P$=$\{r\in R \mid r\vartriangleleft P \}$.
In order to evaluate \toolName, we intend to verify whether there are two patterns in $S^*$ that identify $R_{A_1}$ and $R_{A_2}$, respectively. 
To this aim, we first identify the best matching patterns $P_{A_1},P_{A_2}\in S^*$,
where $P_{A_1}$ (respectively $P_{A_2}$) is 
 chosen by selecting the pattern that maximizes F1-score while considering requests in $R_{A_1}$ (respectively $R_{A_2}$) as positives and all other requests as negatives.
 Once $P_{A_1}$ and $P_{A_2}$ are identified, precision, recall and F1-score can be derived as follows:
\begin{equation}\label{eq:qprec}
		Q_{prec}  = \frac{\mid C_{P_{A_1}} \cap R_{A_1}\mid + \mid C_{P_{A_2}} \cap R_{A_2}\mid}{\mid C_{P_{A_1}}\mid + \mid C_{P_{A_2}}\mid }
\end{equation}
\begin{equation}\label{eq:qrec}
		Q_{rec}  = \frac{\mid C_{P_{A_1}} \cap R_{A_1}\mid + \mid C_{P_{A_2}} \cap R_{A_2}\mid}{\mid R_{A_1} \cup R_{A_2}\mid }
\end{equation}
\begin{equation}\label{eq:qfscore}	
		Q_{F1} =  2\cdot\frac{Q_{prec}\cdot Q_{rec}}{Q_{prec}+Q_{rec}}
\end{equation}
These measures can be also applied to evaluate baseline techniques, since F1-score-based techniques and DeCaf return set of patterns which identify clusters of requests (similarly to \toolName), while clustering algorithms directly return clusters of requests.

Besides comparing effectiveness measures of techniques, we also compare their ``stability'' across different scenarios, \ie how effectiveness measures vary across different scenarios. In order to measure stability, we use interquartile range (IQR). A high IQR indicate that the quality measure highly vary across different scenarios. On the other hand, a low IQR indicates more stability across scenarios.

Both \toolName and baselines involve randomness (except for KrSa which uses a deterministic algorithm),
thus to mitigate effectiveness variability we execute these techniques on each dataset 20 times.

It is worth to notice that workloads used in our experiments may not be representative of real-world workloads, as the latter ones often change over time and involve complex mixtures of loads~\cite{Ardelean2018}.
However, we used such workloads for a specific reason. In fact, a wrong choice of workloads can severely affect the validity of our study. In particular, we observed that more elaborate workloads (\eg involving different APIs) may severely reduce the proportion of ``ADC requests'' in requests that violate SLO (\ie with latency $L > L_{SLO}$ ). For example, when using  \texttt{PPTAM}~\cite{Avritzer2019}, which is an elaborated Train Ticket workload used in prior work~\cite{Avritzer2020,Janes2019}, we observed that only $\sim$19\% of requests that violate SLO are assigned to an ADC. Such behavior have significant implications on the validity of our study, as our evaluation is based on the assumption that ADCs represent relevant causes of latency degradation.
We carefully considered these critical aspects when defining workloads for our experiments.
Eventually, we managed to ensure a reasonable degree of concurrency (\ie 20 parallel simulated users) while keeping our evaluation ``clean'' (\ie on average $\sim$86\% of requests that violate SLO are assigned to an ADC).

 \begin{figure}
 	\center
   	\includegraphics[width=0.85\linewidth]{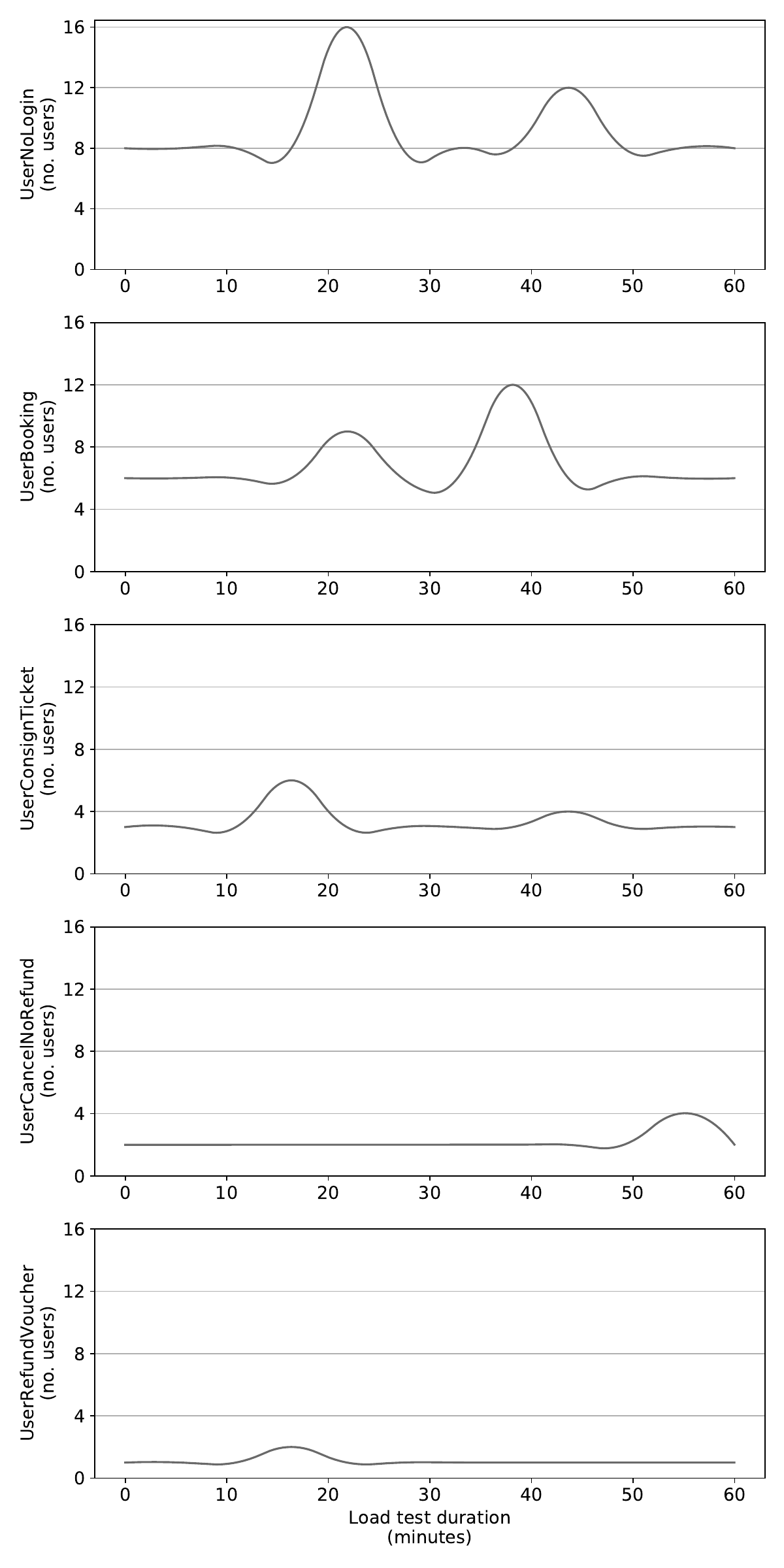}
   	\caption{Train Ticket load mixture. Each plot represents the number of users of a specific type (\eg UserNoLogin) over time. The x-axis represents time in minutes, while the y-axis represents the number of users.}
   	\label{fig:loadshape}
 \end{figure}
 
 \begin{figure}
	\centering
     \begin{subfigure}[b]{0.95\linewidth}
		\centering
         \includegraphics[width=0.8\linewidth]{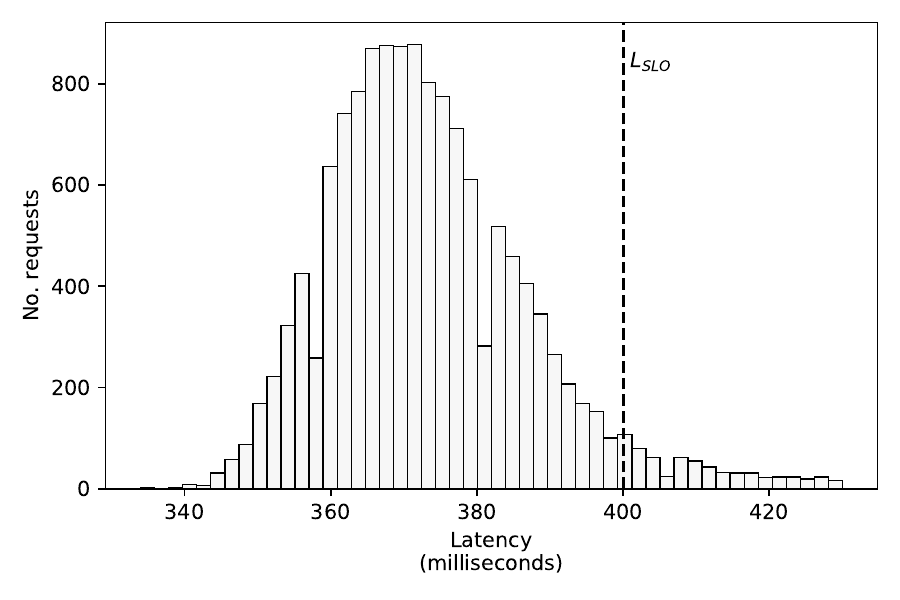}
         \caption{\footnotesize $RC_1$. \texttt{web}$\diamond$\texttt{home} (E-Shopper)}
         \label{fig:workload_latency_home}
     \end{subfigure}
     \hfill
     \begin{subfigure}[b]{0.95\linewidth}
         \centering
         \includegraphics[width=0.8\linewidth]{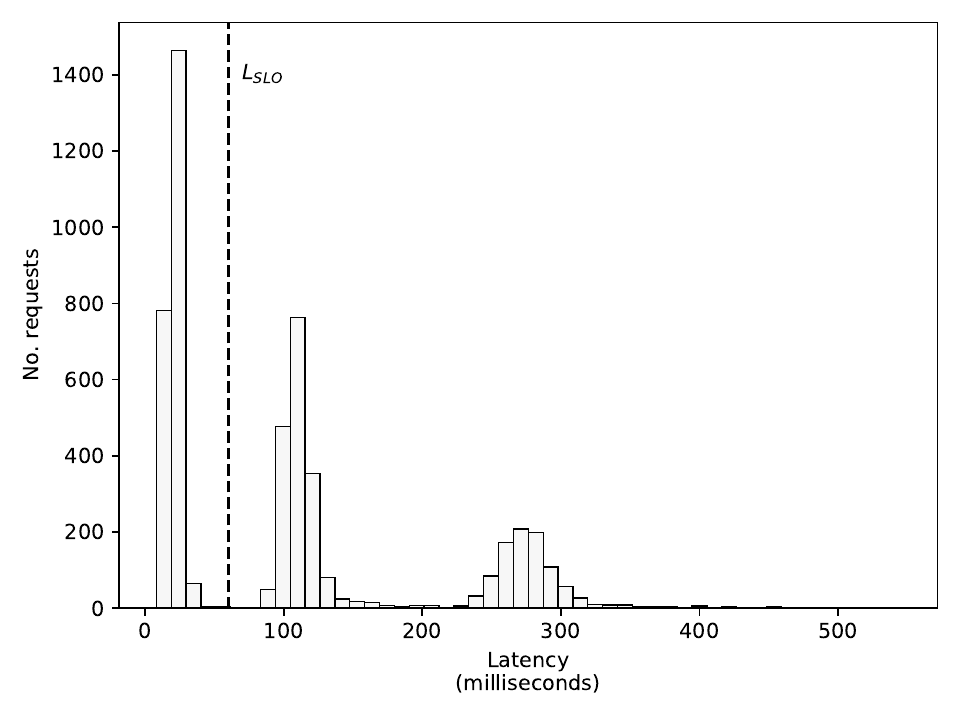}
         \caption{\footnotesize $RC_2$. \texttt{travel-service}$\diamond$\texttt{queryInfo} \\(Train Ticket)}
         \label{fig:workload_latency_queryInfo}
     \end{subfigure}
	\hfill
     \begin{subfigure}[b]{0.95\linewidth}
         \centering
         \includegraphics[width=0.8\linewidth]{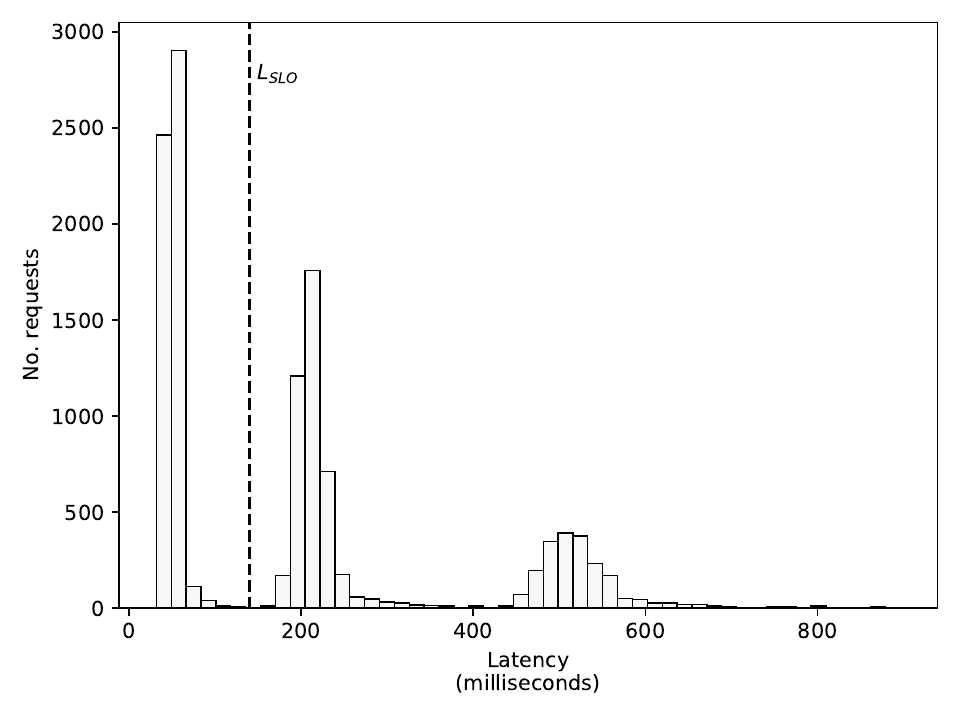}
         \caption{\footnotesize $RC_3$. \texttt{travel-plan-service}$\diamond$\texttt{getByCheapest} \\(Train Ticket)}
         \label{fig:workload_latency_getByCheapest}
     \end{subfigure}
	\hfill
     \begin{subfigure}[b]{0.95\linewidth}
         \centering
         \includegraphics[width=0.8\linewidth]{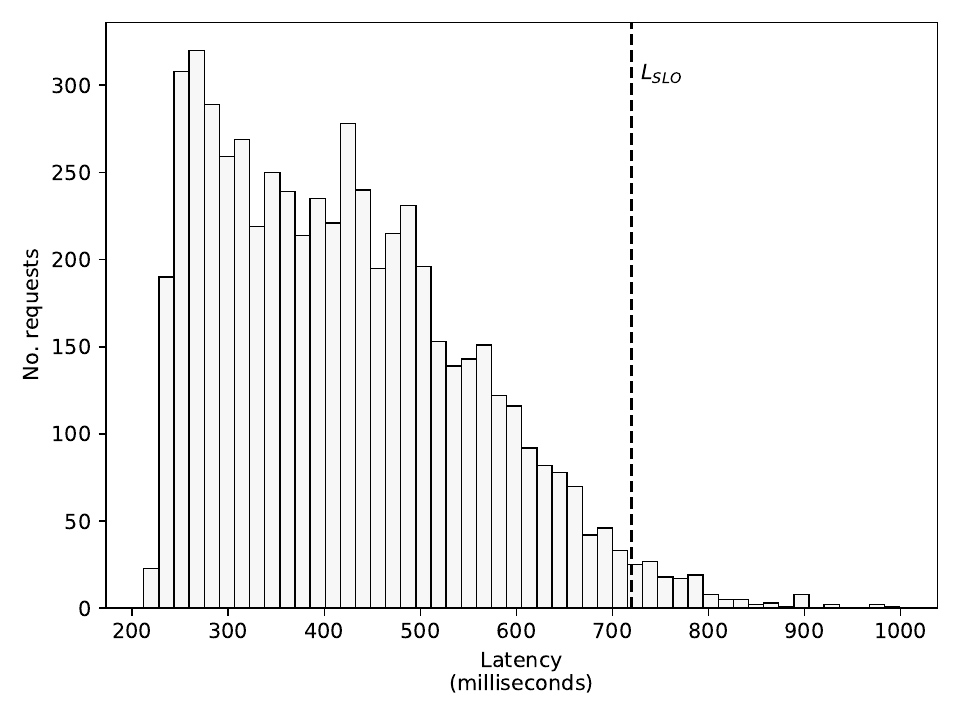}
         \caption{ \footnotesize  $RC_4$. \texttt{preserve-service}$\diamond$\texttt{preserve} \\(Train Ticket)}
         \label{fig:workload_latency_preserve}
     \end{subfigure}

     \caption{Latency distributions for each request class. The subcaption provides the name of the request class, followed by the name of the root RPC denoted as ``\texttt{service}$\diamond$\texttt{api}'', where \texttt{service} denotes the microservice name and \texttt{api} the name of the exposed API. The name of the case study system is reported between brackets. The grey dotted lines depict $L_{SLO}$ thresholds used in our evaluation.}
     \label{fig:workload_latency}
\end{figure} 

To further mitigate this limitation, we perform an additional analysis using more elaborated workloads.
Due to the aforementioned challenges, we do not use ADCs in these cases. Instead, we assess the effectiveness of DeLag in characterizing specific latency behaviors when no artificial delays are injected into the system (\ie no known LDPs/ground-truth), and when mixtures of continuously changing workloads are involved.
This complementary analysis is helpful to (i) demonstrate that DeLag is able to detect patterns correlated with latency deviations even on complex workloads, and (ii) to give a better idea to the reader about the capabilities of DeLag in supporting the analysis of specific latency behaviors.
Similarly to recent studies~\cite{Liao2021,Liao2020}, we use load mixtures that involve multiple types of simulated users (\ie load drivers), where each user type performs different classes of requests on the system.
For example, in the Train Ticket case of study, some types of user may only visit the homepage and subsequently search trains for some random locations, while others may first login and then book random tickets.
Besides this, we also ensure that the number of simulated users per type keeps changing over time.
In this way, workloads will more closely resemble real-world ones, as they generate mixtures of different classes of requests that change over time~\cite{Ardelean2018}.
For the Train Ticket case study, we use \texttt{PPTAM}~\cite{Avritzer2019} as a load generator, which involves 5 different user types (we refer to our replication package~\cite{replication} for details on each user type).
We slightly modified \texttt{PPTAM} to continuously change the number of users of each type at run-time.
\figref{fig:loadshape} shows the number of simultaneous users over time for each \texttt{PPTAM} user type. Overall the number of simultaneous users ranges from a minimum of 20 to a maximum of 31.
 
For the E-Shopper case study,  we use a load generator from our previous work \cite{Cortellessa2022}. In this case, the workload involves three different user types. 
Again, we modified the load generator to ensure that the number of users of each type keeps changing over time. The overall number of simultaneous users over time ranges from a minimum of 17 to a maximum of 31.

We perform a load testing session of 1 hour for each case study. At the end of each session, we transform the operation data in tabular format (as previously explained). 
Since each load testing session involves multiple classes of requests (\eg loading the homepage, booking a ticket), we process each class individually. That is, we create a table like \figref{fig:dataset} for each class of request, which we analyze independently by the other classes. In particular, we deem two requests as of the same class if the chain of RPCs is triggered by the same root RPC, and we define the overall $Latency$ of each request as the execution time of its root RPC.
At the end of this process, we obtain 16 separate datasets (13 for Train Ticket and 3 for E-Shopper), \ie one for each request class. We then discard datasets involving less than 6 unique RPCs, as we consider them too trivial for the sake of the study.
After the filtering step, we obtain 4 datasets (\ie 3 for Train Ticket and 1 for E-Shopper) for 4 distinct request classes, namely $RC_1$, $RC_2$, $RC_3$, $RC_4$. \figref{fig:workload_latency} shows the $Latency$ distribution for each request class.
For the evaluation of \toolName, we set $L_{SLO}$ based on the latency distribution shape. In particular, we visually check the latency distribution and we manually set $L_{SLO}$ as follows: If the distribution is multi-modal (\ie multiple peaks), we set the $L_{SLO}$ to the end of the first distribution ``mode''(see Figures \ref{fig:workload_latency_queryInfo} and \ref{fig:workload_latency_getByCheapest}); on the other hand, if the distribution is uni-modal (\ie one peak), we focus our analysis on the right tail and we set $L_{SLO}$ accordingly (see Figures \ref{fig:workload_latency_home} and \ref{fig:workload_latency_preserve}).
 By doing so, we can investigate the effectiveness of \toolName in two different tasks: (i) characterizing recurrent latency behaviors that deviate from the first mode, and (ii) characterizing tail latency\footnote{Note that, despite the small portion of requests involved, tail latency is often considered critical in distributed systems \cite{Dean2013,Jialin2014}.}.
We assess the effectiveness of \toolName in two ways. The first one is to visually check that patterns characterize specific latency behaviors. For this task we leverage histogram plots, where requests satisfied by patterns are highlighted using darker bars (see \figref{fig:workload_res}). The second one is to quantitatively assess each \emph{pattern set} using precision, recall and F1-score. Note that these metrics are different from the ones defined in Equations (11), (12) and (13) (\ie $Q_{prec}$, $Q_{rec}$ and $Q_{F1}$), since, in this case, there is no prior knowledge on the patterns that affect requests (\ie ADCs). For this reason, in this complementary analysis, precision, recall and F1-score are computed by deeming as \emph{positives} those requests that show latency $L>L_{SLO}$ and as \emph{negatives} the ones with latency $L\leq L_{SLO}$.
In other words, the precision and recall of a \emph{pattern set} are defined as in our optimization objectives (see \secref{sec:obj}), while the F1-score is computed using the usual formula (see \eqqref{eq:fscore}). We also compute the proportion of overlap  within each pattern set to assess the extent to which patterns identify distinct system behaviors. We formally define the overlap of a pattern set $S$ as follows:
\begin{equation}
	overlap = \frac{\mid\bigcup_{P_1, P_2 \in S}(R_{P_1}\cap R_{P_2})\mid}{\mid\bigcup_{P\in S}R_P\mid}
\label{eq:overlap} 	
\end{equation}
where $R_P$ denotes the set of requests that satisfy pattern $P$.
The numerator of \eqqref{eq:overlap} denotes the number of requests that are satisfied by at least two distinct patterns $P_1, P_2\in S$, while the denominator denotes the number of requests that are satisfied by at least one pattern $P\in S$.
 An overlap of 1 indicates redundancy as all the requests are satisfied by at least two patterns.
 On the other hand, an overlap of 0 means that each pattern in the set identifies a distinct subset of requests, and therefore a distinct system behavior.
 Note that we do not consider overlap in our prior analysis since $Q_{prec}$, $Q_{rec}$, $Q_{F1}$ use as ground-truth the set of requests assigned to ADCs, which (by construction) is composed by two distinct subsets, namely $R_{A_1}$ and $R_{A_2}$.

\subsection{RQ$_1$: Effectiveness}\label{sec:rq1}
LDPs provide a useful starting point for the diagnosis and debugging of performance issues in service-based systems. In this RQ, we want to assess the effectiveness of \toolName in detecting LDPs. We compared the effectiveness of \toolName to the one of baseline techniques, using precision ($Q_{prec}$), recall ($Q_{rec}$) and F1-score ($Q_{F1}$) (see \secref{sec:methodology} for a detailed description of these quality measures). In our evaluation, we consider a technique more effective than another if it provides a higher $Q_{F1}$.
In order to answer RQ$_1$, we generated 50 random scenarios for each case study.
Each ADC $A$ is randomly generated as follows.
Firstly, a total delay $\mathfrak{d}$ associated with $A$ is chosen;
that is, every request assigned to $A$ will have an overall slowdown of $\mathfrak{d}$.
Secondly, 1 to 3 RPCs are randomly selected among those executed in the critical path (\ie 5 for E-Shopper and 13 for Train Ticket).
We explicitly chose RPCs in the critical-path to ensure that every delay introduced by $A$ causes latency degradation.
Thirdly, the total delay $\mathfrak{d}$ is evenly split among selected RPCs.
Note that if the RPC $j$ is called a number of $\mathfrak{i}_j$ times in the request, then the whole delay assigned to $j$ is divided by $\mathfrak{i}_j$.
At the end of this process, the ADC $A$ is composed by a set of pairs $\langle j, d\rangle$,
where $d$ denotes the delay in milliseconds that is introduced in each execution of RPC $j$,
and it is such that:
\begin{equation*}
		\mathfrak{d}  = \sum_{\langle j, d\rangle \in A} \mathfrak{i}_j\cdot d
\end{equation*}
The total delay $\mathfrak{d}$ is randomly chosen in the $[L_\mu\cdot 0.2, L_\mu\cdot 0.4]$ range,
where $L_\mu$ is the average request latency for the system without any ADCs.
$L_\mu$ is equal to 116ms for Train ticket and to 393ms for E-Shopper; these values are derived by performing load testing sessions on the non-altered systems using the setup defined in \secref{sec:methodology}.

Modern service-based systems involve many asynchronous interactions \cite{Zhou2018}, therefore many RPCs execution times variations could not produce any degradation on request latency.
In order to reproduce this behavior, we also inject a random delay $\widehat{\mathfrak{d}}$ in one non-critical RPC that does not produce any effect on request latency.
$\widehat{\mathfrak{d}}$ is injected in both non-altered requests and in requests assigned to ADCs (with probability 0.5).
Thus 50\% of requests on each scenario will manifest execution time variations on the selected non-critical RPC.
Similarly to $\mathfrak{d}$,  $\widehat{\mathfrak{d}}$ is randomly chosen, for each scenario, in the $[L_\mu\cdot 0.2, L_\mu\cdot 0.4]$ range.
The non-critical RPCs are carefully chosen by ensuring that every delay $\widehat{\mathfrak{d}}$ in the $[L_\mu\cdot 0.2, L_\mu\cdot 0.4]$ range does not cause slowdown to requests.
This selection required three steps.
First, we ran load testing sessions without injecting any delay into the system.
Then, we visually inspected the traces to identify non-blocking asynchronous RPCs through the trace visualization tool provided by Jaeger or Zipkin (\ie  Gantt chart).
Finally, we validated our experimental setup by ensuring that $\widehat{\mathfrak{d}}$ does not produce any effect on the overall request latency.
In particular, we performed additional load testing sessions on each case study while altering each system with $\widehat{\mathfrak{d}}=L_\mu\cdot 0.4$ and $\mathfrak{d}=0$,
and we verified that the observed average latency of requests does not show notable deviation from $L_\mu$.
In this way, we ensured that, even when injecting the maximum allowed delay $\widehat{\mathfrak{d}}$, the execution time degradation of the non-critical RPC does not cause any notable deviations.
At the end of this process, we selected one non-critical RPC per case study.

Datasets for every scenario and case study are generated by performing load testing sessions by 20 minutes each.
The generation of the 50 datasets for both case studies took $\sim$47 hours,
while experimentation of \toolName and baselines on the generated datasets lasted $\sim$61 hours,
which leads to an overall time of $\sim$4.5 days spent for RQ$_1$ experiments.

We complement the investigation for RQ$_1$ with a quantitative analysis of \toolName on continuously-varying load mixtures (see \secref{sec:methodology} for details). This  additional analysis is helpful to demonstrate that DeLag is able to detect patterns correlated with latency deviations even on elaborated workloads.
The generation of these datasets took approximately 2 hours overall.

\subsubsection*{Results}
\begin{figure*}
  	\includegraphics[width=\linewidth]{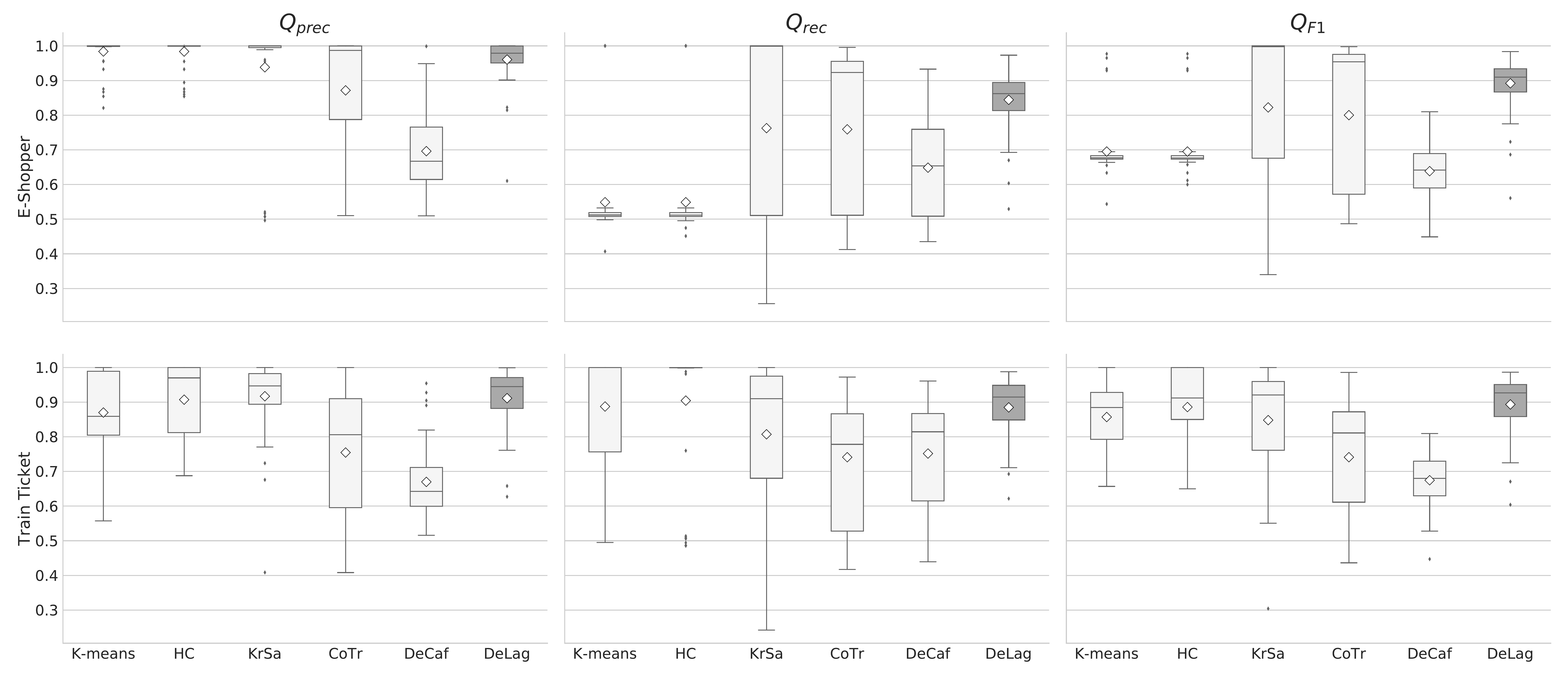}
  	\caption{RQ$_1$. Precision ($Q_{prec}$), Recall ($Q_{rec}$) and F1-score ($Q_{F1}$) for \toolName and baselines methods for each case study. The central box represents the values from the lower to upper quartile (\ie 25 to 75 percentile). The middle line represents the median while the white diamond represents the mean.}
  	\label{fig:rq1}
\end{figure*}

\input{tables/effsize}

\input{tables/rq1}

\begin{figure*}
         \centering
     \begin{subfigure}[b]{0.3\linewidth}
         \centering
         \includegraphics[width=\linewidth]{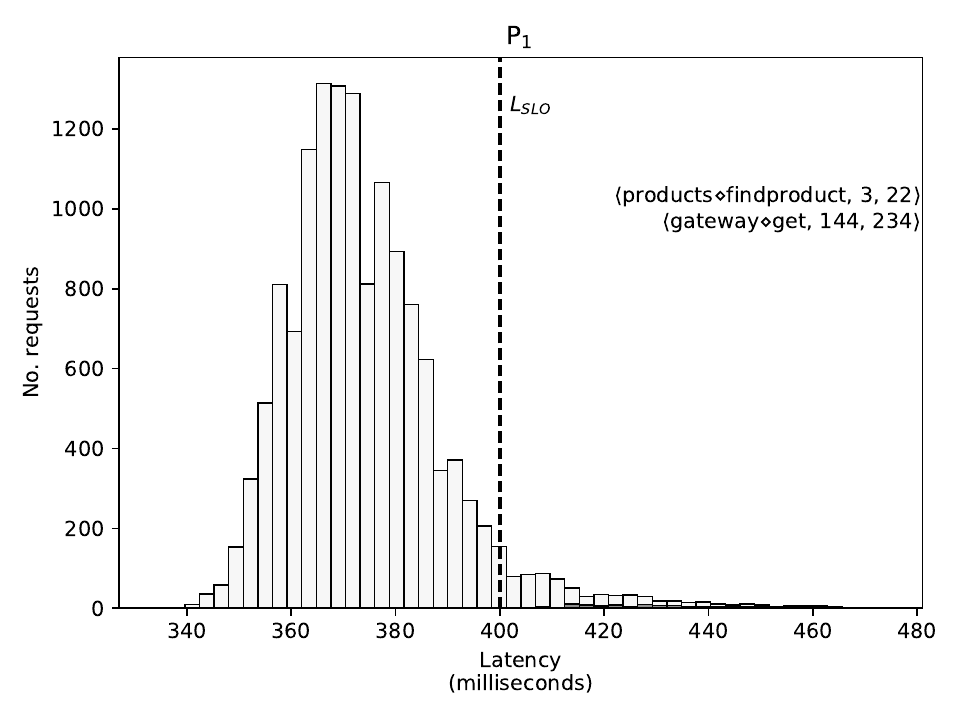}
         \caption{\footnotesize $RC_1$. \texttt{web}$\diamond$\texttt{home} (E-Shopper)}
         \label{fig:workload_res_home}
     \end{subfigure}
     \vfill
     \begin{subfigure}[b]{0.6\linewidth}
         \centering
         \includegraphics[width=\linewidth]{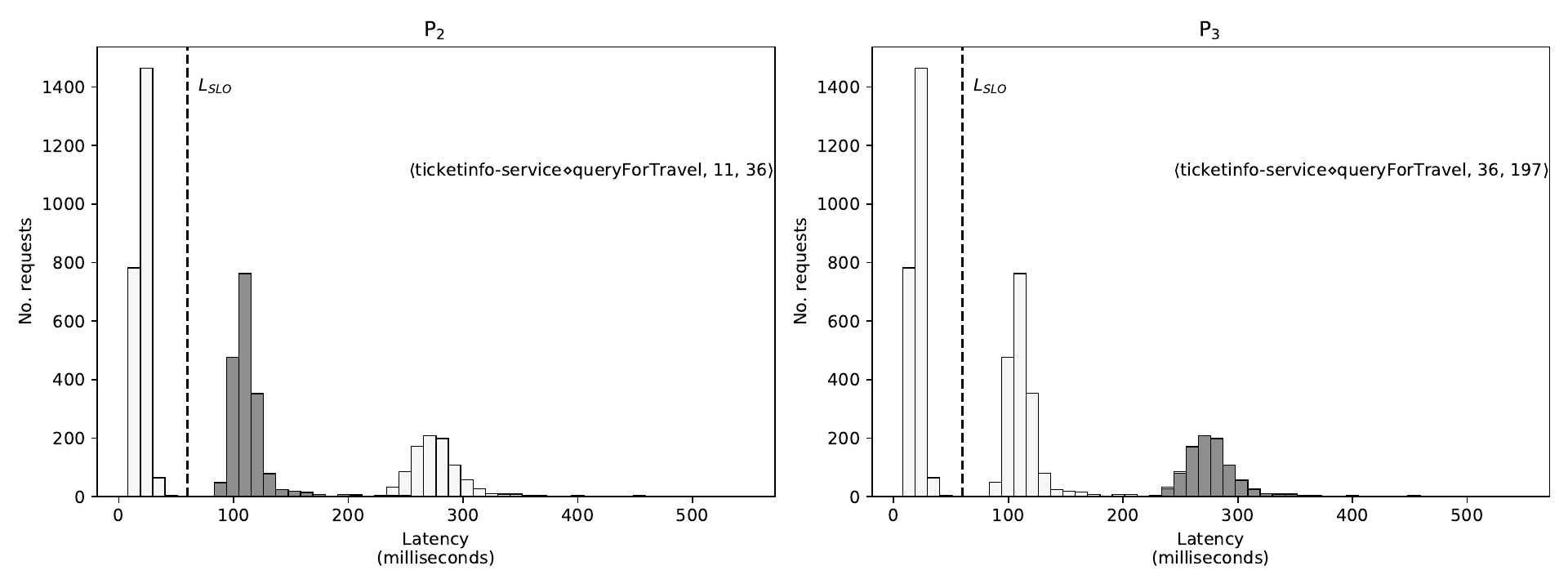}
         \caption{\footnotesize $RC_2$. \texttt{travel-service}$\diamond$\texttt{queryInfo} (Train Ticket)}
         \label{fig:workload_res_queryInfo}
     \end{subfigure}
	\hfill
     \begin{subfigure}[b]{0.6\linewidth}
         \centering
         \includegraphics[width=\linewidth]{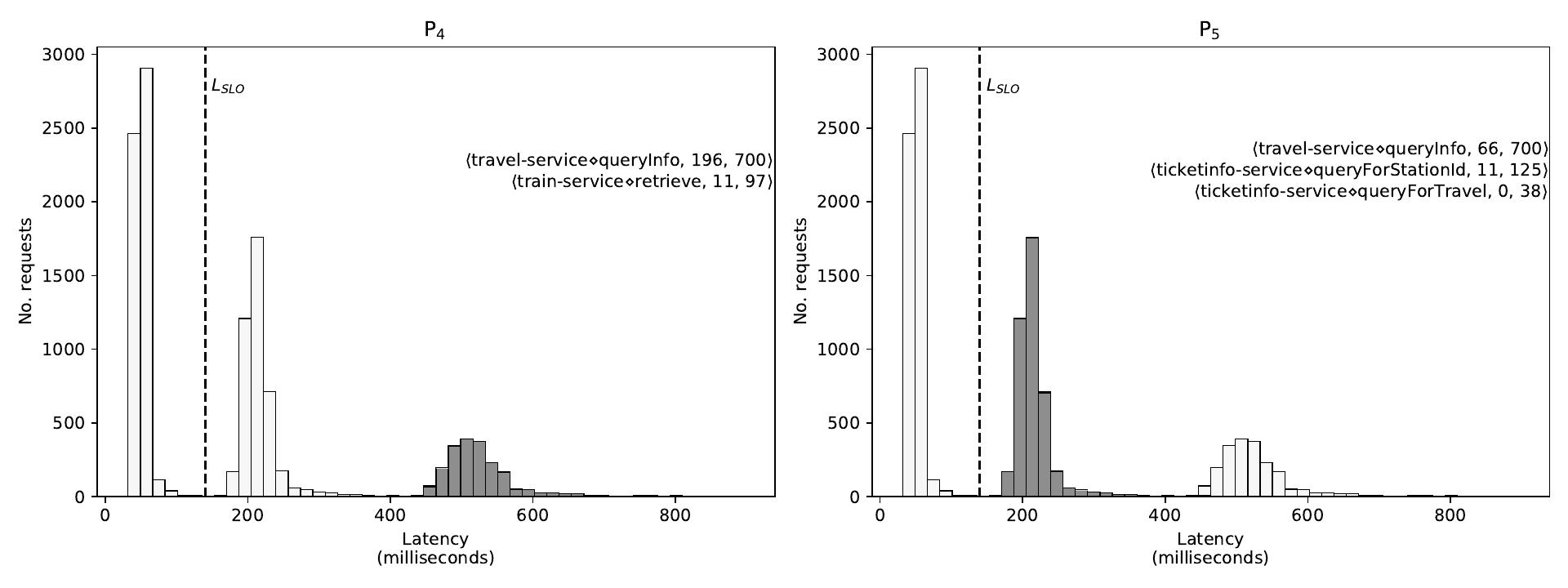}
         \caption{\footnotesize $RC_3$. \texttt{travel-plan-service}$\diamond$\texttt{getByCheapest} (Train Ticket)}
         \label{fig:workload_res_getByCheapest}
     \end{subfigure}
	\hfill
     \begin{subfigure}[b]{0.6\linewidth}
         \centering
         \includegraphics[width=\linewidth]{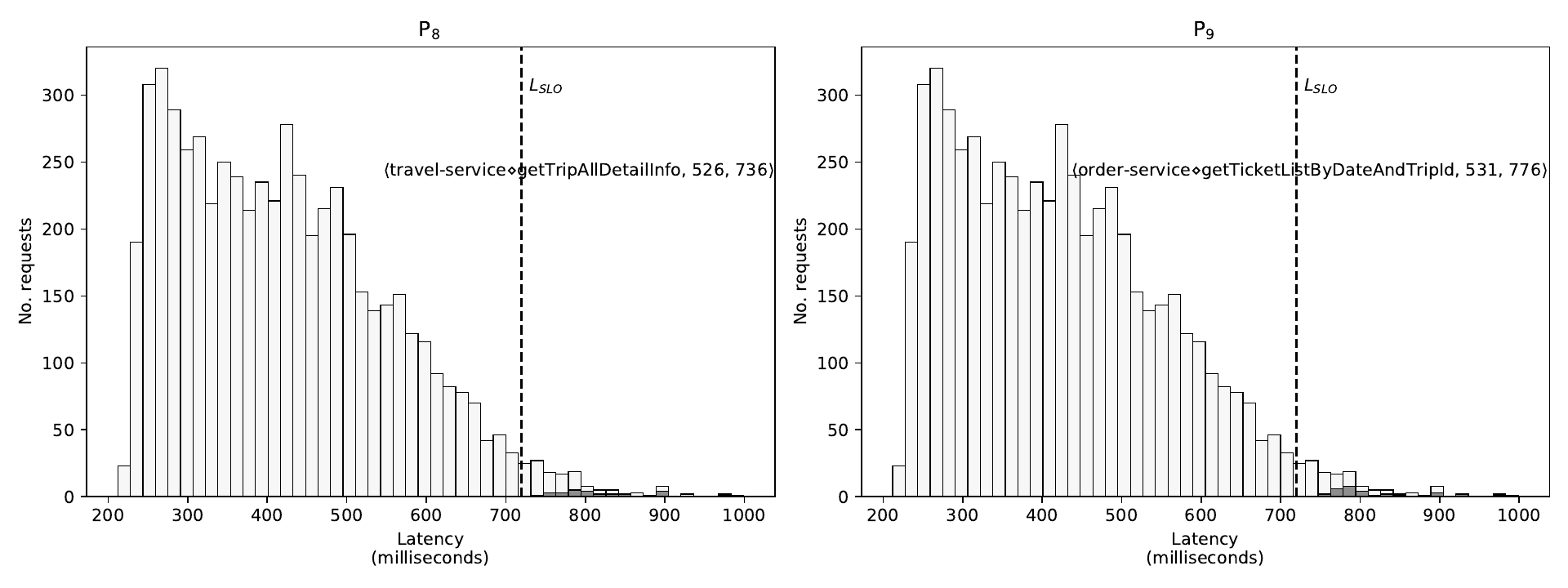}
         \caption{\footnotesize $RC_4$. \texttt{preserve-service}$\diamond$\texttt{preserve} (Train Ticket)}
         \label{fig:workload_res_preserve}
     \end{subfigure}

     \caption{RQ$_1$. Patterns detected by \toolName. Each subfigure illustrates the best two \emph{patterns} detected by \toolName for a specific request class (in terms of F1-score). Each histogram plot highlights in dark grey requests satisfied by a specific pattern $P_i$ . Predicates that constitute the pattern are reported in the upper right part of the plot using the notation ``$\langle$\texttt{service}$\diamond$\texttt{api}, $e_{min}$, $e_{max}$$\rangle$'', where \texttt{service}$\diamond$\texttt{api} denotes the RPC and, $e_{min}$ and $e_{max}$ the execution time range in milliseconds.} 
     \label{fig:workload_res}
\end{figure*}

\input{tables/rq1_workload}

For each scenario, we calculated the mean value of each quality measure, namely precision ($Q_{prec}$), recall ($Q_{rec}$) and F1-score ($Q_{F1}$), over 20
runs for each technique. Note that a single run is performed for KrSa, given its deterministic behavior.
By doing so, we obtained (for each quality measure) 100 values per technique, \ie 50 for each case study. In the case of KrSa, each one of these values represents the result for an individual execution, while for \toolName and baselines it represents the mean value across 20 runs.
\figref{fig:rq1} shows the distribution of these values, where each boxplot contains the mean values
obtained from all scenarios of a given case study for a particular technique.
From a bird's eye view of \figref{fig:rq1}, we can observe that the effectiveness provided by \toolName is more "stable" compared to those provided by other techniques.
The $Q_{F1}$ first and third quartile are 0.86 and 0.93 respectively for E-Shopper, and 0.85 and 0.95 for Train Ticket,
thus leading to an interquartile range (IQR) smaller than any other technique in the Train Ticket case study, and smaller than three out of the five baselines in the E-Shopper case study.
The plot shows that the variation of F1-score provided by \toolName is smaller, compared to other approaches, not only within each case study, but also across them.
For example, precision and recall of clustering algorithms (\ie K-means and HC) show tiny dispersions in the E-Shopper case study,
but their distributions are completely different in the Train Ticket case study,
while those of \toolName just slightly change in the two case studies.

By observing \figref{fig:rq1}, we also note that the mean F1-score (showed as white diamond) of \toolName is higher than those of baselines in both case studies.
The median, instead, is greater than those of all the other baseline techniques in the Train Ticket case study,
and is greater than those of 3 out of the 5 baselines in the E-Shopper case study.
We report results of the Wilcoxon test (together with the corresponding Cliff's delta effect size) in Table \ref{tab:rq1} to compare the statistical significance and effect size of the improvements in terms of precision, recall and F1-score over the baselines.
The effect sizes are interpreted using thresholds provided in prior research \cite{Wan2021, Hassani2018, Ding2020} (see Table \ref{tab:effsize}).
In the E-Shopper case study, \toolName outperforms ($p<0.05$) all the other techniques in terms of F1-score (3 of them with large effect size, one with small effect size and another one with negligible effect size),
while, in the Train Ticket case study, it outperforms ($p<0.05$) 3 out of 5 baseline methods (2 with large effect size and one with small effect size).
However, even in case where comparison leads to $p>0.05$ or where effect sizes are small or negligible,
\toolName is preferable, since it provides more stable effectiveness both within the same case study and across them.
For example, box plots for KrSa, whose F1-score comparison with \toolName reports $p>0.05$ in the Train Ticket case study,
show higher variability with respect to \toolName, especially in terms of recall.
Another example is the comparison with HC,
since it provides a similar effectiveness to \toolName in the Train Ticket case study,
but F1-scores provided by HC are clearly worse ($Q_{F1}<$0.7) than those provided by \toolName for most of the E-Shopper scenarios.

\figref{fig:workload_res} and \tabref{tab:workload_res} report results of the additional analysis on continuously-varying load mixtures.
The first observation that we can make, by looking at \figref{fig:workload_res}, is that \toolName is more effective in characterizing recurrent latency behaviors than tail latency.  This result is reasonable as \toolName is specifically designed to detect patterns correlated to recurrent latency behaviors rather unfrequent ones. In particular, if we look at \figref{fig:workload_res_queryInfo}, we can observe that the two patterns detected by \toolName for $RC_2$ effectively characterize two distinct latency behaviors.
The first one ($P_2$) characterizes requests with latency ranging from 80ms to 220ms, and the second one ($P_3$) characterizes requests with latency larger than 220ms. 
A similar behavior can be also observed in \figref{fig:workload_res_getByCheapest}, where $P_4$ and $P_5$ effectively characterize two distinct latency ranges for $RC_3$. 
The quality measures reported in \tabref{tab:workload_res} further confirm the effectiveness of \toolName in these cases. The pattern set detected for $RC_2$ provides a 0.999 F1-score, a 1 precision and a 0.999 recall, while the set detected for $RC_3$ reports F1-score of 0.999, precision of 1, recall of 0.998.
The optimal precision achieved by \toolName for $RC_2$ and $RC_3$ suggests that patterns are peculiar to requests with latency larger than $L_{SLO}$, while the high recall indicates that \toolName is able to detect patterns that effectively characterize almost all the requests in the targeted latency range, \ie only 0.1\% ($RC_2$) and 0.2\% ($RC_3$) of requests with latency $L>L_{SLO}$ do not satisfy any pattern.
Furthermore, \toolName reports zero overlap for $RC_2$, thereby indicating that patterns characterize distinct subsets of requests, and, therefore, distinct system behaviors, while it reports moderate overlap for $RC_3$, \ie 0.329.

By contrast, looking at Figures \ref{fig:workload_res_home} and \ref{fig:workload_res_preserve}, we can observe that \toolName reports only modest effectiveness when used to characterize tail latency.
The pattern set detected for $RC_1$, which is composed by a single pattern $P_1$, achieves a recall of only 0.158, hence suggesting that $P_1$ characterizes only a small portion of requests with latency $L>L_{SLO}$ (15.8\%).
In $RC_4$, instead, \toolName achieves a recall of 0.37, but it still leaves without any characterization more than half of requests with latency larger than $L>L_{SLO}$.
On the other hand, \toolName reports relatively high precisions in both these cases (0.95 for $RC_1$ and 1.0 for $RC_4$), which lead to F1-score of respectively 0.27 and 0.54 for $RC_1$ and $RC_4$. The overlap is moderate for $RC_4$ (0.269), while it is zero for $RC_1$ since the set involves only an individual pattern.

Summing up, we answer RQ$_1$ as follows:
\toolName can effectively detect LDPs,
since it provides a (very often) improved and always more stable effectiveness compared to those of baseline techniques. In addition, our complementary analysis suggests that \toolName can support performance analysis of service-based systems, even when workloads involve mixtures of loads that change over time.

\subsection{RQ$_2$: Overlapping Patterns}\label{sec:rq2}
\input{tables/rq2}

In their paper, Krushevskaja and Sandler~\cite{Krushevskaja2013} highlighted a major limitation of F1-score-based techniques: They are less effective when distinct patterns lead to partially (or entirely) overlapping latency distributions. We provided a more detailed description of this problem at the beginning of \secref{sec:model} . With this RQ, we want to assess whether \toolName overcomes this limitation.
To answer RQ$_2$, we evaluate how proximity of latency distributions (related to distinct LDPs) affects \toolName's effectiveness (\ie $Q_{F1}$).
We generated datasets (for both case studies), while varying the distance between the total delay introduced by $A_1$ and the one introduced by $A_2$.
We defined 5 different experimental setups of delays assigned to $A_1$ and $A_2$ respectively.

Table \ref{tab:rq2} reports these setups, which range from scenarios where latency distributions related to ADCs completely overlap (\ie $Distance=L_\mu\cdot 0$) to scenarios where distributions are clearly separated (\ie $Distance=L_\mu\cdot 0.2$).
For each setup, we generated 20 scenarios (\ie datasets),
where total delays introduced by $A_1$ and $A_2$ are fixed by the setup,
but RPCs involved in each scenario are different.
Then, we avoid influence on effectiveness due to non-critical RPC execution time variation by fixing
$\widehat{\mathfrak{d}}$ to $L_\mu\cdot 0.3$ in all the generated scenarios.
In order to avoid extremely long experimentation time,
we decreased the duration of load testing sessions to 5 minutes.
Nevertheless, we expect that this does not affect the validity of our results, as the number of requests involved in each dataset is still relevant (more than 2.5k requests), and dataset size mainly affects efficiency of techniques (see RQ$_4$) rather than their effectiveness.
Overall, the duration for experiments related to RQ$_2$ took $\sim$5 days.
The dataset generation process lasted $\sim$43 hours,
while the execution of \toolName and baseline techniques on the 200 generated datasets took $\sim$74 hours.

\subsubsection*{Results}
\begin{figure}
  	\includegraphics[width=\linewidth]{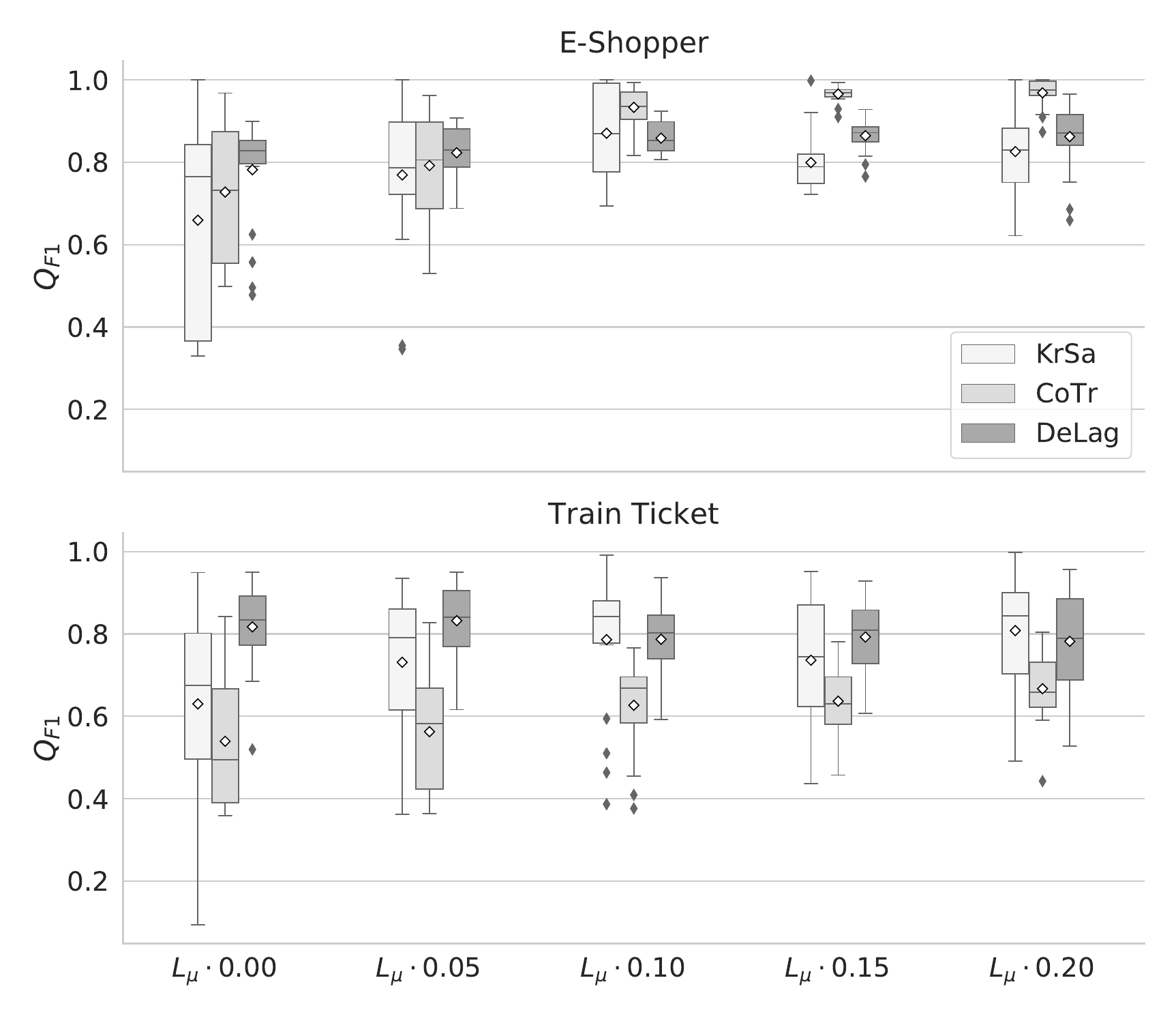}
  	\caption{RQ$_2$. F1-scores ($Q_{F1}$) for KrSa, CoTr and \toolName under different experimental setups (see Table \ref{tab:rq2}).
  	The x-axis labels report the expected distance between the average latency of requests in $R_{A_1}$ and the one of those in $R_{A_2}$.
  	}\label{fig:rq2_box}
\end{figure}
\begin{figure}
  	\includegraphics[width=\linewidth]{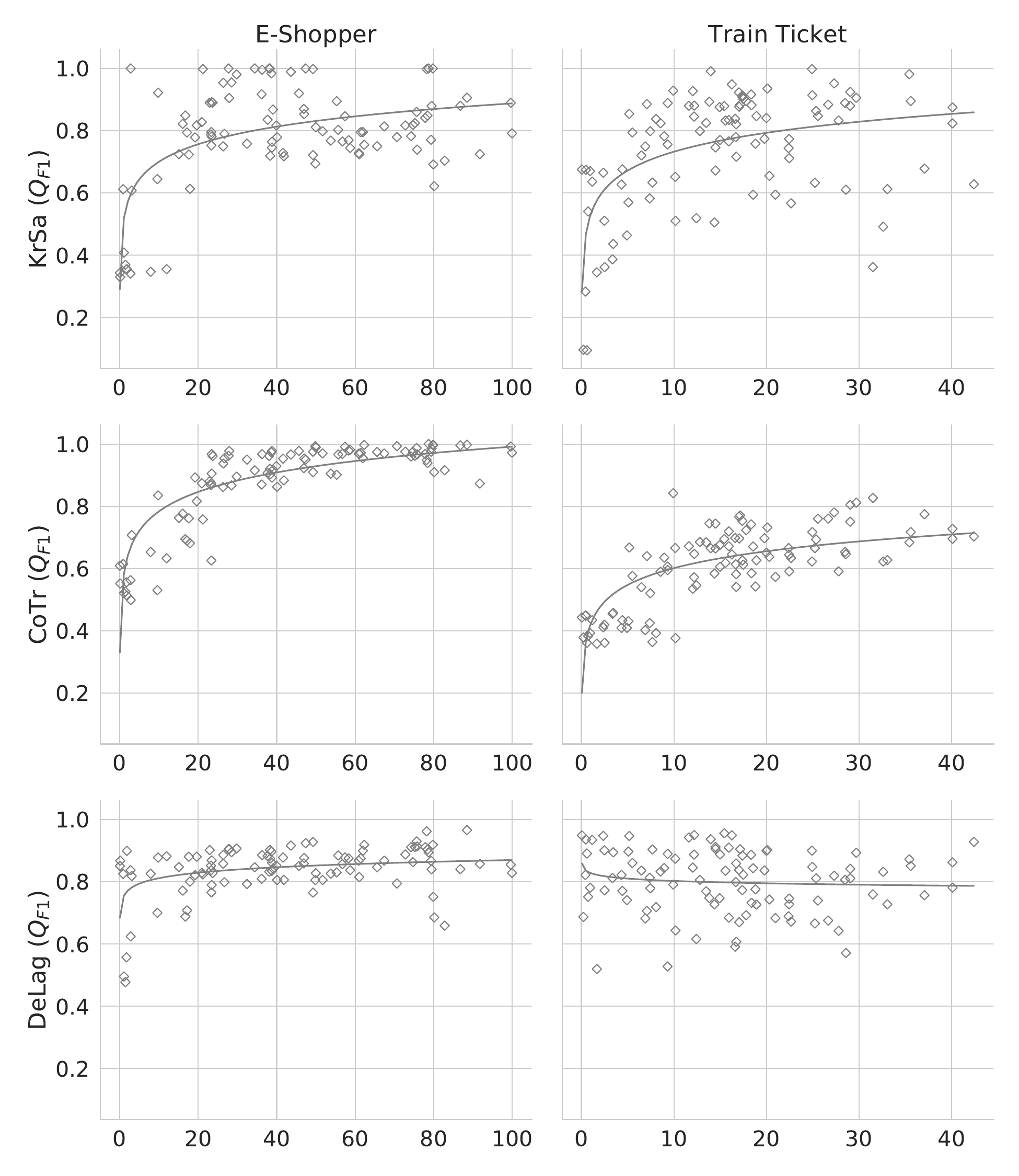}
  	\caption{RQ$_2$. F1-scores provided by KrSa, CoTr and \toolName as function  of the distance (in milliseconds) between the observed average latency of requests in $R_{A_1}$ and the one of those in $R_{A_2}$. Each point of the plot represents the mean F1-score for the method on a particular scenario.
  	  }
  	\label{fig:rq2_scatter}
\end{figure}

We calculated the mean F1-score ($Q_{F1}$) among 20 runs for each technique (except for KrSa).
\figref{fig:rq2_box} depicts the distributions of these means for \toolName, CoTr and KrSa under different experimental setups.
Given the goal of RQ$_2$, in \figref{fig:rq2_box} we do not include results for K-means, HC and DeCaf.
For completeness, results for other techniques are reported in our online appendix \cite{replication}.
\figref{fig:rq2_box} shows that both KrSa and CoTr are less effective as far as the distance between total delays introduced by $A_1$ and $A_2$ decreases (\ie from right to left on the x-axis).
The mean, the median, the first and the third quartile for KrSa and CoTr always decrease starting from $Distance=L_\mu\cdot 0.10$ until $Distance=L_\mu\cdot 0.0$.
This confirms the evidence provided by \cite{Krushevskaja2013} about the inadequacy of F1-score-based approaches on patterns leading to similar latency behaviors.
The same behavior cannot be observed on \toolName, which instead seems to improve for the same range of setups in Train Ticket and does not show a relevant decrease in those in E-Shopper.
From \figref{fig:rq2_box} we can assert that the F1-score provided by \toolName is more stable across different setups than those of F1-score-based methods.
This finding is further confirmed by \figref{fig:rq2_scatter},
 which shows mean F1-scores provided on each scenario as function of the distance between the observed average latency of requests in $R_{A_1}$ and the one of those in $R_{A_2}$.
Logarithmic regression lines in plots clearly show a trend towards lower effectiveness for both KrSa and CoTr when distance between latency related to different patterns decreases.
The same trend does not show up in \toolName.

Summing up, we answer RQ$_2$ as follows:
closeness of latency distributions related to different patterns does not affect the effectiveness of \toolName, 
therefore our approach overcomes this limitation of F1-score-based methods.

\subsection{RQ$_3$: Non-critical RPCs}\label{sec:rq3}
Modern service-based systems involve many asynchronous interactions \cite{Zhou2018}. Hence, many RPCs execution time variations may not produce any degradation on request latency. With this RQ,  we intend to evaluate whether the effectiveness of \toolName is affected by execution time variations on non-critical RPCs,	\ie RPCs whose execution time variations do not cause latency degradation.
We generated datasets for different scenarios while controlling the magnitude of delay introduced on these RPCs.
Similarly to \secref{sec:rq1}, one asynchronous RPC is selected for each case study.
We used different experimental setups,
where each setup is defined by a different value assigned to $\widehat{\mathfrak{d}}$, that is the amount of delay introduced in the non-critical RPC.
We used 5 different experimental setup that are defined by the following values of $\widehat{\mathfrak{d}}$: $L_\mu\cdot 0.0$, $L_\mu\cdot 0.1$,  $L_\mu\cdot 0.2$, $L_\mu\cdot 0.3$ and $L_\mu\cdot 0.4$.
For each setup we generated 20 different scenario.
A delay of $\widehat{\mathfrak{d}}$ is introduced in the non-critical RPC with 0.5 probability on each request performed to the altered system.
Complementarily to what we have done for RQ$_2$, in order to avoid influence on the effectiveness due to closeness of latency distributions related to distinct ADCs,
here we fix delays introduced by $A_1$ and $A_2$ to $L_\mu\cdot 0.25$ and $L_\mu\cdot 0.35$ respectively.
Datasets are generated by performing load testing sessions of 5 minutes, as done for RQ$_2$.

Overall, the duration for experiments related to RQ$_3$ took $\sim$5 days.
The dataset generation process lasted $\sim$44 hours.
The execution of \toolName and baseline techniques on the 200 generated datasets
 took $\sim$74 hours.

\subsubsection*{Results}
\begin{figure}
  	\includegraphics[width=\linewidth]{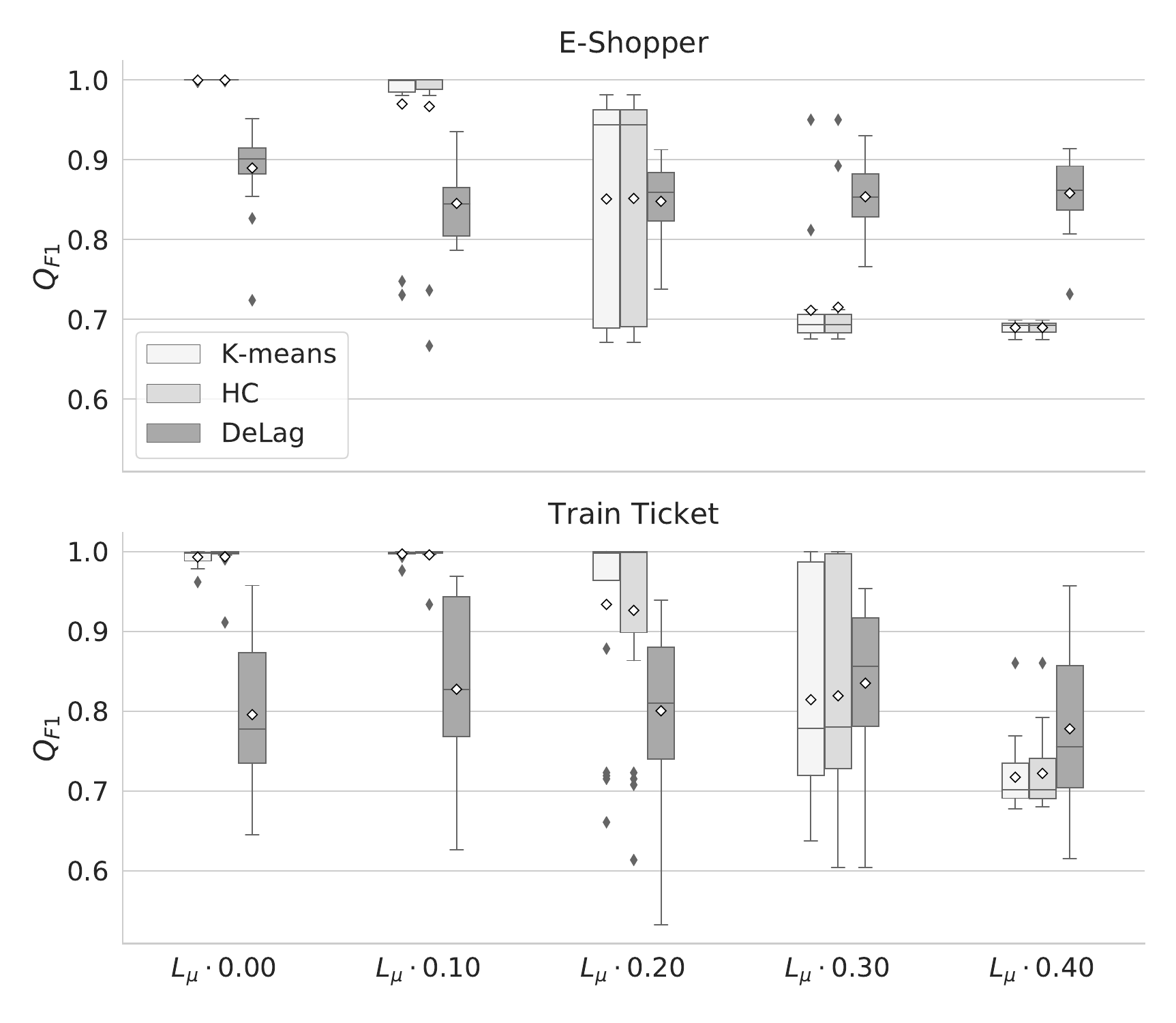}
  	\caption{RQ$_3$. F1-scores ($Q_{F1}$) for K-means, HC and \toolName under different experimental setups.
The x-axis labels report the amount of delay introduced in non-critical RPCs ( $\widehat{\mathfrak{d}}$ ) for each experimental setup.
  	}
  	\label{fig:rq3_box}
\end{figure}

\figref{fig:rq3_box} shows distributions for F1-scores provided by K-Means, HC and \toolName for each experimental setup.
Given the goal of RQ$_3$, we avoided to show results for all techniques, which may make figures less readable. Instead, we showed \toolName's results and compared them against those of techniques that seem to be affected by non-critical execution time deviation, namely HC and K-means.
For completeness, results for other techniques (KrSa, CoTr and DeCaf) are reported in our online appendix \cite{replication}.
\figref{fig:rq3_box} does not suggest a correlation between $\widehat{\mathfrak{d}}$ and \toolName effectiveness.
For example, in the E-Shopper case study, both median and mean of F1-scores decrease from scenarios with $\widehat{\mathfrak{d}}=L_\mu\cdot 0.0$ to scenarios with $\widehat{\mathfrak{d}}=L_\mu\cdot 0.1$, but they both slightly increase in all the subsequents setups.
In the Train Ticket case study, instead, both mean and median of F1-scores show an alternating behavior when $\widehat{\mathfrak{d}}$ increases.
On the contrary, the effectiveness of some other techniques seems to be monotonically affected by execution time variation on non-critical RPCs.
For example, both K-Means and HC provide near-optimal effectiveness when $\widehat{\mathfrak{d}}=L_\mu\cdot 0.0$, \ie there is no execution time variation on non-critical RPC,
but their F1-scores significantly decrease as far as $\widehat{\mathfrak{d}}$ increases.
Summing up, we answer RQ$_3$ as follows:
the effectiveness of \toolName is not monotonically affected by execution time variations in non-critical RPCs.
This is a fundamental step towards the adoption of \toolName in real world settings,
since asynchronous interactions are pervasive in today's service-based systems.

\subsection{RQ$_4$: Efficiency}\label{sec:rq4}
\input{tables/rq4_setup}

As we explained in \secref{sec:scenario}, \toolName is thought to be periodically executed (\eg every day) to automatically detect LDPs. Unfortunately, the detection of LDPs may be computationally expensive, as it can involve the processing of large volume of traces (\eg modern service-based systems collect thousands of traces per day or even more).
For this reason, efficiency is a major concern for \toolName. With this RQ, we aim to assess the efficiency of \toolName, by comparing it to the efficiency of baseline approaches.
In order to analyze the efficiency of \toolName and baseline approaches, we record the elapsed time to complete the entire end-to-end pattern detection process on different datasets with varying sizes.
We generated datasets of different sizes for both case studies by using 5 different experimental setups,
which control the duration of load testing sessions for each scenario (see Table \ref{tab:rq4_setups}).

Longer sessions obviously lead to higher number of requests to analyze, thus to more computationally expensive runs.
For each setup we considered 20 different scenarios,
which are randomly generated by using the same approach as for RQ$_1$.
Note that datasets generated for Train Ticket are more computationally expensive than those of E-Shopper, 
since the number of RPCs under analysis is higher in the former case (25 unique RPCs compared to 7).

Unlike for effectiveness assessment, we do not expect that efficiency of techniques is influenced by randomness,
therefore we perform a single run of each technique on each dataset to reduce the time effort.

The overall data generation process for RQ$_4$ took $\sim$10 days,
while the execution of \toolName and baseline techniques took $\sim$4 days and a half.

\subsubsection*{Results}
\input{tables/rq4_res}

\begin{figure*}
  	\includegraphics[width=\linewidth]{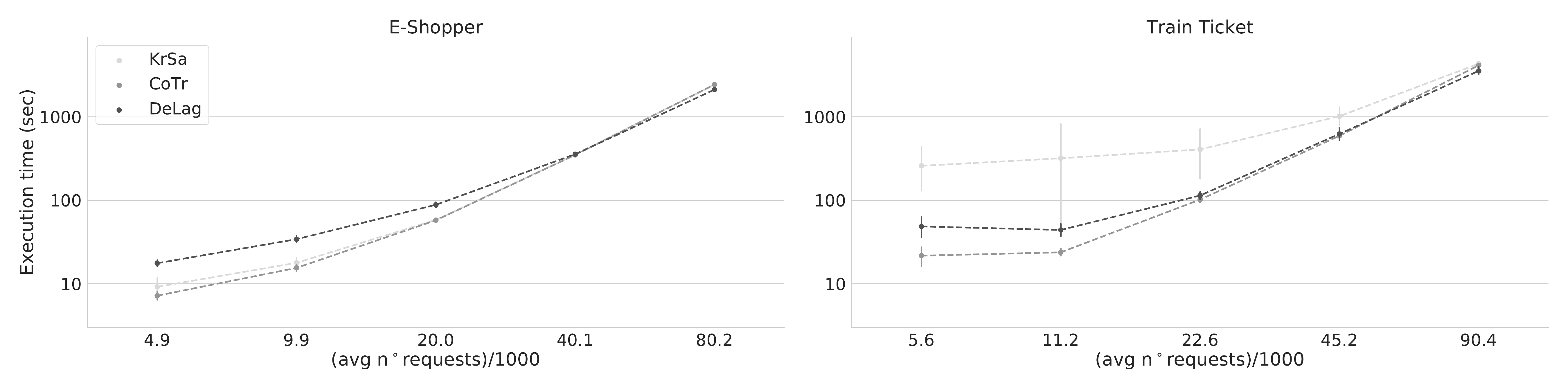}
  	\caption{RQ$_4$. Execution times in seconds of KrSa, CoTr and \toolName for datasets of different sizes. Each point represent the mean execution time on 20 different scenarios within the same experiment setup, \ie similar number of involved requests. The y axis represents the execution time,
  	while x axis labels report the average number of requests for datasets of each experiment setup (see Table \ref{tab:rq4_setups}). The x and y axes are in log scale.
  	}
  	\label{fig:rq4_trend}
\end{figure*}

\begin{figure}
  	\includegraphics[width=\linewidth]{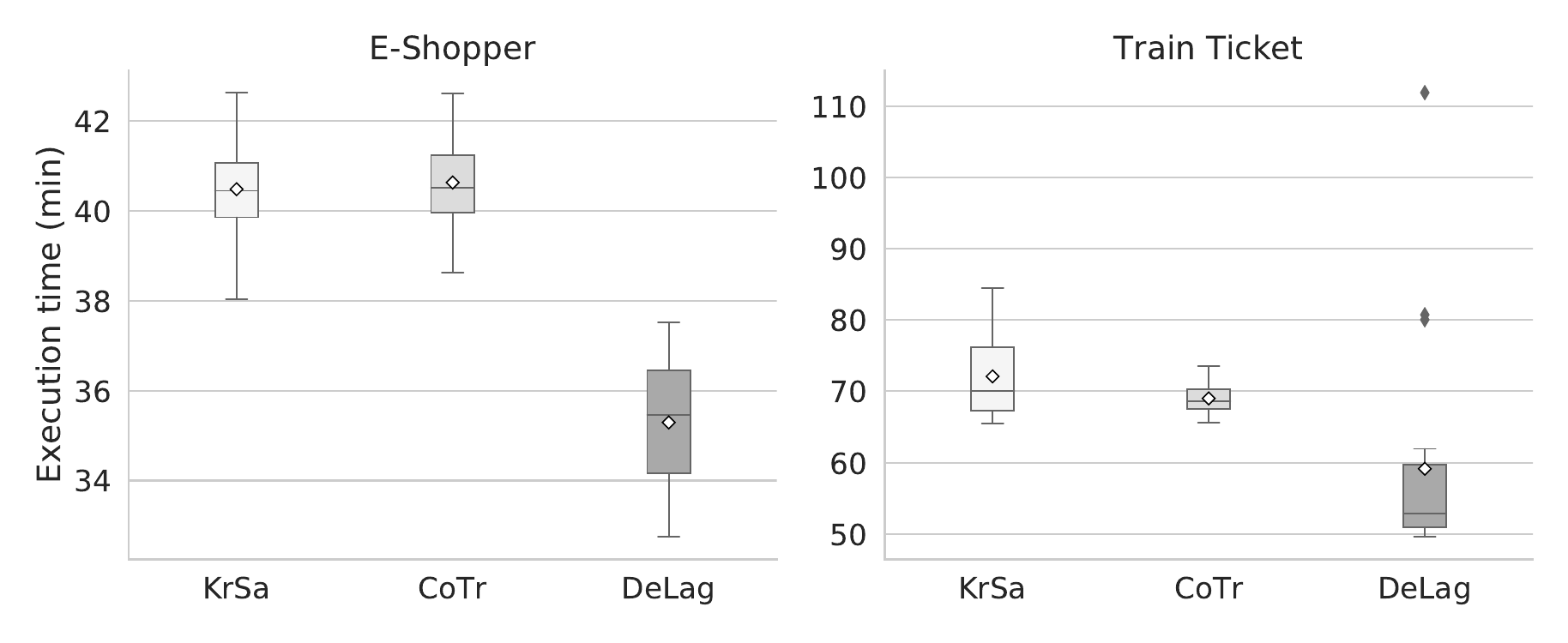}
  	\caption{RQ$_4$. Execution times in minutes of KrSa, CoTr and \toolName on the largest datasets, i.e datasets generated by load testing sessions of 160 minutes.
  	}
  	\label{fig:rq4_box}
\end{figure}

Table \ref{tab:rq4_res} shows the average execution time of each technique for each experimental setup.
DeCaf, K-means and HC severely outperform \toolName in terms of efficiency.
DeCaf is 531.9 times (E-Shopper) and 580.3 times (Train ticket) more efficient than \toolName on datasets generated by 160 minutes load testing sessions, \ie the largest datasets in our evaluation.
On the same datasets, K-means and HC outperform \toolName by 479.3 and 19.3 times, respectively, in the E-Shopper case study, and by 457.4 and 8.3 times in the Train ticket case study.
Despite their efficiency, these techniques have been shown to be less effective when compared to others.
For example, \figref{fig:rq1} showed that both the mean and the median F1-score provided by DeCaf are below 0.7 in both case studies,
while the mean and the median F1-score provided by clustering methods are below 0.7 in one out of the two case studies.
Moreover, even KrSa and CoTr provide a mean F1-score above 0.7 and a median F1-score above 0.8 in both case studies,
therefore overall they provide better effectiveness when compared to DeCaf and general-purpose clustering algorithms.

\figref{fig:rq4_trend} shows the mean execution time of techniques for each experiment setup (see Table \ref{tab:rq4_setups}) under each case study.
On the x axes the average number of requests contained in datasets within the same experiment setup are reported.
On the E-Shopper case study, both KrSa and CoTr are more efficient than \toolName on smaller datasets.
For example, KrSa and CoTr respectively outperform \toolName by 0.92 and 1.45 times on datasets involving $\sim$4.9k requests.
Nevertheless the figure clearly shows that the difference in terms of execution times between \toolName and F1-score-based techniques decreases as the number of requests increases.
Indeed, KrSa and CoTr outperform \toolName by only 0.01 and 0.02 times, respectively, on datasets involving $\sim$40.1k requests,
while, on the largest datasets ($\sim$80.2 requests),
the mean execution time of \toolName is smaller than those of KrSa and CoTr.

On datasets related to Train Ticket case study,
\toolName outperforms KrSa, the most effective baseline method,
in all experiment setups by 0.22 to 6.23 times.
When compared to CoTr, instead, \toolName shows a similar behavior to the one observed in the E-Shopper case study;
that is,  CoTr is more efficient than \toolName when dealing with smaller datasets, but as number of requests grows, the gap between the efficiency of our approach and CoTr decreases until \toolName becomes faster than CoTr.
For example, CoTr is more efficient than \toolName by 1.24 times on smallest datasets ($\sim$5.6k requests).
The improvement of CoTr over \toolName becomes 0.85 on datasets with $\sim$11.2k requests, 0.12 on datasets with $\sim$22.6k requests and 0.05 on datasets with $\sim$45.2k requests.
On the largest datasets, instead, \toolName is faster than CoTr.

\figref{fig:rq4_box} shows execution times in minutes of each technique on the largest datasets used in our evaluation ($\sim$80.2k requests for E-Shopper and $\sim$90.4k requests for Train Ticket).
The mean execution time of \toolName is 35 minutes on E-Shopper and 59 minutes on Train Ticket.
\toolName outperforms KrSa in efficiency by 0.15 times (E-Shopper) and by 0.22 times (Train Ticket),
while CoTr is outperformed by 0.15 times on E-Shopper and by 0.17 times on Train Ticket.

Summing up, we answer RQ$_4$ as follows:
\toolName is consistently less efficient than DeCaf and general-purpose clustering algorithms.
Nevertheless, these techniques provide lower effectiveness on the majority of considered scenarios when compared to \toolName.
\toolName is also less efficient than the second and the third most effective technique when dealing with smaller datasets,
but the efficiency of \toolName improves (when compared to these baselines) as the size of dataset increases.
Moreover, \toolName clearly outperforms both KrSa and CoTr on the largest datasets used in our evaluation.

\subsection{RQ$_5$: Performance analysis}\label{sec:rq5}

To give a better idea to the reader on how \toolName can support performance analysis, we qualitatively analyze \toolName results in the context of continuously varying workloads. 
In particular, we manually analyze the \emph{patterns} provided by \toolName to assess if/how they are helpful when trying to understand latency behavior of case study systems.
Patterns providing an F1-score lower than 0.4 are excluded by this analysis.
For each pattern, we randomly sample 10 requests identified by the pattern and compare them against 10 random requests showing latency smaller than $L_{SLO}$.
Requests are compared using the trace comparison tool\footnote{Jaeger trace comparison \url{https://bit.ly/3Jn6p2L}} and the trace visualization tool\footnote{Jaeger trace page \url{https://bit.ly/3u5Bt0s}} of Jaeger.
The traces and patterns are also cross-related to the source code of the system to better understand system behavior.

Besides this qualitative analysis, we also report the scope reduction provided by \toolName both in terms of requests and RPCs to inspect. Finally, we compare the outcomes of \toolName with those of two general-purpose techniques for performance issue diagnosis \cite{Liao2021, Xiong2013} to assess the advantages (and/or disadvantages) of our pattern detection approach. Both these baselines leverage machine learning algorithms, such as linear regression, to train performance models of software systems (\eg in our evaluation, independent variables are RPCs, and the dependent variable is request latency).  These models are then post-processed to derive metrics (in our case, RPCs) correlated to performance deviation.
For example, the technique proposed by Liao et al.~\cite{Liao2021} (which we name here as LLR) evaluates the effect of each RPC by systematically increasing the associated execution time while assessing how this increase impacts predicted latency. Eventually, LLR returns a ranking where RPCs are sorted based on their impact on latency deviation. For the sake of our evaluation, we only consider the top-5 RPCs.
The technique proposed by Xiong et al.~\cite{Xiong2013} (which we name here as XLR), instead, leverages a hill climbing algorithm to identify the best combination of RPCs that gives the best prediction accuracy measured by average $R^2$ (coefficient of determination).
We assess the quality of RPCs identified by baseline methods\footnote{We reimplemented the two baselines using Python while carefully following the specifications provided in the original papers, both of them use linear regression as machine learning models. The source code used in our experiments is available in our replication package \cite{replication}.} by analyzing their correlation with respect to latency degradation. In order to do that, we measure Pearson correlation coefficient (\textit{r}), and we visually inspect scatter plots where the x-axis represents RPC execution time and the y-axis represents request latency.
Finally, we compare the results provided by LLR and XLR with those provided by \toolName to assess the advantages/disadvantages of our pattern detection approach.

Note that all the analyses performed for answering this RQ involve exclusively datasets generated from continuously varying workloads (see \secref{sec:methodology} for details), as they are the ones that more closely resemble realistic system usage.

\subsubsection*{Results}

To give a better idea to the reader on how \toolName can support performance analysis, we qualitatively analyze patterns $P_2$, $P_3$, $P_4$, and $P_5$, \ie the ones showing F1-score higher than 0.4 (see \figref{fig:workload_res} and \tabref{tab:workload_res}).
In particular, we manually analyze traces related to requests identified by patterns, as well as the source code of RPCs occurring in predicates. Through this analysis, we were able to characterize (in most of cases) \emph{why} latency diverges.
For example, we found that the deviation from the first latency mode in $RC_2$ (see \figref{fig:workload_res_queryInfo}) is induced by specific request inputs that trigger some (otherwise not invoked) RPCs (\eg the city departure and destination for the ticket search).
Notably, these ``special'' inputs  produce two distinct system behaviors.
In the first one, the \texttt{ticketinfo$\diamond$queryForTravel} RPC is invoked once, while in the other one it is invoked three times.
\toolName captures this dichotomy through the RPC execution time intervals defined in $P_2$ and $P_3$.
In particular, when \texttt{ticketinfo$\diamond$queryForTravel} has an overall execution time that ranges from 11ms and 36ms ($P_3$), it means it is invoked once.
On the other hand, when \texttt{ticketinfo$\diamond$queryForTravel} has an execution time that ranges from 36ms and 197ms ($P_2$), it is invoked three times.
The pattern $P_4$ characterizes a similar behavior for $RC_3$. 
For example, the  \texttt{train-service$\diamond$retrieve} RPC, which is part of a predicate of $P_4$, is never invoked in requests showing latency smaller than $L_{SLO}$, while it always appears in requests identified by $P_4$, since the predicate involves an overall execution time higher than 11ms (see \figref{fig:workload_res_getByCheapest}).
We were unable to spot any specific system behavior for $P_5$.

Overall, our qualitative analysis suggests that \toolName can fairly support performance analysis of service-based systems.

In order to assess the benefits of using \toolName when compared to other performance issue diagnosis techniques, we compare \toolName results with those of two machine learning-based techniques, \ie XLR \cite{Xiong2013} and LLR \cite{Liao2021}.
Our results\footnote{We do not report XLR and LLR results in the paper due to space concerns. However, we made them publicly available in our replication package~\cite{replication}.} suggest that LLR and XLR are effective in identifying RPCs correlated with latency degradation.
For all the request classes, they provide at least one RPC reporting a strong correlation with request latency, \ie Pearson correlation coefficient $r > 0.5$. The visual inspection of the scatter plots confirmed the tight relation between the identified RPCs and request latency.
This finding indicates that RPCs identified by LLR and XLR provide a useful starting point for performance issues diagnosis, as they identify RPCs showing execution times that behave in close relation with latency degradation.
Interestingly, we found a significant alignment among the results of \toolName and baselines. For example, when considering LLR,  we found that, in all the request classes, \toolName returns pattern sets that involve at least one RPC that is also identified by LLR and which reports a strong correlation with request latency (\ie $r > 0.5$).
XLR reports a similar alignment in terms of returned RPCs (except in one case, \ie $RC_3$).
This convergence might suggest that \toolName, XLR and LLR can be used interchangeably to support performance diagnosis. However, there are significant differences in these approaches.
The most notable one is that \toolName provides additional information on top of the RPCs to inspect.
First, it provides a set of RPC execution time behaviors that are correlated with latency degradation (\ie predicates). Second, it identifies specific sets of requests to investigate. Finally, it provides a set of quantitative indicators on the relevance of each pattern (\ie precision, recall and F1-score).
This additional information can be extremely useful during performance analysis to (i) prioritize performance diagnosis activities and (ii) guide performance analysis.
For example, performance analysts can leverage the F1-score to prioritize which pattern to investigate first, as patterns with high F1-scores are likely to indicate more relevant latency degradations, and therefore more impactful performance issues.
Furthermore, \toolName can significantly reduce the scope of operational data to inspect during performance analysis. In our evaluation, requests satisfied by patterns account for approximately 14.16\% of the overall set of requests on average. Such scope reduction is not provided by XLR and LLR, which conversely require further analyses to identify clusters of requests worth of investigation.
Also, if we compare the reduction of scope in terms of RPCs to inspect, we can observe that patterns provided by \toolName involve only 10.45\% of the overall set of RPCs on average, while the solutions provided by XLR involve 14.28\% of RPCs on average. Note that we cannot assess RPC scope reduction in LLR as this approach returns a ranking rather than a subset of RPCs.

Overall, these analyses suggest that using \toolName provides considerable advantages over general-purpose techniques for performance issue diagnosis. Detected patterns can be used to narrow the scope of the analysis both in terms of requests and RPCs to inspect, as they provide a useful starting point on which requests (and RPCs) to look at when trying to understand why latency behavior deviates from the ``expected'' one.
To further strengthen (or debunk) the validity of our findings, we encourage future user studies to assess if/how \toolName can effectively support engineers in performance analysis in real-world service-based systems.

%% file: tables/effsize.tex
\begin{table}
\center
\scriptsize
\begin{tabular}{ll}
\toprule
Cliff's Delta Value & Intepretation\\
\midrule
$\delta < 0.147$ & Negligible \\
$0.147 \leq \delta < 0.33 $ & Small \\
$0.33 \leq \delta < 0.474$ & Medium \\
$\delta \geq 0.474$ & Large \\
\bottomrule
\end{tabular}
\caption{Interpretation of Cliff's delta value}
\label{tab:effsize}
\end{table}

%% file: tables/rq1.tex
\begin{table}
\scriptsize
\begin{tabular}{llrrr}
\toprule
Case study & Technique &  Q$_{prec}$ & Q$_{rec}$ & Q$_{F1}$ \\
\midrule
E-Shopper & K-means &     1.000(0.47) &  $<$0.001(0.84) &  $<$0.001(0.83) \\
             & HC &     1.000(0.51) &  $<$0.001(0.84) &  $<$0.001(0.83) \\
             & KrSa &     0.966(0.34) &     0.028(0.15) &     0.043(0.17) \\
             & CoTr &     0.016(0.10) &     0.034(0.06) &     0.026(0.09) \\
             & DeCaf &  $<$0.001(0.92) &  $<$0.001(0.75) &  $<$0.001(0.95) \\
\midrule
Train Ticket & K-means &     0.011(0.19) &     0.853(0.46) &     0.002(0.22) \\
             & HC &     0.306(0.16) &     0.975(0.60) &     0.384(0.07) \\
             & KrSa &     0.764(0.10) &     0.083(0.04) &     0.186(0.03) \\
             & CoTr &  $<$0.001(0.58) &  $<$0.001(0.53) &  $<$0.001(0.61) \\
             & DeCaf &  $<$0.001(0.88) &  $<$0.001(0.59) &  $<$0.001(0.92) \\
\bottomrule
\end{tabular}
\caption{RQ$_1$. Results of the Wilcoxon test (Cliff's delta effect size in brackets) performed on the precision ($Q_{prec}$), recall ($Q_{rec}$) and F1-score ($Q_{F1}$),  provided by \toolName compared to those provided by baseline methods.}
\label{tab:rq1}
\end{table}

%% file: tables/rq1_workload.tex
\begin{table*}[h]
\centering
\scriptsize
\begin{tabular}{|l|l|r|r|r|r|}
\toprule
Req. type &       Pattern Set &                      Precision &                                             Recall &                                           F1-score &  Overlap \\
\midrule
      $RC_1$ &                       \{P$_1$\} &                                        0.951 (P$_1$:0.95) &                                                        0.158 (P$_1$:0.16) &                                                        0.271 (P$_1$:0.27) &     0.0 \\
      $RC_2$ &                \{P$_2$, P$_3$\} &                                1 (P$_2$:1 $\mid$ P$_3$:1) &                                      0.999 (P$_2$:0.65 $\mid$ P$_3$:0.35) &                                      0.999 (P$_2$:0.79 $\mid$ P$_3$:0.51) &     0.0 \\
      $RC_3$ &  \{P$_4$, P$_5$, P$_6$, P$_7$\} &  1 (P$_4$:1 $\mid$ P$_5$:1 $\mid$ P$_6$:1 $\mid$ P$_7$:1) &  0.998 (P$_4$:0.33 $\mid$ P$_5$:0.67 $\mid$ P$_6$:0.15 $\mid$ P$_7$:0.18) &  0.999 (P$_4$:0.50 $\mid$ P$_5$:0.80 $\mid$ P$_6$:0.26 $\mid$ P$_7$:0.30) &   0.329 \\
      $RC_4$ &                \{P$_8$, P$_9$\} &                                1 (P$_8$:1 $\mid$ P$_9$:1) &                                      0.369 (P$_8$:0.23 $\mid$ P$_9$:0.23) &                                      0.539 (P$_8$:0.38 $\mid$ P$_9$:0.38) &   0.269 \\
\bottomrule
\end{tabular}
\caption {RQ$_1$. DeLag quality measures for each request class.
The first column reports the name of the requests class and the second reports the \emph{pattern set} found by \toolName. 
The third, fourth and fifth columns report respectively the precision, recall and F1-score of the \emph{pattern set} (quality measures of individual patterns between parenthesis). The last columns reports the proportion of overlap among requests satisfied by each pattern in the set.}
\label{tab:workload_res}

\end{table*}

%% file: tables/rq2.tex
\begin{table}
\center
\scriptsize
\begin{tabular}{| c | c | c | c |}
\hline
 $\mathfrak{d_1}$ & $\mathfrak{d_2}$ & $Distance$ \\
\hline
 $L_\mu\cdot 0.3$ & $L_\mu\cdot 0.3$ & $L_\mu\cdot 0$ \\
\hline
$L_\mu\cdot 0.275$ & $L_\mu\cdot 0.325$ & $L_\mu\cdot 0.05$ \\
\hline
$L_\mu\cdot 0.25$ & $L_\mu\cdot 0.35$ & $L_\mu\cdot 0.1$ \\
\hline
$L_\mu\cdot 0.225$ & $L_\mu\cdot 0.375$ & $L_\mu\cdot 0.15$ \\
\hline
$L_\mu\cdot 0.2$ & $L_\mu\cdot 0.4$ & $L_\mu\cdot 0.2$ \\
\hline
\end{tabular}
\caption {RQ$_2$. Experimental setups.
Each row represents a particular setup,
where $\mathfrak{d_1}$ and $\mathfrak{d_2}$ denote total delays introduced by $A_1$ and $A_2$ respectively, and $Distance$ denotes expected distance between average request latency of $R_{A_1}$ and the one of $R_{A_2}$.}\label{tab:rq2}
\end{table}

%% file: tables/rq4_setup.tex
\begin{table}
\scriptsize
\center
\begin{tabular}{| c | c | c |}
\hline
 \shortstack{Load testing \\ duration (min)} & \shortstack{Avg n$^\circ$ requests \\ (E-Shopper)}  & \shortstack{Avg n$^\circ$ requests \\ (Train ticket)}\\
\hline
10 & 4932 & 5578 \\
\hline
20 & 9945 & 11177\\
\hline
40 & 19981 & 22563 \\
\hline
80 & 40066 & 45175 \\
\hline
160 & 80212 & 90441 \\
\hline
\end{tabular}
\caption {RQ$_4$. Experimental setups.
The first column reports duration in minutes of load testing sessions to generate each dataset.
The second and the third columns report the average number of requests contained in each dataset of each setup within the same case study.}
\label{tab:rq4_setups}
\end{table}

%% file: tables/rq4_res.tex
\begin{table}
\scriptsize
\resizebox{!}{49pt}{
\begin{tabular}{llrrrrrr}
\toprule
 Case study & \shortstack{n$^\circ$ req.} &  K-means &      HC &     KrSa &     CoTr &  DeCaf &  \toolName \\
\midrule
E-Shopper & $\sim$4.9k  &      1.4 &    1.3 &     9.2 &     7.2 &    0.4 &      17.6 \\
             & $\sim$9.9k  &      2.2 &    2.0 &    17.8 &    15.5 &    0.6 &      34.1 \\
             & $\sim$20k &      2.6 &    6.8 &    58.0 &    57.5 &    1.0 &      88.3 \\
             & $\sim$40.1k &      3.1 &   25.3 &   351.0 &   349.2 &    1.8 &     356.1 \\
             & $\sim$80.2k &      4.4 &  104.4 &  2428.5 &  2437.5 &    4.0 &    2117.5 \\
\midrule
Train Ticket & $\sim$5.6k  &      2.5 &    2.0 &   258.5 &    21.7 &    0.5 &      48.5 \\
             & $\sim$11.2k &      2.7 &    3.4 &   318.3 &    23.8 &    0.7 &      44.0 \\
             & $\sim$22.6k &      3.5 &   10.1 &   406.5 &   101.7 &    1.4 &     113.9 \\
             & $\sim$45.2k &      4.7 &   64.7 &  1015.3 &   591.4 &    2.7 &     622.9 \\
             & $\sim$90.4k &      7.7 &  380.0 &  4325.6 &  4140.3 &    6.1 &    3548.5 \\
\bottomrule
\end{tabular}}
\caption{RQ$_4$. Average execution times (in seconds) of techniques for each experimental setup under different case studies.
The second column reports the approximate average number of requests involved in each experimental setup.
}
\label{tab:rq4_res}

\end{table}

%% file: threats.tex
\subsection{Threats to Validity}\label{sec:threats}
\subsubsection{Internal validity}
Both \toolName and baseline techniques are subject to overfitting, \ie patterns involving negligible numbers of requests.
In order to deal with this behavior, each technique provides one or more configurable parameters.
The use of different configurations may lead to different results,
thus causing unfair comparison among the effectiveness provided by different techniques.
On the other hand, the experimentation of techniques for different combinations of parameter values can be impractical due to extremely long execution times.
In order to minimize this threat, we set parameters across different techniques with "similar policies".
Namely, we used a reasonable threshold across different techniques such that each detected pattern must involve at least $|R_{pos}|\cdot 0.05$ requests (\ie 5\% of requests not meeting SLO expectations).
For example, we forced the Mean Shift algorithm, used by KrSa and CoTr for encoding and split point selection, to discard clusters of requests with sizes less or equal to $|R_{pos}|\cdot 0.05$.
Similarly, the Mean Shift algorithm used in the Search Space Construction phase of \toolName is forced to discard clusters smaller than $|R_{pos}|\cdot 0.05$.
The same threshold (\ie $|R_{pos}|\cdot 0.05$) is also used in the Genetic Algorithm of \toolName to penalize solutions with patterns involving small numbers of requests.
Finally, we used the same value to define the minimum number of training data in a leaf node \cite{Breiman2001} for the random-forest model of DeCaf.

Using different genetic algorithm parameters (\eg crossover and mutation rates) may change the results of \toolName.
Nonetheless, a comprehensive investigation of the impact of these parameters is impractical due to the time-consuming nature of our experiments.
We defined crossover and mutation probabilities by fine-tuning them on subset of 5 datasets from the RQ$_1$ experimental setup (see \secref{sec:rq1}). In particular, we started from the parameters we used in our previous work \cite{Cortellessa2020} (\ie crossover=0.8 and mutation=0.2), and we slightly modified them while checking whether they improve or not the effectiveness of \toolName.
Eventually, we fixed crossover and mutation probability respectively to 0.6 and 0.4.

\toolName and baselines use randomized algorithms, therefore each execution may potentially lead to different results.
Guidelines to assess randomized techniques \cite{Arcuri2011} recommend to perform a high number of repeated runs (e.g 1000 repetitions).
However, using such a high number of repetition in our experiments would be unfeasible due to extremely long execution times.
In order to have statistically significant results in a reasonable time, we performed 20 runs per technique.

\subsubsection{Construct validity}
We generated several scenarios to test the effectiveness of \toolName in LDPs detection.
A potential threat to our work is that ADCs do not represent relevant causes of latency degradation within each scenario,
\ie they do not generate LDPs.
In order to minimize this threat, we plot latencies distribution for each scenario and we check that requests assigned to ADCs are prevalent in requests showing degraded latency, \ie $L>L_{SLO}$.

In order to have quantitative measures on the effectiveness of techniques, we chose, among the returned set of patterns, two patterns ($P_{A_1}$ and $P_{A_2}$) that seems to be related to the targeted ADCs.
Selecting different patterns may result in different effectiveness measures ($Q_{prec}$, $Q_{rec}$, $Q_{F1}$).
One option we considered was to select them manually, but this approach has two cons: it takes significant human effort and it can leave room for subjectivity.
Therefore we opted for an automated approach which selects, for each ADC $A$, the pattern with maximum F1-score while considering requests assigned to $A$ as positives and all other requests as negatives.
We are aware that patterns selected using this strategy can be suboptimal,
and that there may be other patterns among those returned by each technique that can provide higher effectiveness,
but overall we expect that our selection strategy is reasonable enough to evaluate the effectiveness of \toolName, and to compare it with the baseline.

Another threat to our work is that we evaluated \toolName only on scenarios where two LDPs are involved, therefore the effectiveness of techniques may change when considering more patterns.
Nevertheless, we showed that our approach is more effective than those of baselines techniques when two distinct LDPs are involved.

The workloads used in our experimental setup may not be representative of real-world workloads.
However, we chose to use these workloads for a specific reason: We want to ensure that requests assigned to ADC are prevalent in requests showing degraded latency.
In particular, we used 20 parallel users that make requests to the targeted API by randomly waiting from 1 to 3 seconds from one request to another. This provided us a reasonable amount of concurrency while ensuring that ADCs represent major causes of latency degradation (see \secref{sec:methodology} for details).
Additionally, our experimental design enabled the evaluation of DeLag (and other techniques) on a wide range of scenarios (\ie 700) involving different combinations of Latency Degradation Patterns (\ie different RPCs and delays involved).

\subsubsection{External validity}
\toolName achieves a high effectiveness in our evaluation.
We cannot ensure that \toolName can achieve the same effectiveness on other datasets outside our experimental setup (\eg real world scenarios).
Nevertheless, through an evaluation on 700 randomly generated scenarios for two case studies, we showed that our approach is more effective than three state-of-the-art approaches and two general-purpose clustering algorithms.
Datasets for scenarios are generated in laboratory since, at best of our knowledge, there are no publicly available datasets suitable to validate our work.
Moreover, it is challenging to find industries that are willing to share their operational data. 
We evaluated \toolName on two case studies involving, respectively, 25 invocations across 7 unique RPC, and 48 invocations across 14 unique RPCs. The effectiveness and efficiency of \toolName may change when used on requests that involve a higher number of RPCs. We did our best to find open-source systems that involve a non-trivial number of RPCs. Before choosing them, we considered a wide range of service-based systems used in prior work (\eg Acme Air \cite{Ueda2016}, Socks Shop \cite{Aderaldo2017}, TeaStore \cite{Eismann2020}). To the best of our knowledge, E-Shopper and Train~Ticket represent the most suitable choices for our purpose, as they involve the highest number of unique RPCs within the same request. As a matter of fact, Train~Ticket was specifically developed by Zhou \etal \cite{Zhou2018} to fill the lack of non-trivial open-source systems that can be used to conduct practice-relevant research on service-based systems. In addition, unlike other benchmark systems, E-Shopper and Train~Ticket both use distributed tracing solutions, which is a prerequisite for the use of \toolName.

We evaluated the efficiency of \toolName on datasets of different sizes, ranging from 4.9k requests to 80.2k requests for E-Shopper and ranging from 5.6k requests to 90.4k requests for Train Ticket.
LDPs detection in real world service-based systems may involve higher number of requests.
Nevertheless, we showed that the efficiency of \toolName improves, when compared to the second and the third most effective technique, as the number of requests increases.
Moreover, \toolName outperforms both these techniques on the largest datasets used in our evaluation.

%% file: discussion.tex
\section{Discussion}\label{sec:discussion}
We found that clustering algorithms (K-means and HC) are significantly more efficient than \toolName.
Nevertheless, these techniques are shown to be less effective than our approach, \ie \toolName outperforms both clustering algorithms in one case study ($p<0.001$ and large effect size) and provides comparable effectiveness in the other one.
In addition, when compared to \toolName, these techniques show the following limitations.
First, they could be ineffective on systems involving asynchronous executions of RPCs (which are nowadays pervasive),
since we showed that execution time variations of non-critical RPCs severely affects their effectiveness.
Second, clustering algorithms output clusters of requests without providing additional information,
while \toolName also provide RPC execution time characteristics, \ie patterns,
which can be very useful for debugging purposes.
We conclude that the use of \toolName is preferable over clustering algorithms.

We also found that \toolName effectiveness significantly outperforms DeCaf on both case studies (p$<0.001$ with large effect size).
Therefore, when effectiveness is the key priority, we suggest the use of \toolName over DeCaf.
However, when higher efficiency is required
and even moderate effectiveness is acceptable, DeCaf should be considered.
Moreover, DeCaf capability goes beyond LDPs detection, since it enables the detection of patterns on high-ordinal categorical trace attributes (i.e categorical attributes with a high number of possible values),
which is out of the scope of this work.
In future work, we plan to extend the capability of our approach also to high ordinal categorical trace attributes.

\toolName also outperforms in terms of effectiveness CoTr in the E-Shopper case study ($p<0.001$ and large effect size), as well as in the Train Ticket case study ($p\leq 0.05$ and small effect size).
When compared to KrSa, instead, \toolName provide better effectiveness in the E-Shopper case study ($p\leq 0.05$ and negligible effect size) while statistical test returns $p>0.05$ in the Train Ticket study.
Moreover, we found that effectiveness provided by F1-scores-based techniques (KrSa and CoTr) is less stable than the one provided by \toolName (IQRs for $Q_{F1}$ provided by \toolName are significantly smaller than those of KrSa and CoTr).
In addition, these techniques are less effective when distinct patterns lead to partially (or entirely) overlapping latency distributions, while \toolName overcomes this limitation.
Our approach also outperforms in terms of efficiency F1-score-based techniques on largest datasets used in our evaluation.
We conclude that the use of \toolName is preferable over F1-score-based techniques.

Overall, \toolName provides better and more stable effectiveness than other techniques.
Moreover, \toolName is more efficient than the second and the third most effective techniques on the largest datasets used in our evaluation.
However, when higher efficiency is required, fastest techniques such as clustering algorithms or DeCaf could be preferable.
Nevertheless, practitioners have to take into account the limitations of these latter techniques. 

%% file: related_work.tex
\section{Related Work}\label{sec:related}
In this section we summarize prior work on automated diagnosis of software systems.
These techniques can be classified into two broad categories: (1) those that detect patterns in time-series metrics and (2) those that detect patterns in traces.
Time series metrics are operational data measured over intervals of time (\eg average response time per minutes of a particular RPC), while traces contain data about causally related events of individual end-to-end requests (\eg RPC execution time for a particular request).
Our approach, \toolName, falls in the second category, since it detects patterns in distributed traces related to individual end-to-end requests of a service-based system.

Cohen~\etal~\cite{Cohen2004} devised the first technique for automated diagnosis of performance issues in software systems.
They used a class of probabilistic models (Tree-Augmented Bayesian Networks) to identify combinations of time-series metrics and threshold values that correlate with compliance with SLOs for average-case response time.
Duan~\etal introduced Fa \cite{Duan2009}, an automated diagnosis technique that uses anomaly-based clustering to clusters time-series metrics based on how they differ from those related to failure and pinpoints metrics linked to failure.
MonitorRank~\cite{Kim2013} uses the historical and current time-series metrics, along with the call graph of the service-based system to build an unsupervised model for ranking.
This technique identifies metrics correlated to system anomalies by using an adaptation of the PageRank algorithm \cite{Page1999}.
Farshchi~\etal~\cite{Farshchi2015} adopts regression-based analysis to find the correlation between operation’s activity logs and the operation activity’s effect on cloud resources.
Syer~\etal~\cite{Syer2013} proposed an automated technique to diagnose memory-related issues. They combine performance counters and log events by discretizing them into time-slices, and, then, use statistical techniques (\eg clustering and correlation analysis) to identify a set of log events corresponding to a memory-related issue.
Other techniques for detecting patterns in time series metrics relies on association rule mining \cite{Brauckhoff2012}, hierarchical detectors \cite{Nair2015} or pairwise-correlation analysis \cite{Malik2010}.

The first work that falls in the second category (i.e, the one based on traces) is the one introduced by Chen~\etal,
which uses decision trees \cite{Chen2004} to identify causes of failures.
In this technique, decision trees are trained on traces, and combinations of trace attributes are ranked according to their degree of correlation with failure.
Han~\etal introduced StackMine~\cite{Han2012}, a technique that mines callstack traces to help performance analysts to effectively discover costly callstack patterns.
StackMine identifies callstack patterns correlated with poor performance by using an adaptation of a classic association rule mining algorithm \cite{Wang2004}.
Unfortunately these techniques \cite{Chen2004, Han2012} are unsuitable to identify patterns in continuous attributes (\eg execution time),
since they specifically target categorical trace attributes.
For example, StackMine specifically targets function names within callstack traces,
while the technique proposed by Chen~\etal~\cite{Chen2004} targets categorical request trace attributes such as host machine names, request types and thread ids.
At the best of our knowledge, the first automated diagnosis technique suitable for pattern detection in continuous trace attributes is the one proposed by Krushevskaja and Sandler \cite{Krushevskaja2013}.
In this technique, the pattern detection problem is modeled as
as a binary optimization problem and solved using combinatorial search algorithms (\ie dynamic programming combined with branch-and-bound algorithm or forward feature selection).
Although this approach works with continuous trace attributes,
a non-automated encoding step is required to transform continuous trace attributes to binary features.
In our recent work (CoTr)~\cite{Cortellessa2020}, we introduced few advancement on top of the work of Krushevskaja and Sandler: (1) an automated approach to discretize continuous attributes based on the Mean Shift algorithm \cite{Cheng1995},
(2) the approach search on a wider search space using a genetic algorithm, and (3) according to our preliminary experimental evaluation \cite{Cortellessa2020}, it is faster and more effective.
Both the approach of Krushevskaja and Sandler \cite{Krushevskaja2013} and ours \cite{Cortellessa2020} are based on a similar technique.
The latency range considered as degraded is divided through a set of split points, and for each sub-interval the pattern with the best F1-score is considered;
the algorithm searches for the split of the latency range that maximizes the sum of the F1-scores.
We remark that \toolName is not a proper extension of CoTr, in that the former involves a novel modeling approach to the problem, a new workflow, and a different set of search algorithms. CoTr searches for the optimal partition of the latency range that maximizes F1-score, whereas \toolName simultaneously searches multiple LDPs for the whole latency range by combining a multi-objective genetic algorithm and a decision-making heuristic.
Recently, Bansal~\etal introduced DeCaf \cite{Bansal2019}, a technique based on random forests,
which can be applied both on categorical and continuous attributes.
Similarly to \cite{Chen2004}, this technique first trains a random forest model and then ranks predicates extracted by the model according to their correlation with system anomalies.
Bansal~\etal demonstrated that DeCaf can be applied on traces with categorical attributes with up to 1M cardinality, by evaluating their approach in two large scale services.

In this paper we specifically focus on Latency Degradation Patterns 	\cite{Cortellessa2020}(\ie RPC execution time patterns correlated with latency degradation), therefore we compared the effectiveness and efficiency of \toolName to those of techniques suitable to this problem, \ie automated diagnosis techniques that can be applied for pattern detection in continuous trace attributes \cite{Krushevskaja2013, Cortellessa2020, Bansal2019}.

Other studies on software diagnosis rely on visualization techniques.
Beschastnikh~\etal~\cite{Beschastnikh2020} introduced ShiViz, which presents distributed system executions as interactive time-space diagrams to help diagnosis and debugging of software issues.
The study of Sambavisan~\etal~\cite{Sambasivan2013} compares three well-known visualization approaches in the context of  presenting the results of one automated performance root cause analysis approach \cite{Sambasivan2011}.
Visualization techniques are time consuming, as they require human intervention and are more useful when performing fine-grained analysis,
while \toolName automatically detect patterns in RPCs execution times to identify potential relevant performance issues.

\toolName detects patterns in traces collected by a distributed tracing infrastructure,
therefore distributed tracing research \cite{Sambasivan2016} is related to our work.
Dapper \cite{Sigelman2010} was the pioneering work in this space, Canopy \cite{Kaldor2017} processes traces in real-time, derives user-specified features, and outputs performance datasets that aggregate across billions of requests. Pivot Tracing \cite{Mace2018} gives users the ability to define traced metrics at runtime, even when crossing component or machine boundaries.

%% file: conclusion.tex
\section{Conclusion}\label{sec:conclusion}
In this work, we propose \toolName, an automated approach to diagnose performance issues in service-based systems. Our approach leverages a search-based algorithm to detect patterns in RPC execution time behaviors correlated with latency degradation of requests, namely Latency Degradation Patterns.
\toolName simultaneously search multiple patterns while optimizing precision, recall and latency dissimilarity, and it uses a heuristic algorithm to select the optimal pattern set from the set of non-dominated solutions.

Through an evaluation of \toolName on 700 datasets with different combinations of LDPs from two case study systems, we demonstrated that \toolName provides (very often) better and (always) more stable effectiveness than three state-of-the-art techniques and two general purpose clustering algorithms.
We also demonstrate that, contrarily to other techniques, the effectiveness of \toolName is affected neither by the proximity of latency distributions related to different patterns, nor by execution time variations in non-critical RPCs.
Finally, we demonstrate that \toolName is more efficient than the second and the third most effective baseline techniques when a high number of requests is involved.

As future work, we plan to put effort on the improvement of the efficiency of our approach, and to extend the scope of our approach to the detection of patterns in categorical trace attributes, especially those with high-cardinality \cite{Bansal2019}.